\documentclass[usenatbib,useAMS,letterpaper]{mn2e}
\usepackage{graphicx}
\usepackage{epstopdf}
\usepackage{ctable}
\usepackage{url}
\usepackage{times}

\newcommand{\lsun}{~\mathrm{L_{\odot}}}

\newcommand{\lir}{L_{\rm IR}}

\newcommand{\msun}{~\mathrm{M_{\odot}}}
\newcommand{\msunperyr}{~\mathrm{M_{\odot} {\rm ~yr}^{-1}}}
\newcommand{\sfr}{\dot{M}_{\star}}

\newcommand{\mgas}{M_{\rm gas}}
\newcommand{\mstar}{M_{\star}}
\newcommand{\mdust}{M_\mathrm{d}}
\newcommand{\mbar}{M_{\rm bar}}

\newcommand{\sunrise}{\textsc{Sunrise}~}

\newcommand{\gadgettwo}{\textsc{Gadget-2}~}
\newcommand{\arepo}{\textsc{Arepo}~}

\newcommand\plotone[1]
 {\centering \leavevmode \includegraphics[width={0.99\columnwidth}]{#1}}

 \newcommand\plottwo[2]{{%
 \typeout{Plottwo included the files #1 #2}
 \centering
\includegraphics[width={0.99\columnwidth}]{#1}%
\\
\includegraphics[width={0.99\columnwidth}]{#2}%
}}%
\newcommand{\acknowledgments}{\begin{small}\section*{Acknowledgments}\end{small}}

\title[SMG number counts, z distributions, \& the IMF]{Submillimetre galaxies in a hierarchical universe: number counts, redshift distribution,
and implications for the IMF}
\author[Hayward et al.]{
\parbox[t]{\textwidth}{
Christopher C. Hayward$^{1,2}$\thanks{E-mail: christopher.hayward@h-its.org},
Desika Narayanan$^3$\thanks{Bart J. Bok Fellow.},
Du\v{s}an Kere\v{s}$^{4}$, 
Patrik Jonsson$^2$\thanks{Present address: Space Exploration Technologies, 1 Rocket Road, Hawthorne, CA 90250, USA.},
Philip F. Hopkins$^5$, T.~J. Cox$^{6}$, \& Lars Hernquist$^{2}$
}
\vspace*{6pt} \\
$^{1}$Heidelberger Institut f\"ur Theoretische Studien, Schloss--Wolfsbrunnenweg 35, 69118 Heidelberg, Germany \\
$^2$Harvard--Smithsonian Center for Astrophysics, 60 Garden Street, Cambridge, MA 02138, USA \\
$^3$Steward Observatory, Department of Astronomy, University of Arizona, 933 North Cherry Avenue, Tucson, AZ 85721, USA \\
$^4$Department of Physics, Center for Astrophysics and Space Science, University of California at San Diego, 9500 Gilman Drive, La Jolla, CA 92093, USA \\
$^5$Department of Astronomy and Theoretical Astrophysics Center, University of California Berkeley, Berkeley, CA 94720, USA \\
$^6$Carnegie Observatories, 813 Santa Barbara Street, Pasadena, CA 91101, USA}

\begin{document}

\date{Accepted 2012 October 15. Received 2012 October 9; in original form 2012 September 10}

\pagerange{\pageref{firstpage}--\pageref{lastpage}} \pubyear{2012}

\maketitle

\label{firstpage}

\begin{abstract}
High-redshift submillimetre galaxies (SMGs) are some of the most rapidly star-forming galaxies in the
Universe. Historically, galaxy formation models have had difficulty explaining the observed number counts of SMGs.
We combine a semi-empirical model with 3-D hydrodynamical simulations and 3-D dust radiative transfer
to predict the number counts of unlensed SMGs. Because the stellar mass functions, gas and dust masses, and
sizes of our galaxies are constrained to match observations, we can isolate uncertainties related to the dynamical evolution of galaxy mergers and the dust radiative
transfer. The  number counts and redshift distributions predicted by our model agree well with observations.
Isolated disc galaxies dominate the faint ($S_{1.1} \la 1$ mJy, or $S_{850} \la 2$ mJy) population. The brighter sources are a mix of merger-induced starbursts
and galaxy-pair SMGs; the latter subpopulation accounts for $\sim 30 - 50$ per cent of all SMGs at all $S_{1.1} \ga 0.5$ mJy ($S_{850} \ga 1$ mJy).
The mean redshifts are $\sim 3.0 - 3.5$, depending on the flux cut, and the brightest sources tend to be at higher redshifts.
Because the galaxy-pair SMGs will be resolved into multiple fainter sources
by ALMA, the bright ALMA counts should be as much as 2 times less than those observed using single-dish telescopes.
The agreement between our model, which uses a Kroupa IMF, and observations
suggests that the IMF in high-redshifts starbursts need not be top-heavy; if the IMF were
top-heavy, our model would over-predict the number counts. We conclude that the difficulty some models have
reproducing the observed SMG counts is likely indicative of more general problems -- such as an under-prediction of
the abundance of massive galaxies or a star formation rate--stellar mass relation normalisation lower than that observed --
rather than a problem specific to the SMG population.
\end{abstract}


\section{Introduction}

Submillimetre galaxies (SMGs; \citealt{Smail:1997,Barger:1998,Hughes:1998,Eales:1999}; see \citealt{Blain:2002} for a review)
are amongst the most luminous, rapidly star-forming galaxies
known, with luminosities in excess of $10^{12} \lsun$ and star formation rates (SFR) of order $\sim10^2-10^3$ $\msunperyr$
\citep[e.g.,][]{Kovacs:2006,Coppin:2008,Michalowski:2010masses,Michalowski:2012,Magnelli:2010,Magnelli:2012,Chapman:2010}.
They have stellar masses of $\sim 10^{11} \msun$, although recent estimates
\citep{Hainline:2011,Michalowski:2010masses,Michalowski:2012} differ by a factor of $\sim 6$, and typical
gas fractions of $\sim 40$ per cent (\citealt{Greve:2005,Tacconi:2006,Tacconi:2008}; but cf. \citealt*{Narayanan:2012gas_frac}).

The most luminous local galaxies, ultra-luminous infrared galaxies (ULIRGs, defined by $\lir > 10^{12} \lsun$), are almost exclusively
late-stage major mergers \citep[e.g.,][]{Lonsdale:2006} because
the strong tidal torques exerted by the galaxies upon one another when they are near coalescence cause significant gas inflows
and, consequently, bursts of star formation \citep[e.g.,][]{Hernquist:1989,Barnes:1991,Barnes:1996,Mihos:1996}.
Thus, it is natural to suppose
that SMGs, which are the most luminous, highly star-forming galaxies at high redshift, are also late-stage major mergers undergoing starbursts.
There is significant observational support for this picture
\citep[e.g.,][]{Ivison:2002,Ivison:2007,Ivison:2010,Chapman:2003, Neri:2003,Smail:2004,Swinbank:2004,Greve:2005,Tacconi:2006,Tacconi:2008,
Bouche:2007,Biggs:2008,Capak:2008,Younger:2008phys_scale,Younger:2010,Iono:2009,Engel:2010,Bothwell:2010,Bothwell:2012,Riechers:2011b,Riechers:2011a,Magnelli:2012}.
However, there may not be enough major mergers of galaxies of the required masses to account for the observed SMG abundances
\citep{Dave:2010}. Consequently, explaining the abundance of SMGs has proven to be a challenge for galaxy formation models.

Much observational effort has been invested to determine the number counts and redshift distribution of SMGs
\citep[e.g.,][]{Chapman:2005,Coppin:2006,Knudsen:2008,Chapin:2009,Weiss:2009,Austermann:2009,Austermann:2010,
Scott:2010,Zemcov:2010,Aretxaga:2011,Banerji:2011,Hatsukade:2011,Wardlow:2011,Roseboom:2012,Yun:2012}
because this information is key to relate the SMG population to their descendants and to understand SMGs in the context of
hierarchical galaxy formation models. Various authors have attempted to explain the observed abundance of SMGs using phenomenological models
\citep[e.g.,][]{Pearson:1996,Blain:1999hist_of_SF,Devriendt:2000,Lagache:2003,Negrello:2007,Bethermin:2012}, semi-analytic models
\citep[SAMs; e.g.,][]{Guiderdoni:1998,Blain:1999hier_mod,Granato:2000,Kaviani:2003,Granato:2004,Baugh:2005,Fontanot:2007,Fontanot:2010,Lacey:2008,
Lacey:2010,Swinbank:2008,LoFaro:2009,Gonzalez:2011}, and cosmological hydrodynamical simulations \citep{Fardal:2001,Dekel:2009nature,Dave:2010,
Shimizu:2012}.

\citet{Granato:2000} presented one of the first SAMs
to self-consistently calculate dust absorption and emission by coupling the {\sc Galform} SAM \citep{Cole:2000} with the {\sc Grasil} spectrophotometric
code \citep{Silva:1998}. This was a significant advance over previous work, which effectively treated the dust temperature as a free parameter.
Self-consistently computing dust temperatures made matching the submm counts significantly more difficult:
the submm counts predicted by the \citeauthor{Granato:2000} model were a factor of $\sim 20-30$ less than those observed \citep{Baugh:2005,Swinbank:2008}.

The work of \citet[][hereafter B05]{Baugh:2005} has attracted significant attention to the field because of its claim that a flat IMF is
necessary to reproduce the properties of the SMG population, which we will discuss in detail here.
B05 set out to modify the \citet{Granato:2000} model so that it would reproduce the properties of both $z \sim 2$ SMGs and
Lyman-break galaxies (LBGs) while also matching the observed $z = 0$ optical and IR luminosity functions.
Adopting a flat IMF\footnote{Specifically, the IMF they use is $dn/d\log M = $ constant for the mass range $0.15 < M < 125 \msun$.
The \citet{Kroupa:2001} IMF has $dn/d\log M \propto M^{-1.3}$ for $M > 1 \msun$, so the difference between the B05 IMF
and that observed locally is considerable.} in starbursts rather than the \citet{Kennicutt:1983} IMF used in \citet{Granato:2000} was the key change that
enabled B05 to match the observed SMG counts and redshift distribution while still reproducing the local $K$-band luminosity function.
A more top-heavy IMF results in both more luminosity emitted and more dust produced
per unit SFR; consequently, the submm flux per unit SFR is increased significantly (see B05 and \citealt{Hayward:2011smg_selection}, hereafter H11, for details).
The B05 modifications increased the $S_{850}$ per unit SFR for starbursts by a factor of $\sim 5$ (G.-L. Granato, private communication),
which caused starbursts to account for a factor of $\sim 10^3$ times more sources at $S_{850} = 3$ mJy than in \citet{Granato:2000}. As a result,
in the B05 model, ongoing starbursts dominate the counts for $0.1 \la S_{850} \la 30$ mJy. Interestingly,
these starbursts are triggered predominantly by minor mergers \citep[B05;][]{Gonzalez:2011}.
\citet{Swinbank:2008} present a detailed comparison of the properties of SMGs in the B05 model with those of observed
SMGs. The far-IR SEDs, velocity dispersions, and halo masses (see also
\citealt{Almeida:2011}) are in good agreement; however, recent observations suggest that the typical redshift of SMGs may be higher than predicted by the
B05 model and, contrary to the B05 prediction, brighter SMGs tend to be at higher redshifts \citep{Yun:2012,Smolcic:2012}.
Furthermore, the rest-frame $K$-band fluxes of the B05 SMGs are a factor of $\sim 10$ lower than observed; the most plausible
explanation is that the masses of the SMGs in the B05 SAM are too low \citep{Swinbank:2008}, but the top-heavy IMF
in starbursts used by B05 makes a direct comparison of masses difficult. These disagreements
are reasons it is worthwhile to explore alternative SMG models.

\citet{Granato:2004} presented an alternate model, based on spheroid formation via monolithic collapse, that predicts submm counts in good agreement
with those observed and reproduces the evolution of the $K$-band luminosity function. However, the typical redshift they predict for SMGs is lower
than recent observational constraints \citep{Yun:2012,Smolcic:2012}, and this model does not include halo or galaxy mergers.

The \citet{Fontanot:2007} model predicts SMG number counts in reasonable agreement with those observed using a standard IMF;
they argue that the crucial difference between their model and that of B05 is the cooling
model used (see also \citealt{Viola:2008} and \citealt{DeLucia:2010}). However, their SMG redshift distribution
peaks at a lower redshift than the redshift distribution derived from recent observations \citep{Yun:2012,Smolcic:2012}. Furthermore,
the \citet{Fontanot:2007} model produces an overabundance of bright galaxies at $z < 1$. However, this problem has been
significantly reduced in the latest version of the model \citep{LoFaro:2009}, which provides a significantly better fit to the galaxy stellar mass
function at low redshift \citep{Fontanot:2009}. In the revised model, the submm counts are reduced by $\sim 0.5$ dex, primarily because of the change in the
IMF from Salpeter to Chabrier, but the redshift distribution is unaffected. Thus, the submm counts for the new model are consistent with the data
for $S_{850} \la 3$ mJy, but they are slightly less than the observed counts at higher fluxes \citep{Fontanot:2010}. No fine-tuning of the dust
parameters has been performed for the new model.

A compelling reason to model the SMG population in an alternative manner is to test whether a top-heavy
IMF is required to explain the observed SMG counts.
Matching the submm counts is the primary reason B05 needed to adopt a flat IMF in starbursts.\footnote{Recent observations suggest
the number counts are as much as a factor of 2 lower than those used by B05. Thus, if \citeauthor{Baugh:2005}
were to attempt to match the revised counts, the required IMF variation would be more modest.}
Using the same model,
\citet{Lacey:2008} show that the flat IMF is necessary to reproduce the evolution of the mid-IR luminosity function.
Others \citep[e.g.,][]{Guiderdoni:1998,Blain:1999hier_mod,Dave:2010} have also suggested that the IMF may be top-heavy in SMGs,
but they do not necessarily require variation as extreme as that assumed in B05.
However, the use of a flat IMF in starbursts remains controversial: though there are some theoretical reasons to believe the IMF is
more top-heavy in starbursts \citep[e.g.,][]{Larson:1998,Larson:2005,Elmegreen:2003,Elmegreen:2004,Hopkins:2012IMF,Narayanan:2012IMF},
there is to date no clear evidence for strong, systematic IMF variation in any
environment (\citealt{Bastian:2010} and references therein). Furthermore, in local massive ellipticals, the probable descendants of SMGs,
the IMF may actually be {\em bottom-heavy} \citep[e.g.,][]{vanDokkum:2010,vanDokkum:2011,Conroy:2012,Hopkins:2012IMF}.
Finally, the large parameter space of SAMs can yield multiple qualitatively distinct solutions that satisfy all
observational constraints \citep{Bower:2010,Lu:2011a,Lu:2012}, so it is possible that a top-heavy IMF in starbursts is not required to match
the observed submm counts even though it enables B05 to match the submm counts.\footnote{However, whether a unique solution, if any solution at
all, can be found when \emph{all} possible observational constraints are included is an open question.} Thus, it is useful to explore other
methods to predict the submm counts and to determine whether a match can be achieved without using a top-heavy IMF.

Another reason to model the SMG population is to investigate whether, like local ULIRGs, they are predominantly merger-induced starbursts.
Some observational evidence suggests that some SMGs may be early-stage mergers in which the discs have not yet coalesced and are
likely not undergoing starbursts \citep[e.g.,][]{Tacconi:2006,Tacconi:2008,Engel:2010,Bothwell:2010,Riechers:2011b,Riechers:2011a},
and massive isolated disc galaxies may also contribute to the population \citep[e.g.,][]{Bothwell:2010,Carilli:2010,Ricciardelli:2010,Targett:2011,Targett:2012}
In H11 and H12, we suggested that the inefficient scaling of (sub)mm flux with SFR in starbursts results in an SMG population that is heterogeneous: major mergers
contribute both as coalescence-induced starbursts and during the pre-coalescence infall stage, when the merging discs are
blended into one (sub)mm source because of the large ($\sim 15$'', or $\sim 130$ kpc at $z \sim 2 - 3$) beams of the single-dish (sub)mm
telescopes used to perform large SMG surveys. We refer to the latter subpopulation as `galaxy-pair SMGs'.
Similarly, compact groups may be blended into one source and can thus also contribute to the population. 
The most massive, highly star-forming isolated discs may also contribute (H11). Finally, it has been observationally
demonstrated that there is a contribution from physically unrelated galaxies blended into one source \citep{Wang:2011}. It is becoming
increasingly clear that the SMG population is  a mix of various classes of sources; if one subpopulation does not dominate the
population, physically interpreting observations of SMGs is significantly more complicated than previously assumed.

In previous work, we demonstrated that major mergers can reproduce the observed 850-\micron ~fluxes and typical
SED \citep{Narayanan:2010smg}; CO spatial extents, line-widths, and excitation ladders \citep{Narayanan:2009};
stellar masses (\citealt{Narayanan:2010smg}; H11; \citealt{Michalowski:2012}); $\lir$--effective dust temperature relation, IR excess,
and star formation efficiency \citep[][hereafter H12]{Hayward:2012smg_bimodality} observed for SMGs. 
In this work, we present a novel method to predict the (sub)mm counts from mergers and quiescently star-forming disc galaxies.
We utilise a combination of 3-D hydrodynamical simulations, on which we perform radiative transfer in post-processing
to calculate the UV-to-mm SEDs, and a semi-empirical model (SEM) of galaxy formation -- both of which have been extensively validated
in previous work -- to predict the number counts
and redshift distribution of SMGs in our model. We  address four primary questions: 1. Can our model reproduce the observed
SMG number counts and redshift distribution? 2. What are the relative contributions of merger-induced starbursts, galaxy pairs,
and isolated discs to the SMG population? 3. How will the number counts and redshift distribution of ALMA-detected SMGs
differ from those determined using single-dish surveys? 4. Does the SMG population provide evidence for a top-heavy IMF
in high-redshift starbursts?

The remainder of this paper is organised as follows: In Section \ref{S:sim_methods}, we present the details of the simulations we use to determine
the time evolution of galaxy mergers and to translate physical properties of model galaxies into observed-frame (sub)mm
flux densities. In Section \ref{S:predicting_counts}, we discuss how we combine the simulations with a SEM
to predict the (sub)mm counts for merger-induced starburst SMGs (Section \ref{S:predicting_counts_starbursts}) and
isolated disc and galaxy-pair SMGs (Section \ref{S:predicting_counts_iso_and_pair}). In Section \ref{S:results},
we present the predicted counts and redshift distribution of our model SMGs and the relative contribution of each subpopulation.
We discuss implications for the IMF, compare to previous work, and highlight some uncertainties in and limitations of our model in
Section \ref{S:discussion}, and we conclude in Section \ref{S:conclusions}.

\section{Summary of the model}

Predicting SMG counts requires three main ingredients: 1.  Because SFR and dust mass are the most important properties for predicting
the (sub)mm flux of a galaxy (H11), one must model the time evolution of those properties for individual discs and mergers.
2. The physical properties of the model galaxies must be used to determine the observed-frame (sub)mm flux density of those
galaxies. 3. One must put the model galaxies in a cosmological context. Ideally, one could combine a cosmological hydrodynamical simulation with dust radiative
transfer to self-consistently predict the (sub)mm counts. However, this is currently infeasible because the resolution required for
reliable radiative transfer calculations cannot be achieved for a cosmological simulation large enough to contain a significant number
of SMGs (see, e.g., \citealt{Dave:2010}).\footnote{Recently, \citet{Shimizu:2012} predicted SMG number counts using a cosmological simulation
with a self-consistent model to calculate the far-IR emission. However, their model assumes a single dust temperature and neglects
dust self-absorption, so the submm fluxes predicted by their model may be significantly ($\sim 0.3 - 0.5$ dex) greater those calculated
using full 3-D dust radiative transfer (H11). An investigation of the effects of this uncertainty on the predicted counts is underway.}

\ctable[
	caption = 		{Summary of methods\label{tab:methods}},
				center,
				star
]{llll}{}{
																				\FL
Ingredient & Isolated discs & Early-stage mergers & Merger-induced starbursts					\ML
Physical properties & semi-empirical & semi-empirical & simulations							\NN
(Sub)mm flux density & H11 relations & H11 relations & simulations								\NN
Cosmological context & observed SMF &  merger rates from SEM & merger rates from SEM			\NN
& & + duty cycle from sims & + duty cycle from sims											\LL
}

Here, we develop a novel method to predict the number counts and redshift distribution of high-z SMGs while still resolving the dusty
ISM on scales of $\sim 200$ pc.
We predict (sub)mm counts using a combination of a simple SEM
\citep{Hopkins:2008red_Es,Hopkins:2008cosm_frame1} and idealised high-resolution simulations
of galaxy mergers. The method we use for each of the three model ingredients depends on the subpopulation being modelled. The physical properties of the
isolated disc galaxies and early-stage mergers are determined using the SEM. For the late-stage mergers, hydrodynamical simulations are used
because of the complexity of modelling a merger's evolution. Dust radiative
transfer is performed on the hydrodynamical simulations to translate the physical properties into observed (sub)mm flux density.
For the isolated discs and early-stage mergers,
fitting functions derived from the simulations are used, whereas for the late-stage mergers, the (sub)mm light curves are taken directly from the
simulations. Finally, the isolated galaxies are put in a cosmological context using an observed stellar mass function (SMF). For the mergers,
merger rates from the SEM and duty cycles from the simulations are used. The methods are summarised in Table
\ref{tab:methods}, and each component of the model is discussed in detail below.

We emphasize that we do not attempt to model the SMG population in an ab initio manner. Instead, we
construct our model so that the SMF, gas fractions, and metallicities are consistent with observations.
This will enable us to test whether, given a demographically accurate galaxy population, we are able to reproduce the SMG counts and redshift distribution.
If we are not able to reproduce the counts and redshift distribution, then our simulations or radiative transfer calculations must be incomplete. If we can
reproduce the counts and redshift distribution, then it is possible that the failure of some SAMs
and cosmological simulations to reproduce the SMG counts may be indicative of a more general problem with those models (e.g.,
a general under-prediction of the abundances of massive galaxies) rather than a problem specific to the SMG population.

In the next two sections, we describe our model in detail. Readers whom are uninterested in the details of the methodology
may wish to skip to Section \ref{S:results}.

\section{Simulation methodology} \label{S:sim_methods}

\subsection{Hydrodynamical simulations}

We have performed a suite of simulations of isolated and merging disc galaxies with \gadgettwo
\citep{Springel:2001gadget, Springel:2005gadget}, a TreeSPH \citep{Hernquist:1989treesph} code that computes gravitational interactions via a
hierarchical tree method \citep{Barnes:1986} and gas dynamics via smoothed-particle hydrodynamics
\citep[SPH;][]{Lucy:1977,Gingold:1977,Springel:2010}.\footnote{Recently, some authors
\citep{Agertz:2007,Springel:2010arepo,Bauer:2012,Sijacki:2012} have noted several significant flaws inherent in the traditional formulation of SPH,
including the artificial suppression of fluid instabilities, artificial damping of turbulent eddies in the subsonic regime, and a lack of efficient gas stripping of infalling structures.
Consequently, the results of cosmological simulations performed using \gadgettwo can differ significantly from those performed with the more accurate
moving-mesh code \arepo \citep{Springel:2010arepo}  even when the physics included in the codes is identical \citep{Vogelsberger:2011,Keres:2012,Torrey:2011}.
Fortunately, a comparison of idealised merger simulations run with \gadgettwo and \arepo suggests that these issues do not significantly alter the
global properties (e.g., star formation histories) of the mergers (Hayward et al., in preparation), so the results presented here should be robust to these numerical issues.}
It explicitly conserves both energy and entropy when appropriate \citep{Springel:2002}. Beyond the core gravitational and gas physics,
the version of \gadgettwo we use includes radiative heating and cooling \citep{Katz:1996}. Star formation is implemented
using a volume-density-dependent Kennicutt-Schmidt (KS) law \citep{Schmidt:1959,Kennicutt:1998}, $\rho_{\rm SFR} \propto
\rho_{\rm gas}^{N}$, with a low-density cutoff. We use $N = 1.5$, which reproduces the global KS law and is consistent with
observations of high-redshift disc galaxies (\citealt{Krumholz:2007KS,Narayanan:2008CO_SFR,Narayanan:2011ks};
but see \citealt{Narayanan:2012X_CO_II}).

Furthermore, our simulations include a two-phase sub-resolution model for the interstellar medium \citep[ISM;][]{Springel:2003}
in which cold dense clouds are in pressure equilibrium with a diffuse hot medium. The division of mass, energy, and
entropy between the two phases is affected by star formation, radiative heating and cooling, and supernova feedback,
which heats the diffuse phase and evaporates the cold clouds \citep{Cox:2006feedback}.
Metal enrichment is calculated assuming each particle behaves as a closed box the yield appropriate for a \citet{Kroupa:2001} IMF.
The simulations also include the \citet{Springel:2005feedback}
model for feedback from active galactic nuclei (AGN), in which black hole (BH) sink particles, initialised with mass $10^5 h^{-1} \msun$, undergo
Eddington-limited Bondi-Hoyle accretion \citep{Hoyle:1939,Bondi:1944,Bondi:1952}. They deposit $5$ per cent of their luminosity ($L = 0.1\dot{m}c^2$, where
$\dot{m}$ is the mass accretion rate and $c$ is the speed of light) to the surrounding ISM. This choice is made so that the
normalisation of the $M_{\rm BH}-\sigma$ relation is recovered \citep{DiMatteo:2005}. Note that our results do not depend crucially on the
implementation of BH accretion and feedback for two reasons: 1. the AGN typically do not dominate (but can still contribute significantly
to) the luminosity of our model SMGs because the SEDs during the phase of strong AGN activity tend to be hotter than during the starburst
phase (e.g., \citealt{Younger:2009}, Snyder et al., in preparation), so the mergers are typically not SMGs during the AGN-dominated phase.
2. Even in the absence of AGN feedback, the SFR decreases sharply after the
starburst simply because the majority of the cold gas is consumed in the starburst.

\ctable[
	caption =		{Progenitor disc galaxy properties\label{tab:disc_properties}},
				center,
				doinside=\small,
				star,
				notespar
]{lcccccc}{
	\tnote[a]{Virial velocity.}
	\tnote[b]{Virial mass.}
	\tnote[c]{Halo concentration.}
	\tnote[d]{Initial stellar mass.}
	\tnote[e]{Initial gas mass.}
	\tnote[f]{Disc scale length.}
}{
																																\FL
		& $V_{200}$\tmark[a] 	& $M_{200}$\tmark[b] 		& c\tmark[c]	& $M_{\star, {\rm init}}$\tmark[d] 	& $M_{\rm gas, init}$\tmark[e]
	& $R_{\rm d}$\tmark[f]	\NN
{Name} 	& {(km s$^{-1}$)} 		& {($10^{12} h^{-1} \msun$)} 	& 			& {($10^{10} h^{-1} \msun$)} 		& {($10^{10} h^{-1} \msun$)}
	& ($h^{-1}$ kpc)		\ML
b6 		& 500				& 6.4				 		& 2.3			& 5.3 						& 22							& 3.3		\NN
b5.5 		& 400				& 3.3				 		& 2.5			& 2.7 						& 11							& 2.6		\NN
b5 		& 320				& 1.7				 		& 2.8			& 1.4 						& 5.6							& 2.0		\NN
b4 		& 260				& 0.60			 		& 3.2			& 0.49 						& 2.0							& 1.7		\LL
}

Each disc galaxy is composed of a dark matter halo with a \citet{Hernquist:1990} profile and an exponential gas and stellar disc
in which gas initially accounts for 80 per cent of the total baryonic mass. At merger coalescence, the baryonic gas fractions are typically
20-30 per cent, which is consistent with the estimates of \citet{Narayanan:2012gas_frac}.
The mass of the baryonic component is 4 per cent of the total.
The galaxies are scaled to $z = 3$ following the method described in \citet{Robertson:2006}. Dark matter particles have gravitational softening
lengths of $200h^{-1}$ pc, whereas gas and star particles have $100h^{-1}$ pc.
We use $6 \times 10^4$ dark matter, $4 \times 10^4$ stellar, $4 \times 10^4$ gas, and 1 BH particle per disc galaxy.
The detailed properties of the progenitor galaxies are given in Table \ref{tab:disc_properties}.
Note that we have chosen galaxy masses such that most of the mergers, based upon
our simulations, will contribute to the bright SMG population (i.e., at some time during the simulation they have observed 850-\micron
~flux density $S_{850} > 3$ mJy). More massive galaxies will also contribute but are increasingly more rare, so our simulations
should be representative of all but the brightest, rarest SMGs \citep{Michalowski:2012}. Note also that we have included some slightly lower mass
mergers for completeness.

\ctable[
	caption = {Merger parameters \label{tab:merger_properties}},
	center,
	notespar%,
	%star
]{lcccc}{
	\tnote[a]{Baryonic mass ratio $M_{\mathrm{b,secondary}}/M_{\mathrm{b,primary}}$.}
	\tnote[b]{Pericentric-passage distance.}
	\tnote[c]{Initial separation of the discs.}
}{
																\FL
		& 				& $R_{\rm peri}$\tmark[b] 	& $R_{\rm init}$\tmark[c]	\NN
Name 	& $\mu$\tmark[a] 	& ($h^{-1}$ kpc) 		& ($h^{-1}$ kpc)		\ML
b6b6 	& 1 		& 6.7 			& 70				\NN
b6b5.5 	& 0.52 	& 6.7 			& 70				\NN
b6b5 	& 0.26 	& 6.7 			& 70				\NN
b6b4 	& 0.09 	& 6.7 			& 70				\NN
b5.5b5.5 	& 1 		& 5.3 			& 57				\NN
b5b5 	& 1 		& 4.0 			& 44				\LL
}

We simulate each disc galaxy listed in Table \ref{tab:disc_properties} in isolation for $1.5h^{-1}$ Gyr and use these isolated disc simulations as part of our simulation suite.
Our suite also includes a number of simulations of major and minor galaxy mergers.
For the merger simulations, two of the progenitor disc galaxies are placed on parabolic orbits (which are motivated by cosmological simulations;
\citealt{Benson:2005,Khochfar:2006}) with initial separation $R_{\rm init}
= 5R_{200}/8$ and pericentric-passage distance equal to twice the disc scale length, $R_{\rm peri} = 2 R_d$ \citep{Robertson:2006}.
The evolution of the system is followed for $1.5h^{-1}$ Gyr, which is sufficient time for the galaxies to coalescence and for significant star formation
and AGN activity to cease.
The details of the merger simulations are given in Table \ref{tab:merger_properties}. For each combination of progenitor discs in Table \ref{tab:merger_properties},
we simulate a subset of the i-p orbits of \citet{Cox:2006}. Specifically, we use the i-p orbits for the major mergers (b6b6, b5.5b5.5, and b5b5) and the i and j
orbits for the unequal-mass mergers (b6b5.5, b6b5, and b6b4) because the latter have shorter duty cycles and the variation in duty cycles amongst orbits
is not a primary source of uncertainty. Consequently, we use a total of 34 \gadgettwo simulations.

\subsection{Dust radiative transfer}

In post-processing, we use the 3-D Monte Carlo radiative transfer code
\sunrise\footnote{\sunrise is publicly available at \url{http://code.google.com/p/ sunrise/}.}
to calculate the UV-to-mm SEDs of the simulated galaxies.
We have previously simulated galaxies with colours/SEDs consistent with
local SINGS \citep{Kennicutt:2003,Dale:2007} galaxies \citep*{Jonsson:2010sunrise}; local ULIRGs \citep{Younger:2009};
massive, quiescent, compact $z \sim 2$ galaxies \citep{Wuyts:2009b,Wuyts:2010}; 
24 \micron-selected galaxies \citep{Narayanan:2010dog}; K+A/post-starburst galaxies \citep{Snyder:2011}; and
extended UV discs \citep{Bush:2010}, among other populations, so we are confident that \sunrise can be used
to model the high-z SMG population. As discussed above, previous work has demonstrated that many properties of
our simulated SMGs agree with observations \citep{Narayanan:2009,Narayanan:2010smg,Hayward:2011smg_selection,Hayward:2012smg_bimodality},
but we have yet to put our simulated SMGs in a cosmological context.
We briefly review the details of \sunrise here, but we refer the reader to
\citet{Jonsson:2006}, \citet{Jonsson:2010sunrise}, and \citet*{Jonsson:2010gpu} for full details of the \sunrise code.

{\sc Sunrise} uses the output of the \gadgettwo simulations to specify the details of the radiative transfer problem to be solved,
specifically the input radiation field and dust geometry.
The star and BH particles from the \gadgettwo simulations are used as sources of emission. Star particles are
assigned {\sc Starburst99} \citep{Leitherer:1999} SEDs according to their ages and metallicities. We conservatively
use the \citet{Kroupa:2001} IMF when calculating the simple stellar population SED templates.
Star particles present at the start of the \gadgettwo simulation are assigned ages by
assuming that their stellar mass was formed at a constant rate equal to the star formation rate of the
initial snapshot, but the results are insensitive to this choice because we discard the early snapshots and the
stars present at the start of the simulation account for a small fraction of the luminosity at later times.
The initial gas and stellar metallicities are $Z = 0.015$ ($\sim Z_{\odot}$; \citealt{Asplund:2009}). We have chosen this value so that the
starbursts lie roughly on the observed mass-metallicity relation; however, the results are fairly robust
to this choice because a factor of 2 change in dust mass changes the (sub)mm flux by only $\sim 50$ per cent
because (sub)mm flux scales approximately as $\mdust^{0.6}$ (Equations \ref{eq:sfr_fitting_function_850}
and \ref{eq:sfr_fitting_function_11}). BH particles are assigned luminosity-dependent templates derived from observations of
un-reddened quasars \citep{Hopkins:2007}, where the luminosity is determined using the accretion rate
from the \gadgettwo simulations as described above.

The dust distribution is determined by projecting the total gas-phase metal density in the \gadgettwo
simulations on to a 3-D adaptive mesh refinement grid assuming a dust-to-metal ratio of 0.4 \citep{Dwek:1998,James:2002}.
We have used a maximum refinement level of 10, which results in a minimum cell size of 55$h^{-1}$ pc. This refinement level
is sufficient to ensure the SEDs are converged to within a few per cent because the structure present in the \gadgettwo
simulations is sufficiently resolved; if the resolution of the \gadgettwo simulations were finer, the radiative transfer would require correspondingly smaller
cell sizes. Note that we assume the ISM is smooth on scales below the \gadgettwo resolution and do not make use of the
\citet{Groves:2008} sub-resolution photodissociation region model. The details of, motivation for, and implications of this choice
are discussed in sections 2.2.1 and 4.6 of H11. We assume the dust has properties given by the Milky Way $R=3.1$ dust model of
\citet{Weingartner:2001} as updated by \citet{Draine:2007}. The (sub)mm fluxes are similar if the LMC or SMC dust models
are used.

Once the star and BH particles are assigned SEDs and the dust density field is specified, \sunrise
performs the radiative transfer using a Monte Carlo approach by emitting photon packets that
are scattered and absorbed by dust as they propagate through the ISM. The energy absorbed by dust
is re-radiated in the IR. Dust temperatures, which depend on both grain size and the local radiation field,
are calculated by assuming the dust is in thermal equilibrium. The ISM of our simulated galaxies can often
be optically thick at IR wavelengths, so \sunrise calculates the effects of dust self-absorption using an
iterative method. This is crucial for ensuring accurate dust temperatures.

The \sunrise calculation yields spatially resolved SEDs (analogous to integral field unit spectrograph data)
of the simulated galaxies viewed from different viewing angles. Here, we use 7 cameras distributed
isotropically in solid angle. We use the SCUBA-2 850-$\micron$, AzTEC 1.1-mm,
and ALMA Bands 6 and 7 filter response curves to calculate the (sub)mm flux densities.
Depending on the mass and the SMG flux cut, the simulations are selected as SMGs
for $\sim 0 - 80$ snapshots, and each is viewed from 7 viewing angles. Consequently,
we have a sample of $\sim 3.7 \times 10^4$ distinct synthetic SMG SEDs that we use to
derive fitting functions for the isolated disc and galaxy-pair SMGs and duty cycles for the starburst SMGs.

\section{Predicting (sub)mm number counts} \label{S:predicting_counts}

To calculate the total SMG number counts predicted by our model, we must account for all subpopulations, including
the infall-stage galaxy-pair SMGs discussed in H11 and H12, late-stage merger-induced starbursts, and isolated discs.
To calculate the counts for the two subpopulations associated with mergers, we must combine the duty cycles of the mergers
[the time the merger has (sub)mm flux greater than some flux cut] with merger rates because the number density is calculated
by multiplying the duty cycles by the merger rates. For the isolated discs, we require the number density of a disc galaxy as a
function of its properties and the (sub)mm flux associated with that galaxy. We describe our methods for predicting the counts
of each subpopulation now.

\subsection{Late-stage merger-induced starbursts} \label{S:predicting_counts_starbursts}

To predict the number counts for the population of late-stage merger-induced starburst SMGs,
we combine merger rates -- which depend on mass, mass ratio, gas fraction, and
redshift -- from the SEM with (sub)mm light curves from our simulations.
We use the (sub)mm light curves from the simulations directly because it is difficult to analytically
model the dynamical evolution of the mergers, which can depend on the galaxy masses, merger mass ratio, progenitor
redshift, gas fraction, and orbital properties.
For the SMG subpopulation attributable to mergers, the number density of sources with flux density greater than $S_{\lambda}$
at redshift $z$ is
\begin{eqnarray} \label{eq:gen_merger_num_density}
n(>S_{\lambda},z) \nonumber &\equiv& \frac{dN(>S_{\lambda},z)}{dV} \nonumber \\
&=& \int \frac{dN}{dV dt d\log \mbar d\mu df_g}(\mbar, \mu, f_g, z) \nonumber \\
&\times& \tau(S_{\lambda}, \mbar, \mu, f_g, z) d\log \mbar d\mu df_g,
\end{eqnarray}
where $dN/dV dt d\log \mbar d\mu df_g(\mbar, \mu, f_g, z)$ is the number of mergers per comoving volume element
per unit time per dex baryonic mass per unit mass ratio per unit gas fraction,
which is a function of progenitor
baryonic mass $\mbar$, merger mass ratio $\mu$, gas fraction at merger $f_g$,
and redshift $z$, and $\tau(S_{\lambda}, \mbar, \mu, f_g, z)$ is the amount of time (duty cycle)
for which a merger with most-massive-progenitor baryonic mass
$\mbar$, mass ratio $\mu$, and gas fraction $f_g$ at redshift $z$ has flux density $>S_{\lambda}$.

\subsubsection{Duty cycles}

We calculate the duty cycles $\tau(S_{850})$ and $\tau(S_{1.1})$ for various $S_{850}$ and $S_{1.1}$ values
for the late-stage merger-induced starburst phase of our merger simulations. 
We neglect the dependence of duty cycle on gas fraction because sampling the range of initial gas fractions
in addition to masses, mass ratios, and orbits is computationally prohibitive. Instead, as described above, we initialise
the mergers with gas fraction $f_g = 0.8$ so that sufficient gas remains at merger coalescence.\footnote{Note, however,
that we calculate the starburst duty cycle using only the snapshots that sample the final starburst induced at merger coalescence,
so the gas fraction is typically less than 40 per cent. We treat the early-stage galaxy-pair SMG contribution
separately below.}

Similarly, because of computational limitations, we scale all initial disc galaxies to $z \sim 3$. We will see below that
all else being equal, the dependence of (sub)mm flux density on $z$ is small ($\la 0.13$ dex) for the redshift range of interest 
($z \sim 1-6$), so we assume the duty cycles are independent of redshift and place the mergers at $z = 3$ (which is
approximately the median redshift for our model SMGs) when calculating the duty cycles. Note, however, that the submm
duty cycles for the starbursts may differ for mergers with progenitor disc properties scaled to different redshifts, but our model does not
capture this effect. However, because most SMGs in our model have $z \sim 2-4$ (see below), this uncertainty should be
subdominant.

For each $S_{\lambda}$, we average the duty cycles for each set of models with identical $(\mbar,\mu)$ and then fit the resulting
$\tau(\mbar,\mu)$ surface with a second-degree polynomial in $\mbar$ and $\mu$ to estimate the duty cycle
for $(\mbar,\mu)$ values not explicitly sampled by our simulations.

\subsubsection{Merger rates}

\begin{figure}
\centering
\plotone{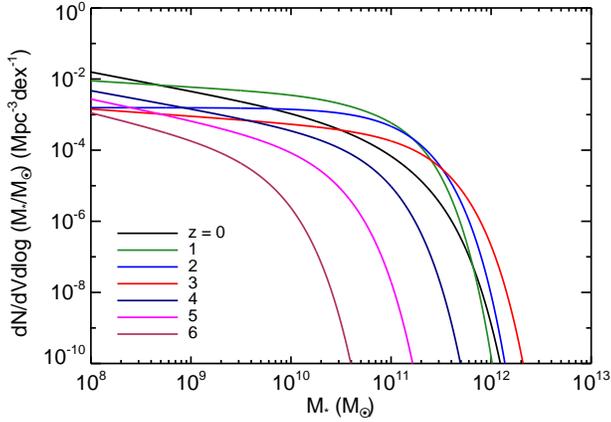}
\caption{Number density of disc galaxies, $dN/dV d\log (\mstar/\mathrm{M_{\odot}})$ (Mpc$^{-3}$ dex$^{-1}$), versus $\mstar (\mathrm{M_{\odot}})$
for integer redshifts in the range $z = 0-6$ for our composite SMF. For $z < 2$, we use the SMF for star-forming galaxies from \citet{Ilbert:2010}. For $2 \le z \le 3.75$, we use the
\citet{Marchesini:2009} SMF, and for $z > 3.75$, we use the \citet{Fontana:2006} parameterisation of the SMF.}
\label{fig:def_smf}
\end{figure}

The other ingredient needed to predict the counts for merger-induced starbursts is the merger rates. We use rates from the SEM
described in detail in \citet{Hopkins:2008cosm_frame1,Hopkins:2010merger_rates,Hopkins:2010merger_rate_uncertainties,Hopkins:2010IR_LF},
which we will briefly summarise here. The model starts with
a halo mass function that has been calibrated using high-resolution $N$-body simulations. Galaxies are assigned
to haloes using an observed SMF for star-forming galaxies and the halo occupation formalism \citep{Conroy:2009}.
We use a fiducial SMF that is a combination of multiple observed SMFs, in which each covers a subset of the total redshift range.
For $z < 2$, we use the SMF of star-forming galaxies from \citet{Ilbert:2010}.
For $2.0 \le z \le 3.75$, we use the SMF of \citet{Marchesini:2009} because their survey is amongst the widest and deepest available and because they
have performed a thorough analysis of the random and systematic uncertainties affecting the SMF determination. For $z > 3.75$, we use
the \citet{Fontana:2006} SMF parameterization; though they only constrained the SMF out to $z \sim 4$,
the extrapolation agrees reasonably well with the $4 < z < 7$ constraints from \citet{Gonzalez:2011}, so this extrapolation is not unreasonable.
Because the SMF at $z \ga 4$ is uncertain, it may be possible to constrain the SMF at those redshifts by using the observed SMG redshift distribution
and relative contributions of the subpopulations; we discuss
these possibilities below. The interested reader should see \citet{Hayward:2012thesis} for a detailed exploration of how the choice of SMF affects the predictions of our model.
We do not correct for the passive galaxy fraction beyond $z > 2$, but this fraction is relatively small at $z \sim 2$ and decreases rapidly at higher redshifts
\citep[e.g.,][]{Wuyts:2011b,Brammer:2011}.
Our composite SMF at integer redshifts in the range $z = 0-6$ is plotted in Fig. \ref{fig:def_smf}.
Finally, we use halo-halo merger rates from high-resolution $N$-body simulations and translate to galaxy-galaxy merger rates by
assuming the galaxies merge on a dynamical friction time-scale.

The merger rates, SMF, and observed gas fractions are all uncertain.
The merger rates are uncertain at the factor of $\sim 2$ level; the various sources of uncertainty and effects of modifying the model
assumptions are discussed in detail in \citet{Hopkins:2010merger_rates,Hopkins:2010merger_rate_uncertainties}.
At the redshifts of interest the random and systematic uncertainties in the SMF are comparable
to the total uncertainty in the merger rates.

\subsubsection{Predicted counts}

Using the above assumptions, Equation (\ref{eq:gen_merger_num_density}) becomes
\begin{eqnarray} \label{eq:merger_num_density}
n(>S_{\lambda},z) &=& \int \frac{dN}{dV dt d\log \mbar d\mu}(\mbar, \mu, z) \nonumber \\
&\times& \tau(S_{\lambda},\mbar,\mu) d\log \mbar d\mu.
\end{eqnarray}

To calculate the observable cumulative counts (deg$^{-2}$), we must multiply by $dV/d\Omega dz$, the comoving volume element in solid angle
$d\Omega$ and redshift interval $dz$, and integrate over redshift:
\begin{equation} \label{eq:counts}
\frac{dN(>S_{\lambda})}{d\Omega} = \int \frac{dN(>S_{\lambda},z)}{dV} \frac{dV}{d\Omega dz}(z) dz,
\end{equation}
where
\begin{equation}
\frac{dV}{d\Omega dz}(z) = \frac{c}{H_0} \frac{(1+z)^2 D_A^2(z)}{E(z)}.
\end{equation}
Here, $D_A(z)$ is the angular diameter distance at redshift $z$ and
$E(z) = \sqrt{\Omega_m(1+z)^3 + \Omega_k^2(1+z)^2 + \Omega_\Lambda}$.

\subsection{Isolated discs and early-stage mergers} \label{S:predicting_counts_iso_and_pair}

We treat the isolated discs and early-stage mergers, which are dominated by quiescent star formation, in a semi-empirical manner,
in which we assign galaxy properties based off observations. To calculate the observed (sub)mm flux densities using scaling relations similar to those
of H11, we must determine the SFR and dust mass of a galaxy as a function of stellar mass and redshift. We then
use SMF and merger rates to calculate the (sub)mm counts for these populations.

\subsubsection{Assigning galaxy properties}

Following \citet{Hopkins:2010merger_rates,Hopkins:2010IR_LF}, we assign gas fractions and sizes as a function of stellar mass
using observationally derived relations.
We present the relevant relations below, but we refer the reader to \citet{Hopkins:2010merger_rates,
Hopkins:2010merger_rate_uncertainties,Hopkins:2010IR_LF} for
full details, including the list of observations used to derive the relations and justifications for the forms used. \citet{Hopkins:2010IR_LF}
have shown that this model reproduces global constraints, such as the IR luminosity function at various redshifts and the star formation
history of the Universe, among others; these results support the application of the model in this work.

The baryonic gas fraction, $f_{\rm gas} = \mgas/(\mgas+\mstar)$, of a galaxy of stellar mass $\mstar$ and redshift $z$, as
determined from the observations listed in \citet{Hopkins:2010IR_LF}, is given by
Equation (1) of \citet{Hopkins:2010IR_LF},
\begin{eqnarray} \label{eq:f_gas}
f_{\rm gas}(\mstar | z = 0) & \equiv & f_0 \approx \frac{1}{1 + (\mstar/10^{9.15} \msun)^{0.4}}, \nonumber \\
f_{\rm gas}(\mstar, z) & = & f_0 \left[1 - \tau(z)\left(1-f_0^{3/2}\right)\right]^{-2/3},
\end{eqnarray}
where $\tau(z)$ is the fractional look-back time to redshift $z$.
At a given mass, galaxy gas fractions increase with redshift. At fixed redshift, they decrease with stellar mass.
Using $f_{\rm gas}(\mstar,z)$, we can calculate the gas mass as a function of $\mstar$ and $z$,
\begin{equation} \label{eq:M_gas}
M_{\rm gas}(\mstar, z) = \frac{f_{\rm gas}(\mstar, z)}{1 - f_{\rm gas}(\mstar, z)} \mstar.
\end{equation}

Similarly, we parameterize the radius of the gas disc as a function of mass and redshift using the observations listed in \citet{Hopkins:2010IR_LF}.
(Note that the stellar disc radii are significantly smaller.) The relation (Equation 2 of \citealt{Hopkins:2010IR_LF}; see also
\citealt{Somerville:2008disk_sizes}) is
\begin{eqnarray}
R_e(\mstar | z = 0) & \equiv & R_{0} = 5.28 {\rm ~kpc} \left(\frac{\mstar}{10^{10} \msun}\right)^{0.25}, \\
R_e(\mstar,z) & = & R_0 (1+z)^{-0.6}.
\end{eqnarray}

\begin{figure}
\centering
\plotone{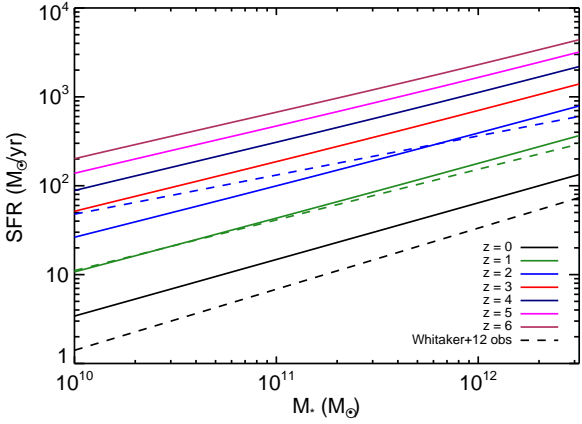}
\caption{SFR ($\msunperyr$) versus $\mstar (\msun)$ for model
disc galaxies at integer redshifts in the range $z = 0-6$ (solid lines) and from the 
observationally derived fitting function of \citet{Whitaker:2012} at $ z = 0, 1,$ and 2 (dashed lines).
The normalisation of the relation increases with redshift both because gas fractions are higher and
galaxies are more compact. The model agrees reasonably well with the observations except at $z \sim 0$,
but we shall see that the $z \sim 0$ contribution to our model SMG population is small. If the SFRs from \citet{Whitaker:2012}
were used instead of those calculated from Equation (\ref{eq:SFR}), the $z \sim 0$ contribution would be even less, so the discrepancy is unimportant.}
\label{fig:sfr-mstar}
\end{figure}

We assume the quiescently star-forming discs obey the KS relation \citep{Schmidt:1959,Kennicutt:1998},
\begin{equation}
\dot{\Sigma}_{\star} = 1.3 \times 10^{-4} \msunperyr {\rm kpc}^{-2} \left(\frac{\Sigma_{\rm gas}}{\msun {\rm pc}^{-2}}\right)^{n_K},
\end{equation}
where $\dot{\Sigma}_{\star}$ and $\Sigma_{\rm gas}$ are the SFR and gas surface densities, respectively and $n_K = 1.4$ \citep{Kennicutt:1998},
at all redshifts, as is supported by observations (e.g., \citealt{Narayanan:2008SFR_dense_gas,Daddi:2010,Genzel:2010,Narayanan:2011ks};
but c.f. \citealt{Narayanan:2012X_CO_II}). We normalise the relation assuming a \cite{Kroupa:2001} IMF.
Assuming $\Sigma_{\rm gas} \approx M_{\rm gas}/(\pi R_e^2)$ and $\dot{\Sigma}_{\star} \approx \dot{M}_{\star}/(\pi R_e^2)$,
where $\dot{M}_{\star}$ is the SFR, we find
\begin{eqnarray} \label{eq:SFR}
\sfr(\mstar,z) & = & 1.3 \left(\frac{10^4}{\pi}\right)^{n_K-1} \left(\frac{M_{\rm gas}(\mstar, z)}{10^{10} \msun}\right)^{n_K} \nonumber \\
& \times & \left(\frac{R_e(\mstar, z)}{\rm kpc} \right)^{-2(n_K-1)} \msunperyr,
\end{eqnarray}
which can be recast in terms of $\mstar$ rather than $\mgas$ using Equations (\ref{eq:f_gas}) and (\ref{eq:M_gas}).
Fig. \ref{fig:sfr-mstar} shows the SFR-$\mstar$ relation given by Equation (\ref{eq:SFR}) for integer redshifts in the range $z = 0-6$
and the observed relations from \citet{Whitaker:2012} for $z \sim 0$, 1, and 2. The agreement between our relation and those observed for $z \sim 1$ and 2
is reasonable for the masses ($\mstar \ga 10^{11} \msun$) relevant to the SMG population. The agreement is less good for $z \sim 0$, but, as we
shall see below, this is unimportant because the fraction of SMGs at $z \la 1$ is small. If the \citet{Whitaker:2012} SFRs were used instead
of those calculated from Equation (\ref{eq:SFR}), the $z \sim 0$ contribution would be even smaller.

In addition to the SFR, we require the dust mass to calculate the (sub)mm flux densities. To determine the dust mass, we must
know the gas-phase metallicity. Observations have demonstrated that metallicity increases with stellar mass;
this relationship has been constrained for redshifts $z \sim 0 - 3.5$ \citep{Tremonti:2004,Savaglio:2005,Erb:2006,Kewley:2008,Maiolino:2008}.
\citet{Maiolino:2008} parameterized the evolution of the mass-metallicity relation (MMR) with redshift using the form
\begin{eqnarray}
12 + \log ({\rm O/H}) &=& -0.0864 [\log \mstar - \log M_0(z)]^2 \nonumber \\
& + & K_0(z).
\end{eqnarray}
They determine the values of $\log M_0$ and $K_0$ at redshifts $z = $ 0.07, 0.7, 2.2, and 3.5 using the observations of \citet{Kewley:2008},
\citet{Savaglio:2005}, \citet{Erb:2006}, and their own work, respectively. To crudely capture the evolution of the MMR with redshift, we
fit the values of $\log M_0$ and $K_0$ given in Table 5 of \citet{Maiolino:2008} as power laws in $(1+z)$; the result is $\log M_0(z) \approx
11.07 (1+z)^{0.094}$ and $K_0(z) \approx 9.09 (1+z)^{-0.017}.$

Using $12 + \log({\rm O/H})_{\odot} = 8.69$ \citep{Asplund:2009}, we have
\begin{eqnarray}
\log({\rm O/H}) &-& \log({\rm O/H})_{\odot} = \nonumber \\
&-& 0.0864 \left[\log \mstar - 11.07 (1+z)^{0.94}\right]^2 \nonumber \\
&+& 9.09 (1+z)^{-0.017} - 8.69.
\end{eqnarray}
The solar metal fraction is $Z_{\odot} = 0.0142$ \citep{Asplund:2009}, so
\begin{equation}
Z(\mstar,z) = 0.0142 \left(10^{\log({\rm O/H}) - \log({\rm O/H})_{\odot}}\right).
\end{equation}
We assume the dust mass is proportional to the gas-phase metal mass, $M_d = \mgas Z f_{\rm dtm}$. Thus,
\begin{equation} \label{eq:M_dust}
M_d(\mstar,z) = \mstar \left(\frac{f_{\rm gas}(\mstar, z)}{1 - f_{\rm gas}(\mstar, z)}\right) \times Z(\mstar, z) f_{\rm dtm},
\end{equation}
where we use a dust-to-metal ratio $f_{\rm dtm} = 0.4$ \citep{Dwek:1998,James:2002}.

Motivated by equation (1) of H11, we fit the (sub)mm flux densities of our simulated galaxies as
power laws in SFR and $\mdust$.
We find that
\begin{eqnarray} \label{eq:sfr_fitting_function_850}
S_{850} &=& 0.81 {\rm ~mJy} \nonumber \\
&\times& \left(\frac{\sfr}{100 ~\msunperyr}\right)^{0.43} \left(\frac{\mdust}{10^8 \msun}\right)^{0.54} \\
S_{1.1} &=& 0.35 {\rm ~mJy} \nonumber \\
&\times& \left(\frac{\sfr}{100 ~\msunperyr}\right)^{0.41} \left(\frac{\mdust}{10^8 \msun}\right)^{0.56},\label{eq:sfr_fitting_function_11}
\end{eqnarray}
is accurate to within $0.13$ dex for $z \sim 1-6$. (The flux for galaxies at $z \la 0.5$ is underestimated significantly by these equations,
but such galaxies contribute little to the overall counts because of the smaller cosmological volume probed and the significantly
lower gas fractions and SFRs, so this underestimate is unimportant for our results.) The (sub)mm flux is insensitive to redshift in this redshift range because
as redshift increases, the decrease in flux caused by the increased luminosity distance is almost exactly cancelled by the increase in
flux caused by the rest-frame wavelength moving closer to the peak of the dust emission
(this effect is referred to as the negative K correction; see, e.g., \citealt{Blain:2002}).
By combining Equations (\ref{eq:SFR}) and (\ref{eq:M_dust}) with Equations (\ref{eq:sfr_fitting_function_850}) and (\ref{eq:sfr_fitting_function_11}),
we can calculate $S_{850}(\mstar, z)$ and $S_{1.1}(\mstar, z)$,
\begin{eqnarray} \label{eq:S_850_m_z}
S_{850} &=& 0.81 {\rm ~mJy} \nonumber \\
&\times& \left[0.013 \left(\frac{10^4}{\pi}\right)^{0.4} \left(\frac{\mgas}{10^{10}\msun}\right)^{1.4} \left(\frac{R_e}{\rm kpc}\right)^{-0.8} \right]^{0.43} \nonumber \\
&\times& \left[\left(\frac{\mgas}{10^8 \msun}\right) Z(\mstar, z) f_{\rm dtm}\right]^{0.54},\\
S_{1.1} &=& 0.35 {\rm ~mJy} \nonumber \\
&\times& \left[0.013 \left(\frac{10^4}{\pi}\right)^{0.4} \left(\frac{\mgas}{10^{10}\msun}\right)^{1.4} \left(\frac{R_e}{\rm kpc}\right)^{-0.8} \right]^{0.41}
\nonumber \\
&\times& \left[\left(\frac{\mgas}{10^8 \msun}\right) Z(\mstar, z) f_{\rm dtm}\right]^{0.56}, \label{eq:S_11_m_z}
\end{eqnarray}
where we can substitute the appropriate expressions for $\mgas$, $R_e$, and $Z$ to express $S_{850}$ and $S_{1.1}$ in terms of $\mstar$ and $z$ only.

\begin{figure}
%\epsscale{0.25}
\centering
\plottwo{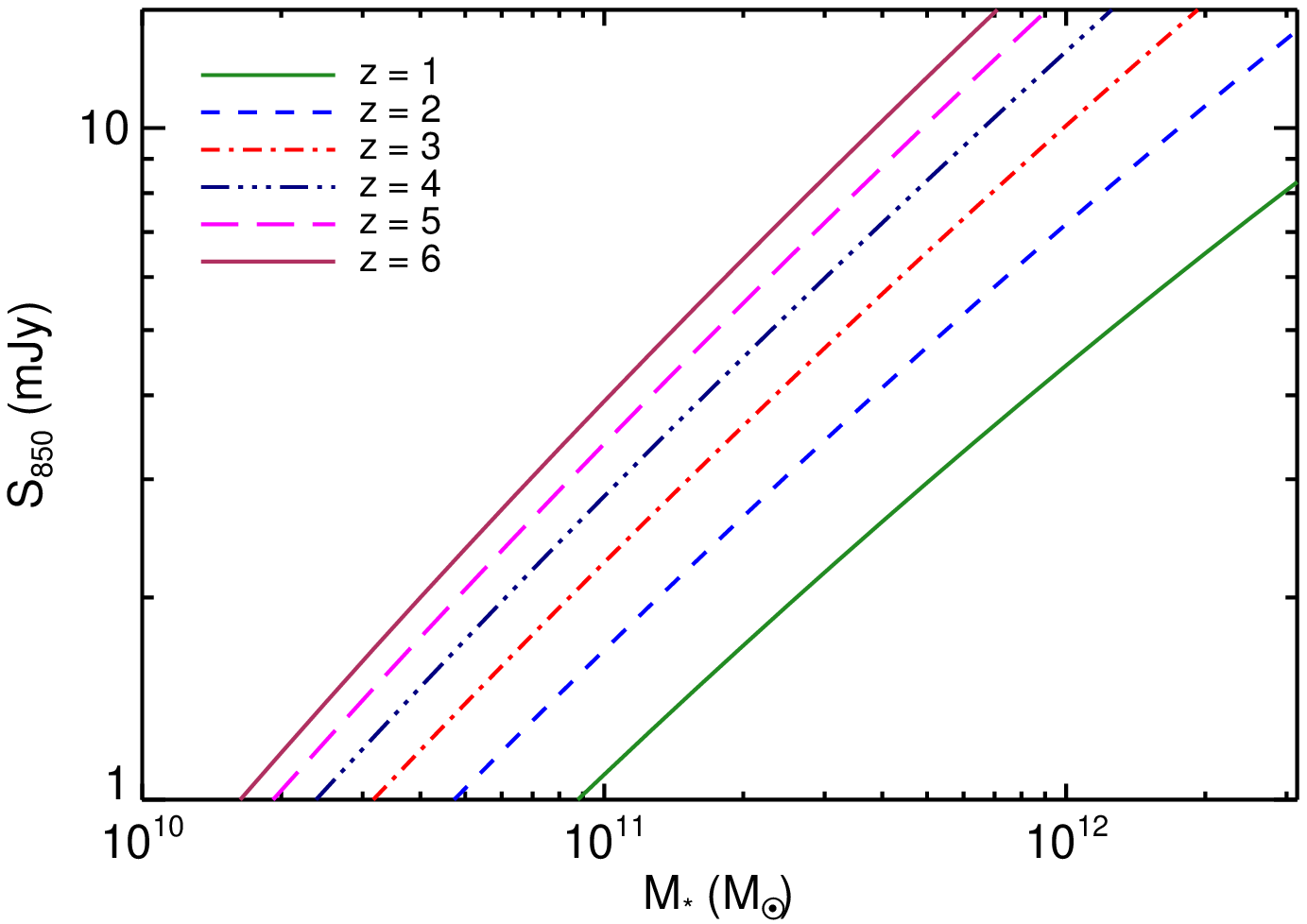}{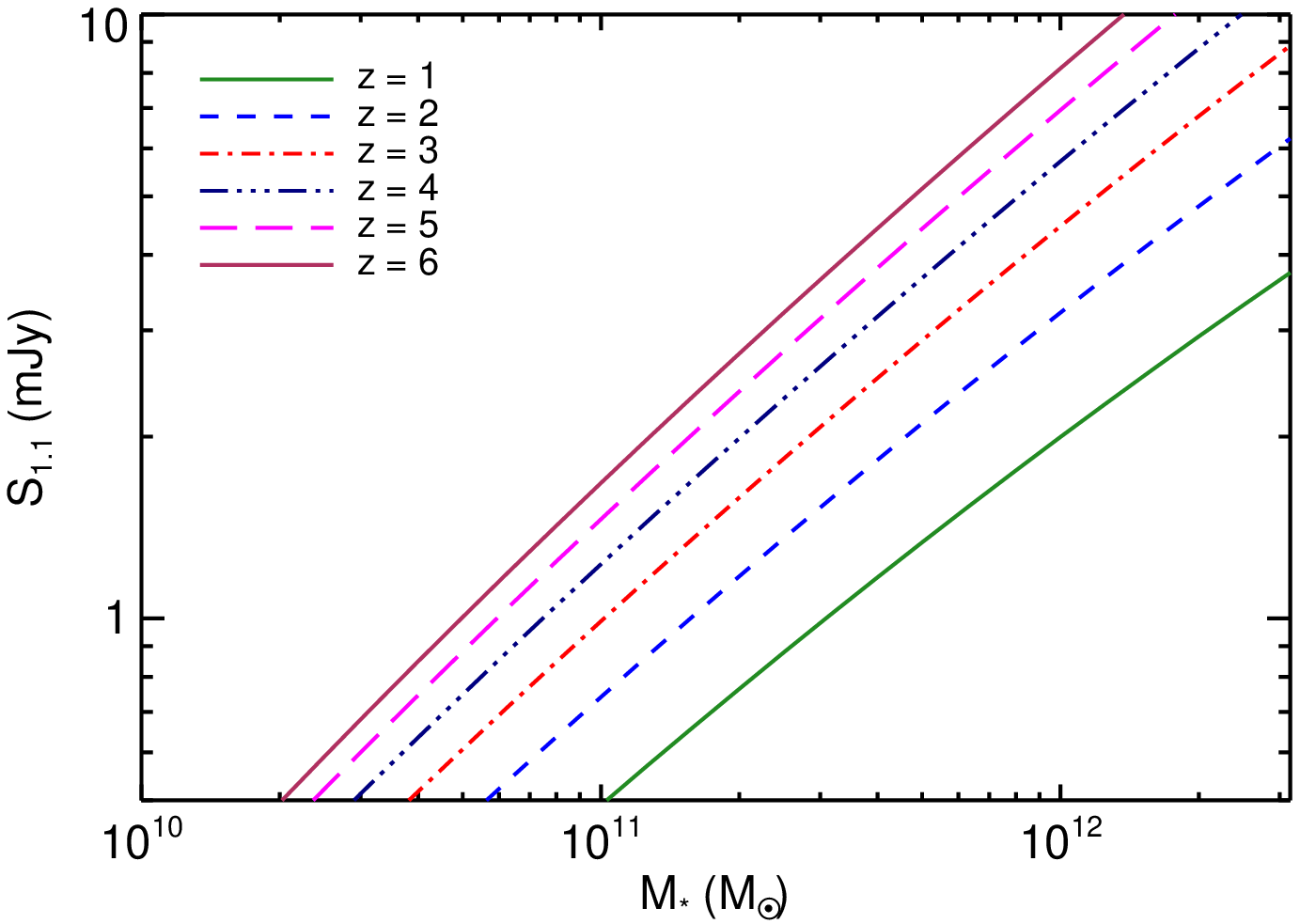}
\caption{Observed-frame 850-$\micron$ ($S_{850}$; top) and 1.1-mm ($S_{1.1}$; bottom) flux density in mJy versus
$\mstar (\msun)$ for isolated discs at integer redshifts in the range $z = 1-6$
(see Equations \ref{eq:S_850_m_z} and \ref{eq:S_11_m_z}). The (sub)mm flux of a disc of fixed $\mstar$ increases with redshift for two reasons: 1. As shown
in Fig. \ref{fig:sfr-mstar}, the normalisation of the SFR-$\mstar$ relation increases with redshift. 2. For fixed $\mstar$, gas fraction increases with redshift.
For fixed $Z$, a higher gas fraction corresponds to a higher gas-phase metal mass. However, because the normalisation of the MMR decreases as $z$
increases, the increase of the gas-phase metal mass with gas fraction is partially mitigated. Both the increased SFR and increased dust mass cause the (sub)mm flux to increase.}
\label{fig:flux_vs_mstar}
\end{figure}

Fig. \ref{fig:flux_vs_mstar} shows the $S_{850}-\mstar$ and $S_{1.1}-\mstar$ relations given
by Equations (\ref{eq:S_850_m_z}) and (\ref{eq:S_11_m_z}), respectively,
for isolated discs at integer redshifts in the range $z = 1-6$. As redshift increases, galaxies become more gas-rich and compact;
both effects cause the SFR for a given $\mstar$ to increase (see Fig. \ref{fig:sfr-mstar}).
For fixed $Z$, a higher gas fraction corresponds to a higher gas-phase metal mass. However, because the normalisation of the MMR decreases as $z$
increases, the increase of the gas-phase metal mass with gas fraction is partially mitigated.\footnote{For very high redshifts, at which there should be very few metals
and thus very little dust, the (sub)mm flux of a galaxy of a given mass should decrease sharply. However, the precise redshift dependence depends
on still-uncertain details of dust production; thus, it is possible that constraints on the fraction of SMGs with $z >> 4$ will yield insight into the physics of
dust production.}
The increased SFR and $\mdust$ both result in a higher (sub)mm flux for a given $\mstar$. To produce
an isolated disc SMG ($S_{850} \ga 3-5$ mJy, or $S_{1.1} \ga 1-2$ mJy) at $z \sim 2-3$, we require $\mstar \ga 10^{11} \msun$.
This value is consistent with the results of \citet{Michalowski:2010masses,Michalowski:2012}.

Note also that we can use these relations to calculate the expected $S_{850}/S_{1.1}$ ratio, $S_{850}/S_{1.1} \approx 2.3$ (this is similar
to observational estimates; e.g., \citealt{Austermann:2010}).
For simplicity, we will use this ratio to derive approximate $S_{850}$ values to also show an $S_{850}$ axis on the relevant plots.

\subsubsection{Infall-stage galaxy-pair SMGs}

During the infall stage of a merger, the discs are dominated by quiescent star formation that would occur even if they were not merging. Only for nuclear separation
$\la 10$ kpc\footnote{This value is derived for the $z \sim 3$ simulations presented here, and it may differ for $z \sim 0$ simulations because of differences in structural
properties and gas fractions.} do the discs have SFRs that are significantly elevated by the mutual tidal interactions (H12); this result is consistent with observed SFR elevations
in mergers \citep[e.g.,][]{Scudder:2012}. Thus, during the infall stage, we
assume the discs are in a steady state (i.e., they have constant SFR and dust mass); even without a source of additional gas, this is a reasonable approximation
for the infall stage to within a factor of $\la 2$ (see fig. 1 of H11).
For a merger of two progenitors with stellar masses $M_{\star,1}$ and $M_{\star,2}$,
the total flux density is $S_{\lambda} = S_{\lambda}(M_{\star, 1}) + S_{\lambda} (M_{\star, 2})$.
The typical beam sizes of single-dish (sub)mm telescopes are 15'', or $\sim130$ kpc at $z \sim 2 - 3$; schematically,
when the projected separation is less than this distance, the sources are blended
into a single source.\footnote{Because of the computational expense involved, we do not create synthetic maps, add noise, convolve the maps with
a Gaussian beam, and then calculate the fluxes of the sources. However, the uncertainty caused by our simple method of calculating
the total flux is subdominant to other uncertainties inherent in the model.}
To predict single-dish counts, we assume the galaxies should be treated as a single source if the physical separation is
$<100$ kpc. From our simulations, which use cosmologically motivated orbits, we find that this time-scale is of order $\sim 500$ Myr.
Though the time-scale depends slightly on the most-massive-progenitor mass, we neglect this dependence because it is subdominant to
various other uncertainties. However, this time-scale is derived from the $z \sim 2-3$ simulations and thus may be too long for mergers at higher $z$.
Given the above assumptions, the duty cycle for a given $S'_{\lambda}$ and merger described by more-massive progenitor mass $M_{\star,1}$ and
stellar mass ratio $\mu = M_{\star,2}/M_{\star,1}$ is 0.5 Gyr if $S_{\lambda}(M_{\star,1}) + S_{\lambda}(M_{\star,1}\mu) > S'_{\lambda}$ and 0 otherwise.
With the duty cycle in hand, we can use Equations (\ref{eq:merger_num_density}) and (\ref{eq:counts}) to calculate the predicted number density and counts.

To predict counts for ALMA, we simply assume that the two discs are resolved into individual sources and thus treat them as two isolated disc galaxies,
as described below.

\subsubsection{Isolated disc counts}

For a given $S_{\lambda}$ and $z$, we invert the $S_{\lambda}(\mstar, z)$ functions (Equations \ref{eq:S_850_m_z} and \ref{eq:S_11_m_z})
to calculate the minimum $\mstar$ required for a galaxy at redshift $z$ to have (sub)mm flux density $> S_{\lambda}$, $\mstar(S_{\lambda} | z)$.
To calculate the number density $n(>S_{\lambda},z)$, we then simply use the star-forming galaxy SMF to calculate
\begin{equation}
n(>S_{\lambda},z) = n\left[>\mstar(S_{\lambda} | z),z \right],
\end{equation}
and we use Equation (\ref{eq:counts}) to calculate the predicted counts. To predict counts for single-dish (sub)mm telescopes, where the galaxy-pair SMGs
are blended into a single source, we subtract the fraction of galaxies with $\mstar > \mstar(S_{\lambda} | z)$ that are in mergers
from the isolated disc counts to avoid double counting.

\section{Results} \label{S:results}

Here, we present the key results of this work, the SMG cumulative number counts, the relative contributions of the subpopulations, and the redshift distribution predicted by our model.
We focus on the AzTEC \citep{Wilson:2008} 1.1-mm counts here because to our knowledge, the best-constrained blank-field counts (i.e., those from the deepest and widest surveys)
have been determined using that instrument \citep{Austermann:2010,Aretxaga:2011}. However, because for our simulated SMGs, $S_{850}/S_{1.1} = 2.3$
to within $\sim 30$ per cent, we can easily convert the 1.1-mm counts to 850-$\micron$ counts. Thus, we include both $S_{1.1}$ and $S_{850}$ values
on the relevant plots and convert observed 850-$\micron$ counts to 1.1-mm counts by assuming the same ratio holds for real SMGs.

\subsection{SMG number counts for single-dish observations} \label{S:single-dish_counts}

\begin{figure}
%\epsscale{0.25}
\centering
\plotone{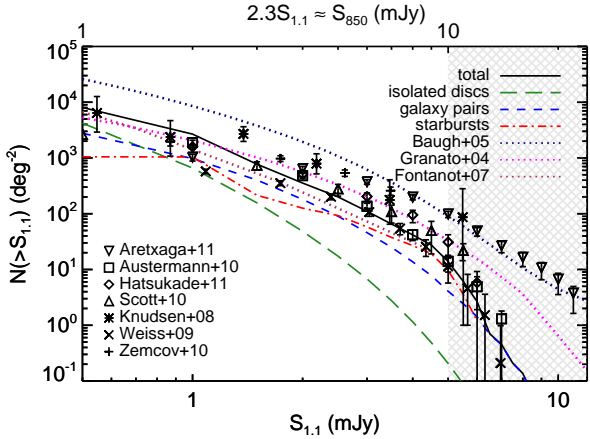}
\caption{Predicted cumulative number counts for the unlensed
SMG population as observed with single-dish (sub)mm telescopes, $N(>S_{1.1})$, in deg$^{-2}$, versus $S_{1.1}$ (mJy). The counts are decomposed into the three SMG subpopulations
we model: the green long-dashed line corresponds to isolated disc galaxies, the blue dashed to galaxy-pair SMGs (i.e., infall-stage pre-starburst mergers), and the red dash-dotted
to merger-induced starbursts. The black solid line is the total for all SMG subpopulations we model. The model predictions of \citet[][navy dotted]{Baugh:2005},
\citet[][magenta dotted]{Granato:2004}, and \citet[][maroon dotted]{Fontanot:2007} are shown for comparison.
The points are observed 1.1-mm and 850- and 870-$\micron$ counts (see the text for details).
The 850- and 870-$\micron$ counts have been converted to 1.1-mm counts by assuming $S_{850}/S_{1.1} \approx S_{870}/S_{1.1} = 2.3$.
This ratio has also been used to show the approximate $S_{850}$ on the top axis.
The hatched area shows the regime where weak lensing is expected to significantly boost the counts for overdense fields \citep{Aretxaga:2011}.
The counts predicted by our model agree very well with the counts that are not thought to be boosted significantly by lensing.
N.B. The steepness of the cutoff in the starburst counts at $S_{1.1} \ga 4$ mJy is artificial; see the text for details.
}
\label{fig:11_counts}
\end{figure}

\ctable[
	caption = 		{Single-dish-detected SMG cumulative number counts \label{tab:counts}},
				center,
				star,
				notespar
]{lccccc}{
	\tnote[a]{1.1-mm flux density.}
	\tnote[b]{Approximate 850-$\micron$ flux density calculated assuming $S_{850}/S_{1.1} = 2.3$.}
	\tnote[c]{Cumulative number counts of SMGs with 1.1-mm flux density greater than the $S_{1.1}$ value given in the first column.}
	\tnote[d-f]{Fractional contribution of each subpopulation to the total cumulative counts for the given $S_{1.1}$.}
}{
																																	\FL
$S_{1.1}$\tmark[a]	&	$\sim S_{850}$	\tmark[b]	&	$N(>S_{1.1})$\tmark[c]	&	\multicolumn{3}{c}{Fractional contribution}								\NN
(mJy)			&	(mJy)				&	(deg$^{-2}$)			&	Isolated discs\tmark[d]	&	Galaxy pairs\tmark[e]	& 	Starbursts\tmark[f]	\ML
 0.5     &    1.1    &       8038  &   0.52    &   0.35 &   0.13    \NN
 1.0     &    2.3    &       2676  &   0.25    &   0.36 &   0.39    \NN
 1.5     &    3.4    &        773  &   0.20    &   0.52 &   0.28    \NN
 2.0     &    4.6    &        354  &   0.14    &   0.51 &   0.35    \NN
 2.5     &    5.7    &        200  &   0.09    &   0.45 &   0.46    \NN
 3.0     &    6.9    &        110  &   0.06    &   0.41 &   0.52    \NN
 3.5     &    8.0    &         65  &   0.04    &   0.38 &   0.58    \NN
 4.0     &    9.2    &         41  &   0.03    &   0.33 &   0.64    \NN
 4.5     &   10.3    &         27  &   0.02    &   0.28 &   0.70    \NN
 5.0     &   11.5    &         13  &   0.01    &   0.29 &   0.69    \NN
 5.5     &   12.6    &          6  &   0.01    &   0.40 &   0.59    \NN
 6.0     &   13.8    &          2  &   0.01    &   0.53 &   0.45    \LL
}

Fig. \ref{fig:11_counts} shows the total cumulative 1.1-mm number counts (black solid line),
which are calculated from the cumulative number density using Equation (\ref{eq:counts}).
We decompose the counts into isolated discs (green long-dashed), galaxy pairs (blue dashed),
and starbursts induced at merger coalescence (red dash-dotted); the relative contribution of each subpopulation is discussed in Section \ref{S:rel_contrib}.
The data points in Fig. \ref{fig:11_counts} are observed counts from various surveys: 1.1-mm counts from \citet[][circles]{Aretxaga:2011},
\citet[][squares]{Austermann:2010}, \citet[][diamonds]{Hatsukade:2011}, and \citet[][triangles]{Scott:2010};
850-$\micron$ counts from \citet[][asterisks]{Knudsen:2008} and \citet[][plus signs]{Zemcov:2010}; and 870-$\micron$ counts
from \citet[][x's]{Weiss:2009}. The 850- and 870-$\micron$ counts have been converted to 1.1-mm counts by assuming $S_{850}/S_{1.1} \approx S_{870}/S_{1.1} = 2.3$.
The model predictions of \citet{Baugh:2005}, \citet{Granato:2004}, and \citet{Fontanot:2007} are shown for comparison.

The predicted and observed counts are in good agreement at the lowest fluxes, but the predicted counts are less than some of those observed at the bright end.
The \citet{Austermann:2010} and \citet{Aretxaga:2011} surveys are the two largest ($\sim0.7$ deg$^{-2}$), so their counts should be least
affected by cosmic variance and thus most robust. Thus, it is encouraging that the agreement between our predicted counts and those of \citet{Austermann:2010}
is very good at all fluxes. The disagreement between our predicted counts and those observed by \citet{Aretxaga:2011} is significant even for the lower flux bins
(a factor of $\sim 2$ for the $S_{1.1} > 2$ mJy bin). However, \citet{Aretxaga:2011} conclude that the excess of sources at
$S_{1.1} \ga 5$ mJy compared with the SHADES field observed by \citet{Austermann:2010} is caused by sources moderately
amplified by galaxy-galaxy and galaxy-group lensing. At higher fluxes, the effect of lensing is more significant
\citep{Negrello:2007,Paciga:2009,Lima:2010}, and it would be incredibly difficult to explain the sources with mm flux density $>> 10$ mJy observed by
\citet{Vieira:2010} and \citet{Negrello:2010} if they are not strongly lensed. We do not include the effects of gravitational lensing in our model,
so it is unsurprising that we significantly under-predict the counts of \citet{Aretxaga:2011} despite the excellent agreement between our counts and
those observed by \citet{Austermann:2010}.

Additionally, it is important to note that the steepness of the cutoff in the starburst counts at $S_{1.1} \ga 4$ mJy is artificial: because we determine
the fluxes of the isolated discs and galaxy pairs in an analytic way, we can extrapolate to arbitrarily high masses for those populations. For the starbursts however,
we are limited by the parameter space spanned by our merger simulations. None of our starburst SMGs reach $S_{1.1} \ge 6.5$ mJy (or $S_{850} \ga 15$ mJy),
so the duty cycle for all starbursts for $S_{1.1} \ge 6.5$ mJy is zero. If we were to simulate a galaxy more massive than our most massive model
(b6), the simulation would reach a correspondingly higher flux, so the predicted counts for $S_{1.1} \ge 6.5$ mJy would no longer be zero. However,
the rarity of such objects does not justify the additional computational expense. Thus for $S_{1.1} \ga 4$ mJy (or $S_{850} \ga 9$ mJy),
the starburst counts should be considered a lower limit. A simple extrapolation from the lower-flux starburst counts suggests that our model may even
over-predict the counts of the brightest sources. However, the observed number density of sources with $S_{1.1} \ga 5$ mJy is highly uncertain because of the effects
of small number statistics, cosmic variance, and lensing, and the uncertainty in the model prediction is significant because of uncertainties
in the abundances and merger rates of such extreme systems. Thus, the counts for the brightest sources should be interpreted with caution.

Furthermore, we do not attempt to model some other potential contributions to the SMG population. In particular, we do not include contributions
from mergers of more than two discs, clusters, or physically unrelated sources blended into a single (sub)mm source
(see \citealt{Wang:2011} for evidence of the last type).

Given these caveats and the modelling uncertainties, our predicted counts are clearly consistent with those observed, and including lensing and the previously
mentioned additional possible contributions to the SMG population would tend to increase the number counts. Also, we stress that our
model is conservative in the sense that it uses a Kroupa -- rather than top-heavy or flat -- IMF and is tied to observations whenever possible.
The consistency of the predicted and observed counts suggests that the observed SMG counts may not provide evidence for IMF
variation; this will be discussed in detail in Section \ref{S:disc_IMF}.

\subsection{Relative contributions of the subpopulations} \label{S:rel_contrib}

\begin{figure}
\centering
\plotone{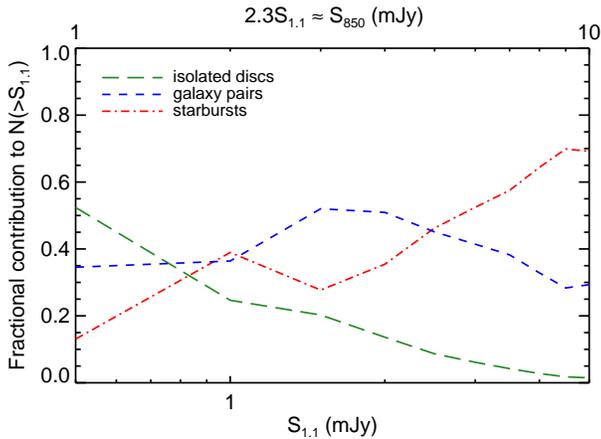}
\caption{Fractional contribution of each subpopulation to the total cumulative counts
versus $S_{1.1}$. The lines are the same as in Fig. \ref{fig:11_counts}. At the lowest fluxes, the
isolated discs dominate, whereas at the highest fluxes, the starbursts dominate. The galaxy pairs are $\sim$ 30-50 per cent of the population
at all fluxes plotted here.}
\label{fig:relative_counts}
\end{figure}

In previous work (H11; \citealt{Hayward:2011num_cts_proc}; H12), we argued that the SMG population is not exclusively late-stage merger-induced starbursts but rather a
heterogeneous collection of starbursts, infall-stage mergers (`galaxy-pair SMGs'), and isolated discs. However, so far we have only
presented the physical reasons one should expect such heterogeneity. It is crucial to quantify the relative importance of each
subpopulation, so we do this now.

The counts shown in Fig. \ref{fig:11_counts} are divided into subpopulations, but the relative contributions can be read more easily from
Fig. \ref{fig:relative_counts}, which shows the fractional contribution of each subpopulation to the total cumulative counts.
At the lowest fluxes, the isolated disc contribution is the most significant. At $S_{1.1} \sim 0.8$ mJy ($S_{850} \sim 2$ mJy), the three subpopulations
contribute almost equally. As expected from conventional wisdom, the starbursts dominate at the highest fluxes.
However, contrary to conventional wisdom, the bright SMGs are not exclusively merger-induced starbursts:
from Fig. \ref{fig:relative_counts}, we see that at all fluxes plotted, the galaxy pairs account for $\sim$ 30-50
per cent of the total predicted counts, so they are a significant subpopulation of SMGs in our model.
As explained in H12, the galaxy-pair SMGs are not physically analogous to the merger-induced starburst SMGs;
thus, their potentially significant contribution to the SMG population can complicate physical interpretation of the observed properties of SMGs.

It is interesting to compare the relative contributions of the isolated disc and galaxy-pair subpopulations because the relative contributions
can be understood -- at least schematically -- in a simple manner. For a major merger of two galaxies with $\mstar = M_{\rm iso}$,
the flux of the resulting galaxy-pair SMG is approximately twice that of the individual isolated discs, $2 S_{1.1}(M_{\rm iso})$. Because $S_{1.1}$ depends sub-linearly on
$\mstar$ (see Fig. \ref{fig:flux_vs_mstar}), for an isolated disc to have $S_{1.1}$ equal to that of the galaxy pair, it must have $\mstar \ga 3 M_{\rm iso}$.
Thus, the relative contribution of the two subpopulations depends on whether the number density of $\mstar = 3 M_{\rm iso}$ discs divided by that of
$\mstar = M_{\rm iso}$ discs, $n(3M_{\rm iso})/n(M_{\rm iso})$, is greater than the fraction of $\mstar = M_{\rm iso}$ discs undergoing a major merger,
which is the merger rate times the duty cycle of the infall phase ($\sim$500 Myr). If the former is larger, the $\mstar = 3 M_{\rm iso}$ discs will dominate the
pairs of $\mstar = M_{\rm iso}$ discs, whereas if the merger fraction is higher than the relative number density, the galaxy pairs will dominate.

The latter scenario is likely for bright SMGs, which are on the exponential tail of the SMF.
For example, at $z \sim 2-3$, a galaxy with $\mstar = 10^{11} \msun$ undergoes $\sim 0.3$ mergers per Gyr. Thus,
if we assume a duty cycle of 500 Myr for the galaxy-pair phase, approximately 15 per cent of such galaxies will be in galaxy pairs. For the \citet{Marchesini:2009} SMF, the number
density of $\mstar = 3 \times 10^{11} \msun$ galaxies is $\sim 8$ per cent that of $\mstar = 10^{11} \msun$ galaxies. Therefore, by the above logic,
the pairs of $\mstar = 10^{11} \msun$ galaxies will contribute more to the submm counts than the isolated $\mstar = 3 \times 10^{11} \msun$ discs.
This simple argument demonstrates why the galaxy pairs become dominant over the isolated discs for $S_{1.1} \ga 0.7$ mJy ($S_{850} \ga 1.6$ mJy).
However, the threshold for dominance depends on both the $S_{1.1}-\mstar$ scaling and the shape of the SMF at the high-mass end. Thus, observationally
constraining the fraction of the SMG population that is galaxy pairs can provide useful constraints on both the (sub)mm flux-$\mstar$ relation and the
shape of the massive end of the SMF.

Unfortunately, the relative contribution of the starburst subpopulation cannot be explained in as simple a manner. The duty cycles for the
merger-induced starbursts depend sensitively on progenitor mass and merger mass ratio, so the mapping from merger
rate to number density is not as simple as it is for the isolated discs and galaxy pairs. Fortunately, the SMF uncertainty, which is
very significant for the overall counts, is relatively unimportant for the relative contribution of starbursts and galaxy pairs. Thus, the
relative contributions of starbursts and galaxy pairs depend primarily on their relative duty cycles. (To achieve a given flux density,
one requires a less massive starburst than galaxy pair because the starburst increases the (sub)mm flux density moderately. Thus, the relative
number density also matters. However, the inefficiency of starbursts at increasing the (sub)mm flux density of the system prevents significantly less massive but
more common starbursts from dominating over more massive and rarer galaxy pairs.) The duty cycles are uncertain, but given that
in our fiducial model the galaxy pairs contribute $\sim 30-50$ per cent of the total counts and the uncertainty in the duty cycles is definitely
less than a factor of $2-3$, the prediction that both the starburst and galaxy pair subpopulations are significant (i.e., more than a few per cent of the
population) is robust.

Though there have been many observational hints suggesting the importance of the galaxy-pair contribution (see H11 and H12 for discussion),
the physical importance of this subpopulation has to date not been fully appreciated, and the
fractional contribution of galaxy-pair SMGs to the total counts remains relatively poorly constrained.
However, clear observational evidence supporting the significance of this subpopulation is accumulating: of the 12 SMGs presented in \citet{Engel:2010},
5 have CO emission that is resolved into two components with kinematics consistent with two merging discs. In two of the cases, the projected separation
of the two components is $>20$ kpc; such objects are prime examples of the galaxy-pair subpopulation.
(See also \citealt{Tacconi:2006,Tacconi:2008,Bothwell:2010,Riechers:2011b,Riechers:2011a}.)
\citet{Smolcic:2012} presented a larger sample of SMGs with (sub)mm-interferometric detections. They found that when observed with interferometers
with $\la 2$'' resolution, $\sim15-40$ per cent of single-dish SMGs were resolved into multiple sources, which is consistent with our prediction for
the relative contribution of the galaxy-pair subpopulation. ALMA observations will significantly increase the number of SMGs observed with $\sim 0.5$''
resolution and thus better constrain the galaxy-pair contribution to the SMG population.

Further evidence for a galaxy-pair contribution consistent with what we predict is the fraction of the SMGs with multiple counterparts at other wavelengths.
One of the earliest observational indications of this population came from the
SCUBA 8-mJy survey: of this sample of 850-$\micron$ sources, \citet{Ivison:2002} found that $\sim25$ per cent have multiple radio counterparts.
Approximately ten per cent of the GOODS-N 850-$\micron$ \citep{Pope:2006}, GOODS-N 1.1-mm \citep{Chapin:2009},
SHADES 850$-\micron$ \citep{Ivison:2007,Clements:2008}, and GOODS-S 1.1-mm \citep{Yun:2012} sources have multiple counterparts.
These fractions are somewhat smaller than the $\sim$ 30-50 per cent contribution shown in Fig. \ref{fig:relative_counts}, but both the
predicted and observed fractions are uncertain. As explained above, the predicted fraction depends sensitively on the shape of the upper-end of the SMF 
and the relation between (sub)mm flux and $\mstar$. Observations, on the other hand, may miss the more widely separated counterparts
and cases when one of the counterparts is significantly more obscured or is radio-quiet.

\subsection{Redshift distribution}

\begin{figure}
%\epsscale{0.25}
\centering
\plottwo{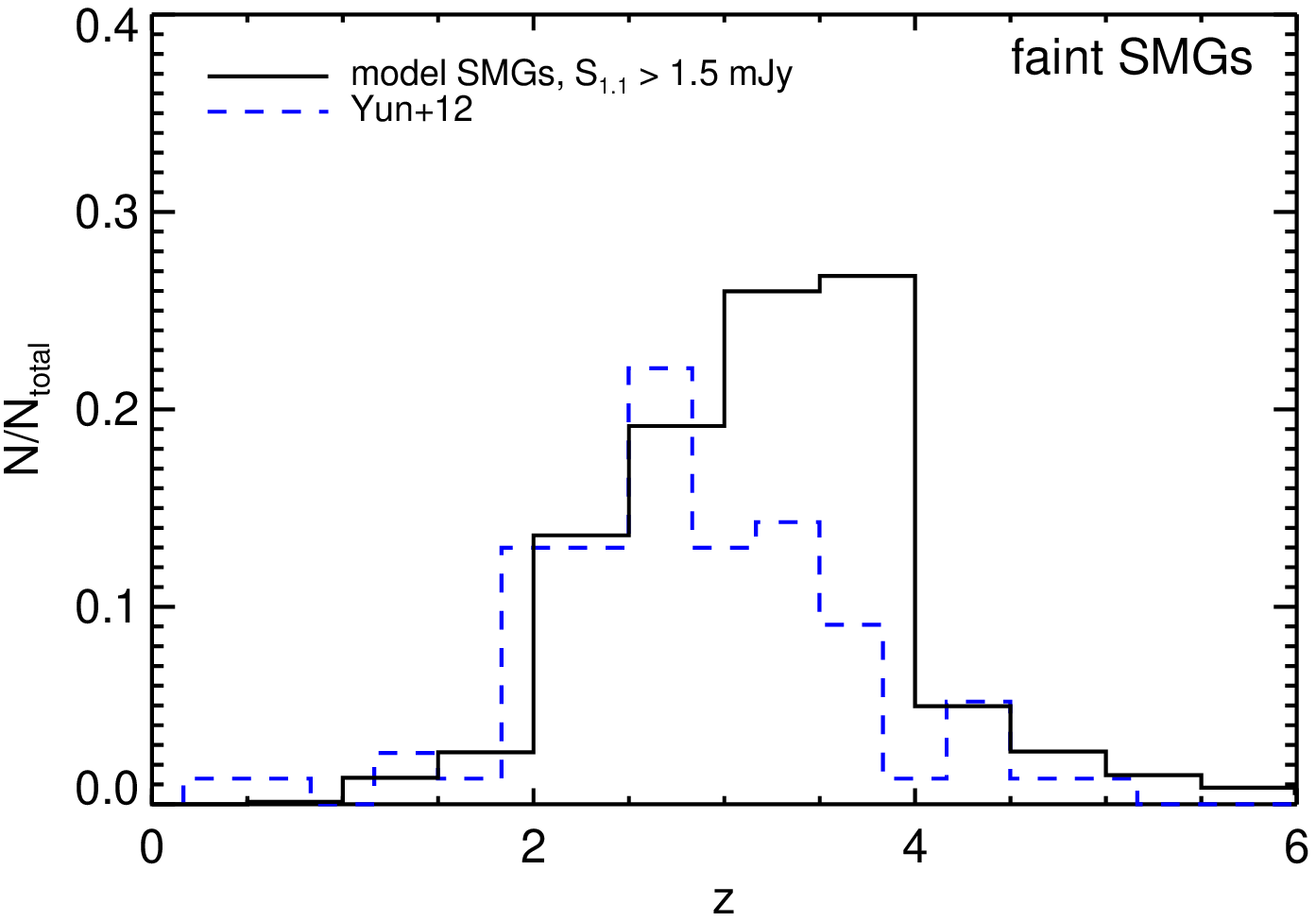}{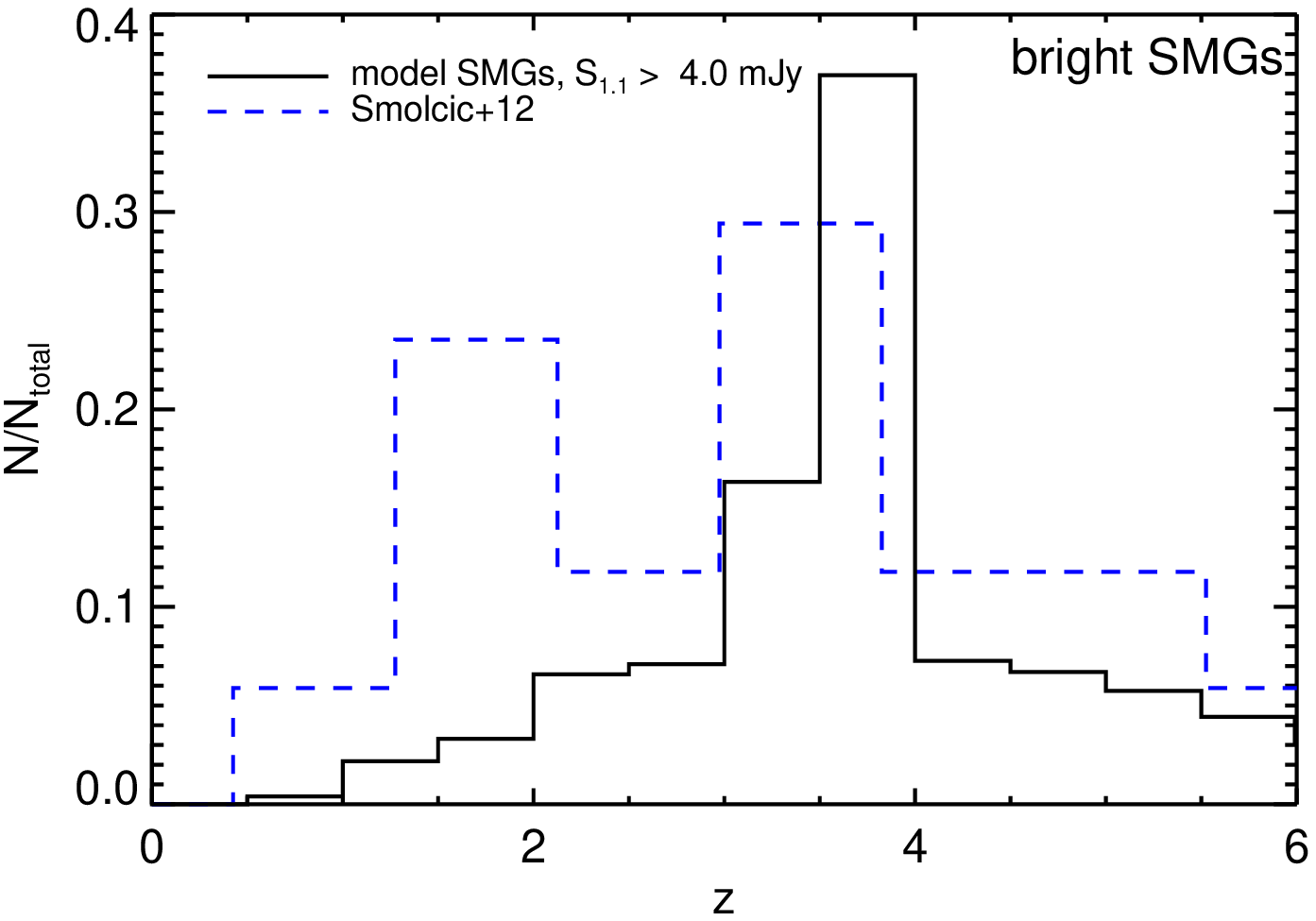}
\caption{Top: the predicted redshift distribution of 1.1-mm sources with $S_{1.1} > 1.5$ mJy ($S_{850} \ga 3.5$ mJy; black solid line)
compared with the observed distribution from \citet{Yun:2012} (blue dashed line).
The mean redshifts are 3.0 and 2.6 for the model and observed SMGs, respectively. Bottom: the predicted redshift distribution
for sources with $S_{1.1} > 4$ mJy ($S_{850} \ga 9$ mJy; black solid line) compared with the observed distribution from \citet{Smolcic:2012}
(blue dashed line). The mean redshifts are 3.5 and 3.1 for the model and observed SMGs, respectively. In both panels,
the flux limits were chosen to approximately match those of the observations. The brighter SMGs tend to be at higher
redshifts.}
\label{fig:11_z_distrib}
\end{figure}

\ctable[
	caption = 		{Single-dish-detected SMG redshift distribution},
				center,
				notespar%,
%				star
]{lcc}{
	\tnote[a]{Redshift.}
	\tnote[b-c]{Fractional contribution of sources in the redshift bin to the total sources with 1.1-mm flux density
			greater than the specified limit.}
}{
																		\FL
$z$\tmark[a]		&	\multicolumn{2}{c}{$N_{\mathrm{in ~bin}}/N_{\mathrm{total}}(>S_{1.1})$}						\NN
				&	1 mJy\tmark[b]		&	4 mJy\tmark[c]		\ML
0.0 -- 0.5  &    0.001  &    0.004  \NN
0.5 -- 1.0  &    0.013  &    0.022  \NN
1.0 -- 1.5  &    0.026  &    0.033  \NN
1.5 -- 2.0  &    0.136  &    0.066  \NN
2.0 -- 2.5  &    0.192  &    0.071  \NN
2.5 -- 3.0  &    0.260  &    0.163  \NN
3.0 -- 3.5  &    0.268  &    0.369  \NN
3.5 -- 4.0  &    0.050  &    0.073  \NN
4.0 -- 4.5  &    0.027  &    0.067  \NN
4.5 -- 5.0  &    0.015  &    0.057  \NN
5.0 -- 5.5  &    0.008  &    0.044  \NN
5.5 -- 6.0  &    0.005  &    0.031  \LL
}

In addition to the number counts, a successful model for the SMG population must reproduce the redshift distribution.
Fig. \ref{fig:11_z_distrib} shows the redshift distribution of 1.1-mm sources predicted by our model for different 1.1-mm flux cuts ($S_{1.1} > 1.5$
mJy, or $S_{850} \ga 3.5$ mJy, in the top panel and $S_{1.1} > 4$ mJy, or $S_{850} \ga 9$ mJy, in the bottom) along with some observed distributions that have similar flux limits.
The redshift distributions are relatively broad, and they peak in the range $z \sim 2-4$ and decline at lower and higher redshifts. 
The $S_{1.1} > 1.5$ mJy sources have mean redshift 3.0, whereas the $S_{1.1} > 4$ mJy sources have mean redshift 3.5, so
there is a tendency for the brighter sources to be at higher redshifts; this trend agrees with observations \citep{Ivison:2002, Yun:2012, Smolcic:2012}.

For the $S_{1.1} > 1.5$ mJy SMGs (top panel of Fig. \ref{fig:11_z_distrib}), compared with the observations, our model predicts a higher mean redshift and a
greater fraction of SMGs at $z \sim 3-4$.
This discrepancy may suggest that the extrapolation of the \citet{Fontana:2006} SMF we use for $z > 3.75$ over-predicts the number of massive galaxies at the highest redshifts.
Furthermore, merger time-scales may be shorter at high redshift, and the dust content may be lower than in our models; both of these effects
would decrease the high-redshift contribution. Additionally, constraining the redshift distribution is complicated by
selection effects, counterpart identification, and, in some cases, a lack of spectroscopic redshifts. Typically, the selection effects and counterpart identification
make identifying higher-redshift SMGs more difficult. Finally, the significant differences amongst observed distributions
(see \citealt{Smolcic:2012} for discussion) demonstrate the difficulty of determining the redshift distribution.
Thus, the differences in the distributions may not be significant; spectroscopic follow-up of ALMA sources should clarify this issue
(but note that the redshift distribution of ALMA sources, which is shown in the next section, should differ slightly).
The typical redshift for the brighter SMGs (bottom panel of Fig. \ref{fig:11_z_distrib}) is also somewhat higher than that observed (3.5 compared with 3.1), and
the peak at $z \sim 1-2$ is not reproduced. However, the observed distribution is based on a sample of 17 sources, so small-number statistics might explain the differences.

\subsection{Predicted ALMA-detected SMG number counts and redshift distribution}

\begin{figure}
%\epsscale{0.25}
\centering
\plotone{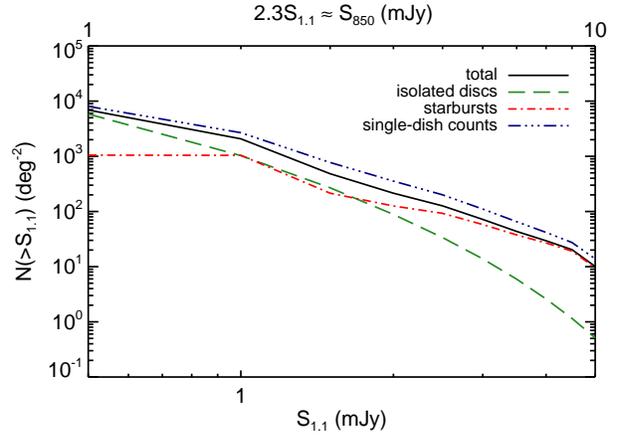}
\caption{Predicted cumulative number counts for the SMG subpopulation observed with telescopes, such as ALMA, that have resolution
sufficient to resolve the galaxy-pair subpopulation into multiple sources. The green long-dashed line corresponds to isolated disc galaxies,
the dash-dotted red to starbursts, and the black solid to the total counts. The single-dish counts from Fig. \ref{fig:11_counts} are
shown for comparison (blue dash dot dot dot). The top axis shows approximate $S_{850}$ values.
Because some of the bright sources would be resolved into multiple sources with ALMA, the counts of bright ALMA sources are
as much as a factor of 2 less than the single-dish counts.}
\label{fig:ALMA_counts}
\end{figure}

\ctable[
	caption = 		{ALMA-detected SMG cumulative number counts \label{tab:ALMA_counts}},
				center,
				star,
				notespar
]{lcccccc}{
	\tnote[a]{1.1-mm flux density.}
	\tnote[b-d]{Approximate 850-$\micron$ and ALMA Bands 6 \& 7 flux densities calculated using the conversion factors $S_{850}/S_{1.1} = 2.3$,
	$S_{\rm ALMA-6}/S_{1.1} = 0.8$, and $S_{\rm ALMA-7}/S_{1.1} = 1.6$.}
	\tnote[e]{Cumulative number counts of SMGs with 1.1-mm flux density greater than the $S_{1.1}$ value given in the first column.}
	\tnote[f-g]{Fractional contribution of each subpopulation to the total cumulative counts for the given $S_{1.1}$.}
}{
																											\FL
$S_{1.1}$\tmark[a]	&	$\sim S_{850}$	\tmark[b]	&	$\sim S_{\rm ALMA-6}$\tmark[c]	& 	$\sim S_{\rm ALMA-7}$\tmark[d]	&	$N(>S_{1.1})$\tmark[e]
&	\multicolumn{2}{c}{Fractional contribution}		\NN
(mJy)			&	(mJy)				&	(mJy)						&	(mJy)						&	(deg$^{-2}$)
&	Isolated discs\tmark[f]	&	Starbursts\tmark[g]	\ML
 0.5		&	 1.1	&	 0.4    &    0.8    &       6897	&	0.85	&	0.15	\NN
 1.0		&	 2.3	&	 0.8    &    1.6    &       2072	&	0.50	&	0.50	\NN
 1.5		&	 3.4	&	 1.2    &    2.4    &        482	&	0.56	&	0.44	\NN
 2.0		&	 4.6	&	 1.6    &    3.2    &        214	&	0.41	&	0.59	\NN
 2.5		&	 5.7	&	 2.0    &    4.0    &        126	&	0.27	&	0.73	\NN
 3.0		&	 6.9	&	 2.4    &    4.8    &         71	&	0.19	&	0.81	\NN
 3.5		&	 8.0	&	 2.8    &    5.6    &         43	&	0.14	&	0.86	\NN
 4.0		&	 9.2	&	 3.2    &    6.4    &         29	&	0.09	&	0.91	\NN
 4.5		&	10.3	&	 3.6    &    7.2    &         20	&	0.06	&	0.94	\NN
 5.0		&	11.5	&	 4.0    &    8.0    &         10	&	0.05	&	0.95	\NN
 5.5		&	12.6	&	 4.4    &    8.8    &          3	&	0.06	&	0.94	\NN
 6.0		&	13.8	&	 4.8    &    9.6    &          1	&	0.08	&	0.92	\LL
}

\begin{figure}
%\epsscale{0.25}
\centering
\plotone{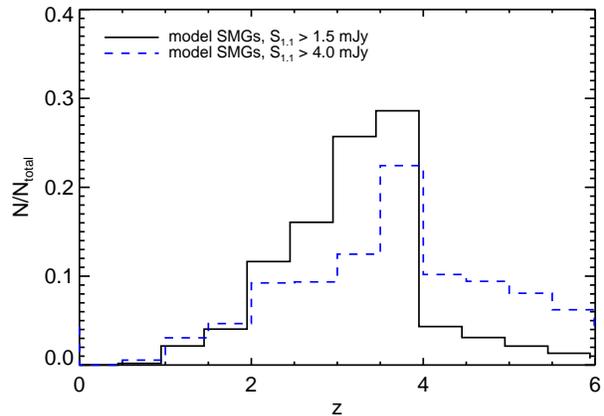}
\caption{Predicted redshift distributions for ALMA-detected SMGs with $S_{1.1} > 1.5$ mJy ($S_{850} \ga 3.5$ mJy; black solid line; $\bar{z} = 3.0$)
and $S_{1.1} > 4$ mJy ($S_{850} \ga 9$ mJy; blue dashed line; $\bar{z} = 3.5$). As for the single-dish sources, the brighter SMGs tend to lie at higher redshift.}
\label{fig:ALMA_z_distrib}
\end{figure}

\ctable[
	caption = 		{ALMA-detected SMG redshift distribution \label{tab:ALMA_z_distrib}},
				center,
				notespar%,
%				star
]{lcc}{
	\tnote[a]{Redshift.}
	\tnote[b-c]{Fractional contribution of sources in the redshift bin to the total sources with 1.1-mm flux density
			greater than the specified limit.}
}{
																		\FL
$z$\tmark[a]		&	\multicolumn{2}{c}{$N_{\mathrm{in ~bin}}/N_{\mathrm{total}}(>S_{1.1})$}						\NN
				&	1 mJy\tmark[b]		&	4 mJy\tmark[c]		\ML
0.0 -- 0.5  &    0.002  &    0.006  \NN
0.5 -- 1.0  &    0.021  &    0.031  \NN
1.0 -- 1.5  &    0.040  &    0.047  \NN
1.5 -- 2.0  &    0.116  &    0.092  \NN
2.0 -- 2.5  &    0.161  &    0.094  \NN
2.5 -- 3.0  &    0.257  &    0.125  \NN
3.0 -- 3.5  &    0.286  &    0.224  \NN
3.5 -- 4.0  &    0.043  &    0.102  \NN
4.0 -- 4.5  &    0.031  &    0.094  \NN
4.5 -- 5.0  &    0.021  &    0.081  \NN
5.0 -- 5.5  &    0.013  &    0.062  \NN
5.5 -- 6.0  &    0.007  &    0.043  \LL
}

In the above sections, we have presented our model predictions for surveys performed with single-dish (sub)mm telescopes
such as the JCMT, for which the $\sim 15$ arcsec beam causes the galaxy-pair SMGs to be blended into a single source for much
of their evolution. If larger single-dish (sub)mm telescopes such as the LMT and CCAT are used, then the galaxy-pair SMGs
will be blended for a smaller fraction of the infall stage of the merger. Thus, the duty cycle for the galaxy-pair phase will be shorter
and the number counts of the bright sources consequently less. For interferometers with angular resolution $\la 0.5$ arcsec (or $\sim 4$ kpc
at $z \sim 2-3$), such as ALMA, essentially all galaxy-pair SMGs will be resolved into multiple sources because for such separations,
the mergers are typically dominated by the starburst mode (H12). Thus, the galaxy pairs would contribute to the number counts
as two quiescently star-forming galaxies rather than one blended source.

To predict the counts and redshift distribution of ALMA sources, we modify our model by removing the galaxy-pair contribution
and re-distributing those galaxies into the isolated disc
subpopulation. The predicted cumulative number counts are shown in Fig. \ref{fig:ALMA_counts}, in which
the total single-dish counts are also plotted for comparison.
The values of the counts and the fractional contributions of the isolated discs
and starbursts for various flux bins are given in Table \ref{tab:ALMA_counts}.
To facilitate comparison with observations, the table includes approximate flux densities for ALMA Bands 6 and 7 calculated using the mean
ratios for our simulated SMGs, $\overline{S_{\rm ALMA-6}/S_{1.1}} = 0.8 \pm 0.06$ and $\overline{S_{\rm ALMA-7}/S_{1.1}} = 1.6 \pm 0.2$.

As for the single-dish counts, the isolated discs
dominate at the lowest fluxes ($S_{1.1} \la 1$ mJy, or $S_{850} \la 2$ mJy), and the brightest sources are pre-dominantly starbursts.
Because the galaxy-pair SMGs are resolved into multiple fainter sources, the cumulative number counts for ALMA-detected SMGs
are lower at all fluxes by $\sim 30 - 50$ per cent, and the differential
counts should be steeper. (At fluxes fainter than those shown,
for which the isolated discs completely dominate and the bright sources contribute negligibly to the cumulative counts,
the single-dish and ALMA cumulative counts will be almost identical.) This effect has also been discussed by \citet{Kovacs:2010} and \citet{Smolcic:2012}.

The redshift distributions for two flux cuts are shown in Fig. \ref{fig:ALMA_z_distrib}, and the values are given
in Table \ref{tab:ALMA_z_distrib}. The mean redshifts for the
$S_{1.1} > $ 1.5 and 4 mJy ($S_{850} \ga 3.5$ and 9 mJy) bins are 3.0 and 3.5, respectively. The mean values are almost identical to those for the
single-dish counts, and there is also the same tendency for the brightest SMGs to be at higher redshifts. The redshift distributions
are similar, but the distribution for the $S_{1.1} > 4$ mJy sources is less strongly peaked than for the single-dish sources.
The latter's redshift distribution is more strongly peaked because the redshift distribution of the bright galaxy pairs peaks
at $z \sim 3.5$.

\section{Discussion} \label{S:discussion}

\subsection{Are modifications to the IMF required to match the observed SMG counts?} \label{S:disc_IMF}

\begin{figure}
%\epsscale{0.25}
\centering
\plotone{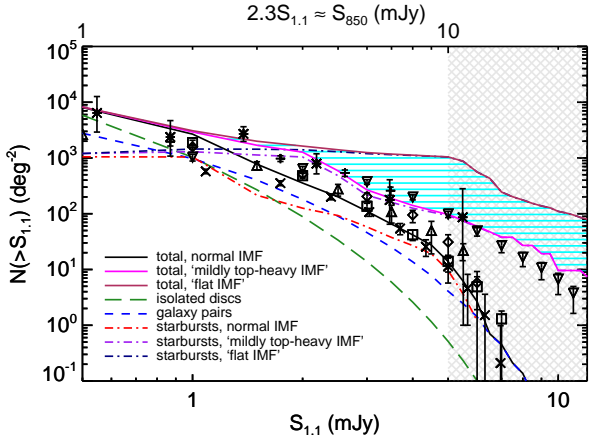}
\caption{Predicted cumulative number counts for the unlensed
SMG population as observed with single-dish (sub)mm telescopes, $N(>S_{1.1})$, in deg$^{-2}$, versus $S_{1.1}$ (mJy),
where we have crudely approximated the effects of using a `mildly top-heavy' (`flat') IMF in starbursts by multiplying the 1.1-mm flux of the 
starburst SMGs by a factor of 2 (5). The region filled with cyan lines indicates the range spanned by IMF boost factors between 2 and 5.
The counts for our standard model, which uses a Kroupa IMF, are shown for comparison.
The points and grey hatched region are the same as in Fig. \ref{fig:11_counts}.
The starburst and total counts are increased significantly by the `top-heavy IMF', but the isolated disc and galaxy pair counts are not affected
because their fluxes are not altered. The starbursts dominate the SMG population for $S_{1.1} \ga 1$ mJy ($S_{850} \ga 2$ mJy).
For $S_{1.1} \ga 2$ mJy ($S_{850} \ga 5$ mJy), the `mildly top-heavy IMF' model over-predicts the counts that are thought not to be boosted by lensing,
and the `flat IMF' model overpredicts all the observed counts, which suggests that a significantly top-heavy IMF
is ruled out. N.B. The lack of a boost in the counts for $S_{1.1} \la 1$ mJy is artificial; see the text for details.}
\label{fig:flat_counts}
\end{figure}

One of the primary motivations for this work is to reexamine the possibility that SMG number counts provide
evidence for a flat IMF (B05, \citealt{Swinbank:2008,Dave:2010}). To test this claim, we have assumed the null hypothesis -- that the IMF
in SMGs does not differ from what is observed locally -- and used a Kroupa IMF. Furthermore, our model is constrained to match the present-day mass function --
or, more accurately, to not over-produce $z \sim 0$ massive galaxies -- by construction; this is an important `litmus test' for putative models of the
SMG population.
The counts predicted by our model agree well with the observed counts for fields believed not to be significantly affected by gravitational lensing (e.g., \citealt{Austermann:2010}).
Thus, our model does not require modifications to the IMF to match the observed counts.
As an additional check, we crudely approximate the effect of varying the IMF in starbursts in our model by multiplying the (sub)mm
fluxes of our starbursts by factors of 2 (to represent a `mildly top-heavy IMF') and 5 (to represent a `flat IMF', because this is the factor appropriate for
the IMF used in B05; G.-L. Granato, private communication).

Fig. \ref{fig:flat_counts} shows the range in number counts predicted for this range of IMF variation.
This modification causes starbursts to completely dominate the counts for $S_{1.1} ~(S_{850}) \ga 1 ~(2)$ mJy (as is the case for the B05 model), and the predicted counts are significantly 
greater than most of the observed counts for unlensed SMGs.\footnote{Amusingly, the counts predicted for the `mildly top-heavy IMF' agree well with those of \citet{Aretxaga:2011}
even though no lensing is included in our model. However, field-dependent IMF variation seems unlikely, so the preferred explanation is still that the \citet{Aretxaga:2011}
counts are boosted by lensing.}
(At the lower fluxes, the similarity between the starbursts counts for the various IMF factors is an artificial effect caused by
the limited parameter space spanned by our simulation suite. If the suite included mergers of lower-mass galaxies, adopting a top-heavy IMF would cause them to contribute
to the counts at these fluxes and thus boost the counts at all fluxes. Furthermore, the strength of the effect at the highest fluxes is underestimated because
of the aforementioned artificial cutoff of our starburst counts at $S_{1.1} \ga 4$ mJy; see \ref{S:single-dish_counts} for details.)
Applying the `flat IMF' factor causes the model to significantly overpredict all the observed counts for $S_{1.1} \ga 2$ mJy ($S_{850} \ga 5$ mJy),
and as explained above, the lack of an effect for lower fluxes is an artificial consequence of our merger simulation suite not including lower-mass
galaxies. The clear conclusion is that in our model, significant modification to the IMF in starbursts is unjustified.

Admittedly, the above arguments against IMF variation depend on the details and assumptions of our model. However, there is a much simpler argument
against the IMF in starbursts being flat or significantly top-heavy: as discussed above (and in detail in H11 and H12), there is growing observational evidence that a significant
fraction of the single-dish-detected SMG population is attributable to multiple sources blended into one (sub)mm source.
In some cases, the SMGs are early-stage mergers with components separated by $\ga 10$ kpc (see \citealt{Engel:2010} and \citealt{Riechers:2011b,Riechers:2011a} for excellent
examples). Hydrodynamical simulations suggest that at such separations, a strong starburst is typically not induced (H12); this
is consistent with observations of local galaxy pairs \citep[e.g.,][]{Scudder:2012}.
When the galaxies eventually merge, their SFRs will increase by a factor of a few at the least (typically significantly more for major mergers, but we
will be conservative) and if the IMF is flat in starbursts, the (sub)mm flux per unit SFR produced will increase by a factor of $\sim 5$. Thus, the (sub)mm flux of the
galaxies should increase by a factor of $\ga 10$ during the starburst. The relative duty cycles of the galaxy-pair and starburst phases are similar (to within a factor of a few),
so for a given mass and mass ratio, the number density of starbursts and mergers in the galaxy-pair phase should be similar. Thus, if the (sub)mm flux is $\ga 10$ times greater in the
starburst phase than in the quiescently star-forming galaxy-pair phase, mergers during the galaxy-pair phase
should contribute negligibly to the bright SMG population, as is the case in Fig. \ref{fig:11_counts}, even when a more modest boost (a factor of 2) is used.
This is in stark contradiction with observational evidence. Indeed, the existence of the galaxy-pair subpopulation is a natural consequence of starbursts being
very inefficient at boosting the (sub)mm flux
of merging galaxies (see the discussion in Section \ref{S:rel_contrib}). The only means to circumvent this argument is to argue that the multiple-component
SMGs observed are all starbursts, but as explained above, this is unlikely.

The above argument does not rule out the possibility of a mildly top-heavy IMF in starbursts or systematic global evolution of the IMF with redshift
(as suggested by, e.g., \citealt{vanDokkum:2008,Dave:2008,Narayanan:2012IMF}); it only requires that the IMFs in starburst
and galaxy-pair SMGs at a given redshift be similar. Furthermore, the argument does not rule out the possibility of a bottom-heavy IMF in starbursts,
which has been suggested recently
for local massive ellipticals, the descendants of high-z starbursts \citep[e.g.,][]{vanDokkum:2010,vanDokkum:2011,Conroy:2012,Hopkins:2012IMF}.
If the IMF were bottom-heavy in starbursts, the scaling between
(sub)mm flux and SFR would be weaker than what we have shown in H11 and here. The weaker scaling would further increase the contribution of galaxy-pair SMGs
to the SMG population. Unfortunately, the model and observational uncertainties are sufficiently large that we cannot use the observed galaxy-pair fraction
to argue for or against a bottom-heavy IMF in starbursts, but the above arguments suggest that a significantly top-heavy IMF is unlikely.

\subsection{Differences between our model and other work} \label{S:diff_from_others}

Because we find that, contrary to some previous suggestions (B05, \citealt{Swinbank:2008,Dave:2010}), a top-heavy IMF is not required to match the observed SMG counts,
it is worthwhile to examine why our results differ from those works.
There are multiple reasons our results may differ: the cosmological context (abundances and merger rates), the evolution of SFR and dust mass for individual
mergers, the radiative transfer calculation, the effects of blending, and differences in the observed counts used. We explore these in turn now.

\subsubsection{Radiative transfer calculation}

In H11, we demonstrated that the (sub)mm
flux density of our simulated galaxies can be well parameterized as a power law in SFR and dust mass (see Equations \ref{eq:sfr_fitting_function_850} and \ref{eq:sfr_fitting_function_11}).
If the same relation does not hold in the B05 model, then differences in the radiative transfer may be one cause of the discrepancy between our counts and theirs. While we have
been unable to compare directly with the B05 model, we have compared our results with those of a SAM that uses a similar radiative transfer treatment
\citep{Benson:2012}. We find that relations similar to Equations (\ref{eq:sfr_fitting_function_850}) and (\ref{eq:sfr_fitting_function_11}) hold for the SMGs in the SAM,
so it appears that the radiative transfer is not the primary cause of the discrepancy even though some aspects of the radiative transfer differ significantly
[e.g., the geometry used in the \sunrise calculations is taken directly from the 3-D \gadgettwo simulations, whereas the geometry assumed by {\sc Grasil} is
analytically specified and azimuthally symmetric; the {\sc Grasil} calculations include a sub-resolution model for obscuration from the birth clouds around young stars,
but we opt not to use the corresponding sub-resolution model that is implemented in \sunrise \citep{Groves:2008} for the reasons discussed in sections 2.2.1 and 4.6 of H11]. 

\citet{Dave:2010} did not perform radiative transfer; instead, they assumed that the most rapidly star-forming galaxies in their simulation were SMGs. However, as noted previously (H11), this
simple ansatz is not necessarily true because of differences in dust mass amongst galaxies and the effects of blending. Thus, if dust radiative transfer were performed
in or our fitting functions were applied to the \citet{Dave:2010} simulations, the conclusions might differ.

\subsubsection{Merger evolution}

Perhaps the time evolution of the SFR or dust mass in the B05 SAM and our model differs.
B05 parameterize the SFR in bursts as $\sfr = M_{\rm gas,c}/\tau_{\star}$, where
$M_{\rm{gas,c}}$ is the cold gas mass and $\tau_{\star}$ is a SFR time-scale given by $\tau_{\star} = \max[f_{\rm dyn}\tau_{\rm dyn},
\tau_{\star {\rm burst, min}}]$. Here, $f_{\rm dyn} = 50$, $\tau_{\rm dyn}$ is the dynamical time of the newly formed spheroid,
and $\tau_{\star \rm burst, min} = 0.2$ Gyr. The major merger shown in fig. 1 of H11 has $\mgas \sim 10^{11} \msun$
when the galaxies are at coalescence. Let us suppose that all the gas is cold. Then, the maximum SFR possible give the B05 prescription is
$10^{11} \msun/0.2$ Gyr = 500 $\msunperyr$, $\sim 9$ times less than that of the simulation. If the dust mass is kept constant,
Equations (\ref{eq:sfr_fitting_function_850}) and (\ref{eq:sfr_fitting_function_11}) imply that a factor of 9 decrease in SFR results in a factor of $\sim 2.5$ decrease in (sub)mm flux,
which would significantly affect the predicted counts. This is of course only a crude comparison, but it demonstrates that the SFHs of starbursts in the B05 model may
disagree with those in our simulations, and differences in the dust content may also be important.

Further evidence that the physical modelling of merger-induced starbursts may account for some of the discrepancy is provided by the different
importance of starbursts in the two models.
In the B05 model, starbursts dominate the submm counts by a large margin and contribute significantly to the SFR density of the universe; they dominate
over quiescently star-forming discs for $z \ga 3$. In our model, isolated discs dominate the (sub)mm counts at the lowest (sub)mm fluxes
and quiescently star-forming galaxy-pair SMGs provide a significant contribution to the bright counts. Furthermore, in our model, merger-induced starbursts account
for $\la 5$ per cent of the SFR density of the universe at all redshifts \citep{Hopkins:2010IR_LF}; this is consistent with the starburst-mode contribution
inferred from the stellar densities, surface brightness profiles, kinematics, and stellar populations of observed merger remnants
\citep{Hopkins:2008extra_light,Hopkins:2009cusps,Hopkins:2009cores,HH:2010}.

Differences in the SFHs of mergers may also contribute to the discrepancy between our model and that of \citet{Dave:2010}. It is not clear that the resolution
of the \citet{Dave:2010} simulation ($3.75 h^{-1}$ kpc comoving) is sufficient to resolve the tidal torques that drive merger-induced starbursts. If it is not, the SFRs
during mergers of their simulated galaxies would be underestimated, and part of the discrepancy between their simulated SMGs' SFRs and those observed
could be attributed to resolution rather than IMF variation. Furthermore, their wind prescription may artificially suppress the SFR enhancement
in merger-induced starbursts \citep{Hopkins:2012mergers}.
Fig. 2 of \citet{Dave:2010} shows that \emph{all} their simulated galaxies lie significantly
under the observed SFR-$\mstar$ relation, so it is perhaps not surprising that they cannot reproduce the SFRs of SMGs, which are
a mix of merger-induced starbursts that are outliers from the SFR-$\mstar$ relation and very massive quiescently star-forming
galaxies that lie near the relation (H12; \citealt{Magnelli:2012,Michalowski:2012}).

\subsubsection{Cosmological context}

The third major component of the models that can disagree is the cosmological context, which includes the SMF and merger rates.
In \citet{Hayward:2012thesis}, we demonstrated that the predicted number densities and redshift distribution of isolated disc SMGs are very sensitive to the assumed SMF.
Thus, it is worthwhile to compare the SMF in the SAMs to the observationally derived SMF we have used. While a direct
comparison of the B05 SMF to those in the literature is complicated by the B05 model's use of the flat IMF in starbursts (see section 4.2 of \citealt{Lacey:2010}),
\citet{Swinbank:2008} have shown that the B05 model under-predicts the rest-frame $K$-band
fluxes of SMGs, which suggests the masses of their model SMGs are lower than observed. This would be a natural result of an under-prediction
of the abundance of massive galaxies.

If the B05 model under-predicts the SMF, then they must compensate by making the starburst contribution significantly higher;
they do this by enabling very gas-rich minor mergers to cause strong starbursts (in their model, minor mergers, which only produce a starburst if the most-massive progenitor
has baryonic gas fraction greater than 70 per cent, account for approximately three-quarters
of the SMG population; \citealt{Gonzalez:2011}) and by modifying the IMF in starbursts such that for a given SFR, they produce significantly greater submm flux. 
An under-prediction of the abundance of all massive galaxies and subsequent need to strongly boost the starburst contribution would
explain why the relative contributions of starbursts and quiescently star-forming galaxies differ so significantly.

\subsubsection{Blending}

An additional reason for the discrepancy is that neither B05 nor \citet{Dave:2010} account for the blending of multiple galaxies
into one (sub)mm source, which can be significant for both merging discs (H11, H12) and physically unrelated galaxies \citep{Wang:2011}.
Our model suggest that galaxy-pair SMGs can account for $\sim 30 - 50$ per cent of the SMG population attributable to isolated
discs and mergers. The types of sources \citet{Wang:2011} observed could also be important. Thus, not accounting
for blending could account for a factor of $\sim 2$ discrepancy between the predicted and observed counts.

\subsubsection{Revised counts}

Finally, B05 compared with and \citet{Dave:2010} utilised number counts that have since been superseded by surveys covering significantly
larger areas. The new counts are as much as a factor of 2 lower than those of, e.g., \citet{Chapman:2005}, so the difference
in counts is another non-trivial factor that can explain part of the discrepancy between our conclusion and that of some previous
works.

\subsection{Uncertainties in and limitations of our model}

Our model has several advantages: the SEM enables us to isolate possible discrepancies between our model and other work
that originate from differences in the dynamical evolution of galaxy mergers and the dust radiative transfer calculation rather than more general
issues, such as an overall under-prediction of the SMF at $z \sim 2-4$. By constraining the SMF and gas fractions to match
observations and including no additional gas in our models, we match the $z \sim 0$ mass function by construction.
Because we use 3-D hydrodynamical simulations combined with 3-D dust radiative transfer calculations, we can more accurately
calculate the dynamical evolution of galaxies and the (sub)mm flux densities than can either SAMs or cosmological simulations.
Furthermore, we conservatively assume a Kroupa IMF rather than invoke ad hoc modifications to the IMF.

However, our model also has several limitations: first, computational constraints prevent us from
running simulations scaled to different redshifts. Instead, we scale all initial disc galaxies to $z \sim 3$. Ideally, we would
run simulations with structural parameters and gas fractions scaled to various redshifts because the variation in the galaxies'
physical properties with redshift may cause the (sub)mm light curves to depend on redshift. In future work, we will run a large
suite of simulations that will more exhaustively sample the relevant parameter space; such a suite could be used to more
accurately predict the SMG counts and redshift distribution. However, because our predicted counts are dominated by
galaxies at $z \sim 3$, our conclusions would likely not differ qualitatively.

Second, to have the resolution necessary to perform accurate radiative transfer on a sufficient number of simulated galaxies, we
must use idealised simulations of isolated disc galaxies and mergers rather than cosmological simulations. In principle,
including gas supply from cosmological scales \citep[e.g.,][]{Keres:2005} could change the results. However, such gas supply is implicitly included
in our model for the isolated discs and galaxy-pair subpopulations because we use observationally
derived gas fractions for these subpopulations. Because the duty cycles for mergers are calculated directly from the
merger simulations, which do not include additional gas supply, the gas contents, and thus SFRs, of the starburst SMGs
could be increased if cosmological simulations were used. However, because the (sub)mm flux density depends only weakly
on SFR [a factor of 2 increase in the SFR increases the (sub)mm flux density by $\sim 30$ per cent; H11],
this would not change our results qualitatively. Similarly, differences in the dust content could affect the results,
but a factor of 2 decrease in the dust mass decreases the (sub)mm flux density by $\la 50$ per cent.

Third, our predictions are affected by uncertainties in the observations used to constrain the model. Of particular importance are uncertainties
in the normalisation and shape of the SMF, which are especially significant at $z \ga 4$. For $z \la 3$, the merger rates
predicted by our model are uncertain by a factor of $\la 3$, but the uncertainties are higher at higher redshifts. A decrease
in the normalisation of the SMF or merger rates would decrease the counts predicted by our model. However, the most interesting
conclusions would remain: 1.Galaxy pairs would still provide a significant contribution to the SMG population. 2.
Depending on the factor by which the abundances are decreased, a mildly top-heavy IMF might not be ruled out. However, a flat
IMF would still over-predict the counts even if the SMF normalisation and merger rates were decreased by $\ga 10$ times,
which is surely an overestimate of the uncertainty.

Fourth, we do not include the contribution from physically unrelated (i.e., non-merging) galaxies blended into a single (sub)mm source.
From observations, it is known that such galaxies contribute to the SMG population \citep{Wang:2011}. Whether such sources contribute
significantly to the population is an open question that should be addressed by obtaining redshifts for SMGs observed with
ALMA. If they do contribute significantly, inclusion of this subpopulation would increase the number counts predicted by our model, change
the relative contributions of the subpopulations, and potentially alter the redshift distributions; however, this would only strengthen the evidence
against a top-heavy IMF.

Finally, we do not model the effects of gravitational lensing. Whereas the fainter sources are likely dominated by un-lensed SMGs,
the bright counts ($S_{1.1} \ga 5$ mJy, or $S_{850} \ga 11.5$ mJy) are likely boosted significantly by lensing \citep{Negrello:2007,Negrello:2010,
Paciga:2009,Lima:2010,Vieira:2010}. Furthermore, the likely cause of the discrepancy between the counts observed by \citet{Austermann:2010} and
\citet{Aretxaga:2011} is that the latter counts are boosted by galaxy--galaxy weak lensing
(see the discussion in \citealt{Aretxaga:2011}), so the discrepancy between our predicted counts and the \citet{Aretxaga:2011}
counts might be resolved if we included the effects of lensing from a foreground overdensity of galaxies. Again, our model
predictions are conservative because inclusion of gravitational lensing would boost the counts; thus, inclusion of lensing would
strengthen the argument against a top-heavy IMF.

\section{Conclusions} \label{S:conclusions}

We have presented a novel method to predict the number counts and redshift distribution of SMGs. We
combined a simple SEM with the results of 3-D hydrodynamical simulations and dust
radiative transfer to calculate the contributions of isolated discs, galaxy pairs (i.e., infall-stage mergers), and
late-stage merger-induced starbursts to the SMG number counts.
Our model is constrained to observations as much as possible; consequently, we are able to isolate
the effects of uncertainties related the dynamical evolution of mergers and the dust radiative transfer -- which are perhaps uniquely
relevant to the SMG population -- from more general issues that affect the high-redshift galaxy population as a whole, such as the SMF.
Furthermore, we have conservatively used a Kroupa -- as opposed to flat or top-heavy -- IMF because we wish to test whether we can match the observed
counts without modifying the IMF from what is observed locally. Our principal results are:

\begin{enumerate}

\item Our fiducial model predicts cumulative number counts that agree very well with those observed for fields thought not to be
significantly affected by lensing.

\item Except at the lowest fluxes ($S_{1.1} \la 1$ mJy, or $S_{850} \la 2$ mJy), merger-induced starbursts account for the bulk
of the population not accounted for by galaxy-pair SMGs, and the brightest sources are pre-dominantly merger-induced starbursts.
Thus, isolated discs contribute negligibly to the bright SMG population.
The contribution of isolated discs to the SMG population is a robust testable prediction of our model.

\item Contrary to the conventional wisdom, bright SMGs are not exclusively merger-induced starbursts; our model predicts that quiescently
star-forming galaxy-pair SMGs account for $\sim30-50$ per cent of SMGs with $S_{1.1} \ga 0.5$ mJy ($S_{850} \ga 1$ mJy).
Though the precise fraction is sensitive to the details of the modelling, the prediction that galaxy pairs contribute significantly to the population
(i.e., tens of per cent rather than a few per cent or less) is robust. The observational diagnostics presented in H12 can be used to determine the subset
of the SMG population that are quiescently star-forming, thereby testing this prediction of our model.

\item The typical redshifts of the model and observed SMGs are similar, but the model may over-predict the number of SMGs at $z \ga 4$.
This may be because the SMF used in our model over-predicts the number of massive galaxies at those redshifts; thus, observations of the
abundance of SMGs at $z \ga 4$ may provide useful constraints on the massive end of the SMF at those redshifts.

\item Because we have not modified the IMF, our results suggest
that we cannot reject the null hypothesis that the IMF in high-redshift starbursts is no different than the IMF in local galaxies.
A crude test suggests that if we use even a mildly top-heavy IMF in our model, the SMG counts are significantly over-predicted.
Thus, we conclude that the observed SMG number counts do not provide evidence for a significantly top-heavy IMF.

\item There are multiple possible reasons our conclusions differ from those of previous work, including differences in the radiative transfer calculations,
the merger evolution, and the cosmological context and the lack of a treatment of blending in previous work.

\end{enumerate}

\acknowledgments

We thank Andrew Benson, Romeel Dav\'{e}, Mark Swinbank, and Naoki Yoshida for comments on the manuscript,
Scott Chapman, Micha{\l} Micha{\l}owski, Pierluigi Monaco, Alex Pope, Isaac Roseboom,
and Josh Younger for useful discussion, Carlton Baugh, Cedric Lacey, Fabio Fontanot, Gian-Luigi Granato, Vernesa Smol\v{c}i\'{c},
Axel Wei\ss, and Min Yun for providing data with which we have compared and for useful discussion, and
Volker Springel for providing the non-public version of \gadgettwo used for this work.
DN acknowledges support from a National Science Foundation Grant (AST-1009452).
DK was supported by NASA through Hubble Fellowship grant HST-HF-51276.01-A.
PJ acknowledges support by a grant from the W. M. Keck Foundation.
The simulations in this paper were performed on the Odyssey cluster supported by the FAS Research Computing Group at Harvard University.
\\

\bibliography{std_citations,smg}

\begin{thebibliography}{203}
\expandafter\ifx\csname natexlab\endcsname\relax\def\natexlab#1{#1}\fi

\bibitem[{{Agertz} {et~al}\mbox{.}(2007){Agertz}, {Moore}, {Stadel}, {Potter},
  {Miniati}, {Read}, {Mayer}, {Gawryszczak}, {Kravtsov}, {Nordlund}, {Pearce},
  {Quilis}, {Rudd}, {Springel}, {Stone}, {Tasker}, {Teyssier}, {Wadsley}, \&
  {Walder}}]{Agertz:2007}
{Agertz} O. {et~al.}, 2007, \mnras, 380, 963

\bibitem[{{Almeida} {et~al}\mbox{.}(2011){Almeida}, {Baugh}, \&
  {Lacey}}]{Almeida:2011}
{Almeida} C., {Baugh} C.~M., {Lacey} C.~G., 2011, \mnras, 1312

\bibitem[{{Aretxaga} {et~al}\mbox{.}(2011){Aretxaga}, {Wilson}, {Aguilar},
  {Alberts}, {Scott}, {Scoville}, {Yun}, {Austermann}, {Downes}, {Ezawa},
  {Hatsukade}, {Hughes}, {Kawabe}, {Kohno}, {Oshima}, {Perera}, {Tamura}, \&
  {Zeballos}}]{Aretxaga:2011}
{Aretxaga} I. {et~al.}, 2011, \mnras, 415, 3831

\bibitem[{Asplund {et~al}\mbox{.}(2009)Asplund, Grevesse, Sauval, \&
  Scott}]{Asplund:2009}
Asplund M., Grevesse N., Sauval A.~J., Scott P., 2009, \araa, 47, 481

\bibitem[{{Austermann} {et~al}\mbox{.}(2009){Austermann}, {Aretxaga}, {Hughes},
  {Kang}, {Kim}, {Lowenthal}, {Perera}, {Sanders}, {Scott}, {Scoville},
  {Wilson}, \& {Yun}}]{Austermann:2009}
{Austermann} J.~E. {et~al.}, 2009, \mnras, 393, 1573

\bibitem[{{Austermann} {et~al}\mbox{.}(2010){Austermann}, Dunlop, Perera,
  Scott, Wilson, Aretxaga, Hughes, Almaini, Chapin, Chapman, Cirasuolo,
  Clements, Coppin, Dunne, Dye, Eales, Egami, Farrah, Ferrusca, Flynn, Haig,
  Halpern, Ibar, Ivison, van Kampen, Kang, Kim, Lacey, Lowenthal, Mauskopf,
  McLure, Mortier, Negrello, Oliver, Peacock, Pope, Rawlings, Rieke, Roseboom,
  Rowan-Robinson, Scott, Serjeant, Smail, Swinbank, Stevens, Velazquez, Wagg,
  \& Yun}]{Austermann:2010}
{Austermann} J.~E. {et~al.}, 2010, \mnras, 401, 160

\bibitem[{{Banerji} {et~al}\mbox{.}(2011){Banerji}, {Chapman}, {Smail},
  {Alaghband-Zadeh}, {Swinbank}, {Dunlop}, {Ivison}, \& {Blain}}]{Banerji:2011}
{Banerji} M., {Chapman} S.~C., {Smail} I., {Alaghband-Zadeh} S., {Swinbank}
  A.~M., {Dunlop} J.~S., {Ivison} R.~J., {Blain} A.~W., 2011, \mnras, 418, 1071

\bibitem[{Barger {et~al}\mbox{.}(1998)Barger, Cowie, Sanders, Fulton,
  Taniguchi, Sato, Kawara, \& Okuda}]{Barger:1998}
Barger A.~J., Cowie L.~L., Sanders D.~B., Fulton E., Taniguchi Y., Sato Y.,
  Kawara K., Okuda H., 1998, \nat, 394, 248

\bibitem[{Barnes \& Hernquist(1991)}]{Barnes:1991}
Barnes J., Hernquist L., 1991, \apjl, 370, L65

\bibitem[{Barnes \& Hernquist(1996)}]{Barnes:1996}
Barnes J., Hernquist L., 1996, \apj, 471, 115

\bibitem[{Barnes \& {Hut}(1986)}]{Barnes:1986}
Barnes J., {Hut} P., 1986, \nat, 324, 446

\bibitem[{{Bastian} {et~al}\mbox{.}(2010){Bastian}, {Covey}, \&
  {Meyer}}]{Bastian:2010}
{Bastian} N., {Covey} K.~R., {Meyer} M.~R., 2010, \araa, 48, 339

\bibitem[{{Bauer} \& {Springel}(2012)}]{Bauer:2012}
{Bauer} A., {Springel} V., 2012, \mnras, 423, 2558

\bibitem[{Baugh {et~al}\mbox{.}(2005)Baugh, Lacey, Frenk, Granato, Silva,
  Bressan, Benson, \& Cole}]{Baugh:2005}
Baugh C.~M., Lacey C.~G., Frenk C.~S., Granato G.~L., Silva L., Bressan A.,
  Benson A.~J., Cole S., 2005, \mnras, 356, 1191 (B05)

\bibitem[{{Benson}(2005)}]{Benson:2005}
{Benson} A.~J., 2005, \mnras, 358, 551

\bibitem[{{Benson}(2012)}]{Benson:2012}
{Benson} A.~J., 2012, \na, 17, 175

\bibitem[{{B{\'e}thermin} {et~al}\mbox{.}(2012){B{\'e}thermin}, {Daddi},
  {Magdis}, {Sargent}, {Hezaveh}, {Elbaz}, {Le Borgne}, {Mullaney}, {Pannella},
  {Buat}, {Charmandaris}, {Lagache}, \& {Scott}}]{Bethermin:2012}
{B{\'e}thermin} M. {et~al.}, 2012, arXiv:1208.6512

\bibitem[{Biggs \& Ivison(2008)}]{Biggs:2008}
Biggs A.~D., Ivison R.~J., 2008, \mnras, 385, 893

\bibitem[{{Blain} {et~al}\mbox{.}(1999{\natexlab{a}}){Blain}, {Jameson},
  {Smail}, {Longair}, {Kneib}, \& {Ivison}}]{Blain:1999hier_mod}
{Blain} A.~W., {Jameson} A., {Smail} I., {Longair} M.~S., {Kneib} J.-P.,
  {Ivison} R.~J., 1999{\natexlab{a}}, \mnras, 309, 715

\bibitem[{{Blain} {et~al}\mbox{.}(1999{\natexlab{b}}){Blain}, {Smail},
  {Ivison}, \& {Kneib}}]{Blain:1999hist_of_SF}
{Blain} A.~W., {Smail} I., {Ivison} R.~J., {Kneib} J.-P., 1999{\natexlab{b}},
  \mnras, 302, 632

\bibitem[{Blain {et~al}\mbox{.}(2002)Blain, Smail, Ivison, Kneib, \&
  Frayer}]{Blain:2002}
Blain A.~W., Smail I., Ivison R.~J., Kneib J.-P., Frayer D.~T., 2002, \physrep,
  369, 111

\bibitem[{Bondi(1952)}]{Bondi:1952}
Bondi H., 1952, \mnras, 112, 195

\bibitem[{Bondi \& Hoyle(1944)}]{Bondi:1944}
Bondi H., Hoyle F., 1944, \mnras, 104, 273

\bibitem[{{Bothwell} {et~al}\mbox{.}(2010){Bothwell}, {Chapman}, {Tacconi},
  {Smail}, {Ivison}, {Casey}, {Bertoldi}, {Beswick}, {Biggs}, {Blain}, {Cox},
  {Genzel}, {Greve}, {Kennicutt}, {Muxlow}, {Neri}, \& {Omont}}]{Bothwell:2010}
{Bothwell} M.~S. {et~al.}, 2010, \mnras, 405, 219

\bibitem[{{Bothwell} {et~al}\mbox{.}(2012){Bothwell}, {Smail}, {Chapman},
  {Genzel}, {Ivison}, {Tacconi}, {Alaghband-Zadeh}, {Bertoldi}, {Blain},
  {Casey}, {Cox}, {Greve}, {Lutz}, {Neri}, {Omont}, \&
  {Swinbank}}]{Bothwell:2012}
{Bothwell} M.~S. {et~al.}, 2012, arXiv:1205.1511

\bibitem[{Bouch{\'e} {et~al}\mbox{.}(2007)Bouch{\'e}, Cresci, Davies,
  Eisenhauer, Schreiber, Genzel, Gillessen, Lehnert, Lutz, Nesvadba, Shapiro,
  Sternberg, Tacconi, Verma, Cimatti, Daddi, Renzini, Erb, Shapley, \&
  Steidel}]{Bouche:2007}
Bouch{\'e} N. {et~al.}, 2007, \apj, 671, 303

\bibitem[{{Bower} {et~al}\mbox{.}(2010){Bower}, {Vernon}, {Goldstein},
  {Benson}, {Lacey}, {Baugh}, {Cole}, \& {Frenk}}]{Bower:2010}
{Bower} R.~G., {Vernon} I., {Goldstein} M., {Benson} A.~J., {Lacey} C.~G.,
  {Baugh} C.~M., {Cole} S., {Frenk} C.~S., 2010, \mnras, 407, 2017

\bibitem[{{Brammer} {et~al}\mbox{.}(2011){Brammer}, {Whitaker}, {van Dokkum},
  {Marchesini}, {Franx}, {Kriek}, {Labb{\'e}}, {Lee}, {Muzzin}, {Quadri},
  {Rudnick}, \& {Williams}}]{Brammer:2011}
{Brammer} G.~B. {et~al.}, 2011, \apj, 739, 24

\bibitem[{Bush {et~al}\mbox{.}(2010)Bush, Cox, Hayward, Thilker, Hernquist, \&
  Besla}]{Bush:2010}
Bush S.~J., Cox T.~J., Hayward C.~C., Thilker D., Hernquist L., Besla G., 2010,
  \apj, 713, 780

\bibitem[{Capak {et~al}\mbox{.}(2008)Capak, Carilli, Lee, Aldcroft, Aussel,
  Schinnerer, Wilson, Yun, Blain, Giavalisco, Ilbert, Kartaltepe, Lee,
  McCracken, Mobasher, Salvato, Sasaki, Scott, Sheth, Shioya, Thompson, Elvis,
  Sanders, Scoville, \& Tanaguchi}]{Capak:2008}
Capak P. {et~al.}, 2008, \apjl, 681, L53

\bibitem[{{Carilli} {et~al}\mbox{.}(2010){Carilli}, {Daddi}, {Riechers},
  {Walter}, {Weiss}, {Dannerbauer}, {Morrison}, {Wagg}, {Dav{\'e}}, {Elbaz},
  {Stern}, {Dickinson}, {Krips}, \& {Aravena}}]{Carilli:2010}
{Carilli} C.~L. {et~al.}, 2010, \apj, 714, 1407

\bibitem[{{Chapin} {et~al}\mbox{.}(2009){Chapin}, {Pope}, {Scott}, {Aretxaga},
  {Austermann}, {Chary}, {Coppin}, {Halpern}, {Hughes}, {Lowenthal},
  {Morrison}, {Perera}, {Scott}, {Wilson}, \& {Yun}}]{Chapin:2009}
{Chapin} E.~L. {et~al.}, 2009, \mnras, 398, 1793

\bibitem[{Chapman {et~al}\mbox{.}(2005)Chapman, Blain, Smail, \&
  Ivison}]{Chapman:2005}
Chapman S.~C., Blain A.~W., Smail I., Ivison R.~J., 2005, \apj, 622, 772

\bibitem[{{Chapman} {et~al}\mbox{.}(2010){Chapman}, {Ivison}, {Roseboom},
  {Auld}, {Bock}, {Brisbin}, {Burgarella}, {Chanial}, {Clements}, {Cooray},
  {Eales}, {Franceschini}, {Giovannoli}, {Glenn}, {Griffin}, {Mortier},
  {Oliver}, {Omont}, {Page}, {Papageorgiou}, {Pearson}, {P{\'e}rez-Fournon},
  {Pohlen}, {Rawlings}, {Raymond}, {Rodighiero}, {Rowan-Robinson}, {Scott},
  {Seymour}, {Smith}, {Symeonidis}, {Tugwell}, {Vaccari}, {Vieira}, {Vigroux},
  {Wang}, \& {Wright}}]{Chapman:2010}
{Chapman} S.~C. {et~al.}, 2010, \mnras, 409, L13

\bibitem[{{Chapman} {et~al}\mbox{.}(2003){Chapman}, {Windhorst}, {Odewahn},
  {Yan}, \& {Conselice}}]{Chapman:2003}
{Chapman} S.~C., {Windhorst} R., {Odewahn} S., {Yan} H., {Conselice} C., 2003,
  \apj, 599, 92

\bibitem[{{Clements} {et~al}\mbox{.}(2008){Clements}, {Vaccari}, {Babbedge},
  {Oliver}, {Rowan-Robinson}, {Davoodi}, {Ivison}, {Farrah}, {Dunlop}, {Shupe},
  {Waddington}, {Simpson}, {Furusawa}, {Serjeant}, {Afonso-Luis}, {Alexander},
  {Aretxaga}, {Blain}, {Borys}, {Chapman}, {Coppin}, {Dunne}, {Dye}, {Eales},
  {Evans}, {Fang}, {Frayer}, {Fox}, {Gear}, {Greve}, {Halpern}, {Hughes},
  {Jenness}, {Lonsdale}, {Mortier}, {Page}, {Pope}, {Priddey}, {Rawlings},
  {Savage}, {Scott}, {Scott}, {Sekiguchi}, {Smail}, {Smith}, {Stevens},
  {Surace}, {Takagi}, \& {van Kampen}}]{Clements:2008}
{Clements} D.~L. {et~al.}, 2008, \mnras, 387, 247

\bibitem[{Cole {et~al}\mbox{.}(2000)Cole, Lacey, Baugh, \& Frenk}]{Cole:2000}
Cole S., Lacey C.~G., Baugh C.~M., Frenk C.~S., 2000, \mnras, 319, 168

\bibitem[{{Conroy} \& {van Dokkum}(2012)}]{Conroy:2012}
{Conroy} C., {van Dokkum} P., 2012, arXiv:1205.6473

\bibitem[{Conroy \& Wechsler(2009)}]{Conroy:2009}
Conroy C., Wechsler R.~H., 2009, \apj, 696, 620

\bibitem[{{Coppin} {et~al}\mbox{.}(2006){Coppin}, {Chapin}, {Mortier}, {Scott},
  {Borys}, {Dunlop}, {Halpern}, {Hughes}, {Pope}, {Scott}, {Serjeant}, {Wagg},
  {Alexander}, {Almaini}, {Aretxaga}, {Babbedge}, {Best}, {Blain}, {Chapman},
  {Clements}, {Crawford}, {Dunne}, {Eales}, {Edge}, {Farrah}, {Gazta{\~n}aga},
  {Gear}, {Granato}, {Greve}, {Fox}, {Ivison}, {Jarvis}, {Jenness}, {Lacey},
  {Lepage}, {Mann}, {Marsden}, {Martinez-Sansigre}, {Oliver}, {Page},
  {Peacock}, {Pearson}, {Percival}, {Priddey}, {Rawlings}, {Rowan-Robinson},
  {Savage}, {Seigar}, {Sekiguchi}, {Silva}, {Simpson}, {Smail}, {Stevens},
  {Takagi}, {Vaccari}, {van Kampen}, \& {Willott}}]{Coppin:2006}
{Coppin} K. {et~al.}, 2006, \mnras, 372, 1621

\bibitem[{Coppin {et~al}\mbox{.}(2008)Coppin, Halpern, Scott, Borys, Dunlop,
  Dunne, Ivison, Wagg, Aretxaga, Battistelli, Benson, Blain, Chapman, Clements,
  Dye, Farrah, Hughes, Jenness, van Kampen, Lacey, Mortier, Pope, Priddey,
  Serjeant, Smail, Stevens, \& Vaccari}]{Coppin:2008}
Coppin K. {et~al.}, 2008, \mnras, 384, 1597

\bibitem[{{Cox} {et~al}\mbox{.}(2006{\natexlab{a}}){Cox}, Dutta, Matteo,
  Hernquist, Hopkins, Robertson, \& Springel}]{Cox:2006}
{Cox} T.~J., Dutta S.~N., Matteo T.~D., Hernquist L., Hopkins P.~F., Robertson
  B., Springel V., 2006{\natexlab{a}}, \apj, 650, 791

\bibitem[{{Cox} {et~al}\mbox{.}(2006{\natexlab{b}}){Cox}, {Jonsson}, {Primack},
  \& {Somerville}}]{Cox:2006feedback}
{Cox} T.~J., {Jonsson} P., {Primack} J.~R., {Somerville} R.~S.,
  2006{\natexlab{b}}, \mnras, 373, 1013

\bibitem[{Daddi {et~al}\mbox{.}(2010)Daddi, {Bournaud}, {Walter},
  {Dannerbauer}, {Carilli}, {Dickinson}, {Elbaz}, {Morrison}, {Riechers},
  {Onodera}, {Salmi}, {Krips}, \& {Stern}}]{Daddi:2010}
Daddi E. {et~al.}, 2010, \apj, 713, 686

\bibitem[{{Dale} {et~al}\mbox{.}(2007){Dale}, {Gil de Paz}, {Gordon}, {Hanson},
  {Armus}, {Bendo}, {Bianchi}, {Block}, {Boissier}, {Boselli}, {Buckalew},
  {Buat}, {Burgarella}, {Calzetti}, {Cannon}, {Engelbracht}, {Helou},
  {Hollenbach}, {Jarrett}, {Kennicutt}, {Leitherer}, {Li}, {Madore}, {Martin},
  {Meyer}, {Murphy}, {Regan}, {Roussel}, {Smith}, {Sosey}, {Thilker}, \&
  {Walter}}]{Dale:2007}
{Dale} D.~A. {et~al.}, 2007, \apj, 655, 863

\bibitem[{Dav{\'e}(2008)}]{Dave:2008}
Dav{\'e} R., 2008, \mnras, 385, 147

\bibitem[{Dav{\'e} {et~al}\mbox{.}(2010)Dav{\'e}, Finlator, Oppenheimer,
  Fardal, Katz, Kere{\v s}, \& Weinberg}]{Dave:2010}
Dav{\'e} R., Finlator K., Oppenheimer B.~D., Fardal M., Katz N., Kere{\v s} D.,
  Weinberg D.~H., 2010, \mnras, 404, 1355

\bibitem[{{De Lucia} {et~al}\mbox{.}(2010){De Lucia}, {Boylan-Kolchin},
  {Benson}, {Fontanot}, \& {Monaco}}]{DeLucia:2010}
{De Lucia} G., {Boylan-Kolchin} M., {Benson} A.~J., {Fontanot} F., {Monaco} P.,
  2010, \mnras, 406, 1533

\bibitem[{{Dekel} {et~al}\mbox{.}(2009){Dekel}, {Birnboim}, {Engel},
  {Freundlich}, {Goerdt}, {Mumcuoglu}, {Neistein}, {Pichon}, {Teyssier}, \&
  {Zinger}}]{Dekel:2009nature}
{Dekel} A. {et~al.}, 2009, \nat, 457, 451

\bibitem[{{Devriendt} \& {Guiderdoni}(2000)}]{Devriendt:2000}
{Devriendt} J.~E.~G., {Guiderdoni} B., 2000, \aap, 363, 851

\bibitem[{{Di Matteo} {et~al}\mbox{.}(2005){Di Matteo}, Springel, \&
  Hernquist}]{DiMatteo:2005}
{Di Matteo} T., Springel V., Hernquist L., 2005, \nat, 433, 604

\bibitem[{Draine \& Li(2007)}]{Draine:2007}
Draine B.~T., Li A., 2007, \apj, 657, 810

\bibitem[{Dwek(1998)}]{Dwek:1998}
Dwek E., 1998, \apj, 501, 643

\bibitem[{Eales {et~al}\mbox{.}(1999)Eales, Lilly, Gear, Dunne, Bond, Hammer,
  F{\`e}vre, \& Crampton}]{Eales:1999}
Eales S., Lilly S., Gear W., Dunne L., Bond J.~R., Hammer F., F{\`e}vre O.~L.,
  Crampton D., 1999, \apj, 515, 518

\bibitem[{{Elmegreen}(2004)}]{Elmegreen:2004}
{Elmegreen} B.~G., 2004, \mnras, 354, 367

\bibitem[{{Elmegreen} \& {Shadmehri}(2003)}]{Elmegreen:2003}
{Elmegreen} B.~G., {Shadmehri} M., 2003, \mnras, 338, 817

\bibitem[{{Engel} {et~al}\mbox{.}(2010){Engel}, {Tacconi}, {Davies}, {Neri},
  {Smail}, {Chapman}, {Genzel}, {Cox}, {Greve}, {Ivison}, {Blain}, {Bertoldi},
  \& {Omont}}]{Engel:2010}
{Engel} H. {et~al.}, 2010, \apj, 724, 233

\bibitem[{{Erb} {et~al}\mbox{.}(2006){Erb}, {Steidel}, {Shapley}, {Pettini},
  {Reddy}, \& {Adelberger}}]{Erb:2006}
{Erb} D.~K., {Steidel} C.~C., {Shapley} A.~E., {Pettini} M., {Reddy} N.~A.,
  {Adelberger} K.~L., 2006, \apj, 646, 107

\bibitem[{{Fardal} {et~al}\mbox{.}(2001){Fardal}, {Katz}, {Weinberg},
  {Dav{\'e}}, \& {Hernquist}}]{Fardal:2001}
{Fardal} M.~A., {Katz} N., {Weinberg} D.~H., {Dav{\'e}} R., {Hernquist} L.,
  2001, astro-ph/0107290

\bibitem[{Fontana {et~al}\mbox{.}(2006)Fontana, Salimbeni, Grazian, Giallongo,
  Pentericci, Nonino, Fontanot, Menci, Monaco, Cristiani, Vanzella, de~Santis,
  \& Gallozzi}]{Fontana:2006}
Fontana A. {et~al.}, 2006, \aap, 459, 745

\bibitem[{{Fontanot} {et~al}\mbox{.}(2009){Fontanot}, {De Lucia}, {Monaco},
  {Somerville}, \& {Santini}}]{Fontanot:2009}
{Fontanot} F., {De Lucia} G., {Monaco} P., {Somerville} R.~S., {Santini} P.,
  2009, \mnras, 397, 1776

\bibitem[{{Fontanot} \& {Monaco}(2010)}]{Fontanot:2010}
{Fontanot} F., {Monaco} P., 2010, \mnras, 405, 705

\bibitem[{{Fontanot} {et~al}\mbox{.}(2007){Fontanot}, {Monaco}, {Silva}, \&
  {Grazian}}]{Fontanot:2007}
{Fontanot} F., {Monaco} P., {Silva} L., {Grazian} A., 2007, \mnras, 382, 903

\bibitem[{{Genzel} {et~al}\mbox{.}(2010){Genzel}, {Tacconi}, {Gracia-Carpio},
  {Sternberg}, {Cooper}, {Shapiro}, {Bolatto}, {Bouch{\'e}}, {Bournaud},
  {Burkert}, {Combes}, {Comerford}, {Cox}, {Davis}, {Schreiber},
  {Garcia-Burillo}, {Lutz}, {Naab}, {Neri}, {Omont}, {Shapley}, \&
  {Weiner}}]{Genzel:2010}
{Genzel} R. {et~al.}, 2010, \mnras, 407, 2091

\bibitem[{{Gingold} \& {Monaghan}(1977)}]{Gingold:1977}
{Gingold} R.~A., {Monaghan} J.~J., 1977, \mnras, 181, 375

\bibitem[{{Gonz{\'a}lez} {et~al}\mbox{.}(2011){Gonz{\'a}lez}, {Lacey}, {Baugh},
  \& {Frenk}}]{Gonzalez:2011}
{Gonz{\'a}lez} J.~E., {Lacey} C.~G., {Baugh} C.~M., {Frenk} C.~S., 2011,
  \mnras, 413, 749

\bibitem[{{Granato} {et~al}\mbox{.}(2004){Granato}, {De Zotti}, {Silva},
  {Bressan}, \& {Danese}}]{Granato:2004}
{Granato} G.~L., {De Zotti} G., {Silva} L., {Bressan} A., {Danese} L., 2004,
  \apj, 600, 580

\bibitem[{{Granato} {et~al}\mbox{.}(2000){Granato}, {Lacey}, {Silva},
  {Bressan}, {Baugh}, {Cole}, \& {Frenk}}]{Granato:2000}
{Granato} G.~L., {Lacey} C.~G., {Silva} L., {Bressan} A., {Baugh} C.~M., {Cole}
  S., {Frenk} C.~S., 2000, \apj, 542, 710

\bibitem[{Greve {et~al}\mbox{.}(2005)Greve, Bertoldi, Smail, Neri, Chapman,
  Blain, Ivison, Genzel, Omont, Cox, Tacconi, \& Kneib}]{Greve:2005}
Greve T.~R. {et~al.}, 2005, \mnras, 359, 1165

\bibitem[{Groves {et~al}\mbox{.}(2008)Groves, Dopita, Sutherland, Kewley,
  Fischera, Leitherer, Brandl, \& van Breugel}]{Groves:2008}
Groves B., Dopita M.~A., Sutherland R.~S., Kewley L.~J., Fischera J., Leitherer
  C., Brandl B., van Breugel W., 2008, \apjs, 176, 438

\bibitem[{{Guiderdoni} {et~al}\mbox{.}(1998){Guiderdoni}, {Hivon}, {Bouchet},
  \& {Maffei}}]{Guiderdoni:1998}
{Guiderdoni} B., {Hivon} E., {Bouchet} F.~R., {Maffei} B., 1998, \mnras, 295,
  877

\bibitem[{{Hainline} {et~al}\mbox{.}(2011){Hainline}, {Blain}, {Smail},
  {Alexander}, {Armus}, {Chapman}, \& {Ivison}}]{Hainline:2011}
{Hainline} L.~J., {Blain} A.~W., {Smail} I., {Alexander} D.~M., {Armus} L.,
  {Chapman} S.~C., {Ivison} R.~J., 2011, \apj, 740, 96

\bibitem[{{Hatsukade} {et~al}\mbox{.}(2011){Hatsukade}, {Kohno}, {Aretxaga},
  {Austermann}, {Ezawa}, {Hughes}, {Ikarashi}, {Iono}, {Kawabe}, {Khan},
  {Matsuo}, {Matsuura}, {Nakanishi}, {Oshima}, {Perera}, {Scott}, {Shirahata},
  {Takeuchi}, {Tamura}, {Tanaka}, {Tosaki}, {Wilson}, \&
  {Yun}}]{Hatsukade:2011}
{Hatsukade} B. {et~al.}, 2011, \mnras, 411, 102

\bibitem[{{Hayward}(2012)}]{Hayward:2012thesis}
{Hayward} C.~C., 2012, PhD thesis, Harvard University

\bibitem[{{Hayward} {et~al}\mbox{.}(2012){Hayward}, {Jonsson}, {Kere{\v s}},
  {Magnelli}, {Hernquist}, \& {Cox}}]{Hayward:2012smg_bimodality}
{Hayward} C.~C., {Jonsson} P., {Kere{\v s}} D., {Magnelli} B., {Hernquist} L.,
  {Cox} T.~J., 2012, \mnras, arXiv:1203.1318 (H12)

\bibitem[{{Hayward} {et~al}\mbox{.}(2011{\natexlab{a}}){Hayward}, {Kere{\v s}},
  {Jonsson}, {Narayanan}, {Cox}, \& {Hernquist}}]{Hayward:2011smg_selection}
{Hayward} C.~C., {Kere{\v s}} D., {Jonsson} P., {Narayanan} D., {Cox} T.~J.,
  {Hernquist} L., 2011{\natexlab{a}}, \apj, 743, 159 (H11)

\bibitem[{{Hayward} {et~al}\mbox{.}(2011{\natexlab{b}}){Hayward}, {Narayanan},
  {Jonsson}, {Cox}, {Kere{\v s}}, {Hopkins}, \&
  {Hernquist}}]{Hayward:2011num_cts_proc}
{Hayward} C.~C., {Narayanan} D., {Jonsson} P., {Cox} T.~J., {Kere{\v s}} D.,
  {Hopkins} P.~F., {Hernquist} L., 2011{\natexlab{b}}, in ASP Conf. Ser. 440,
  Have Observations Revealed a Variable Upper End of the Initial Mass
  Function?, {M.~Treyer, T.~Wyder, J.~Neill, M.~Seibert, \& J.~Lee}, ed., ASP,
  San Francisco, CA, p. 369

\bibitem[{Hernquist(1989)}]{Hernquist:1989}
Hernquist L., 1989, \nat, 340, 687

\bibitem[{Hernquist(1990)}]{Hernquist:1990}
Hernquist L., 1990, \apj, 356, 359

\bibitem[{{Hernquist} \& {Katz}(1989)}]{Hernquist:1989treesph}
{Hernquist} L., {Katz} N., 1989, \apjs, 70, 419

\bibitem[{{Hopkins}(2012)}]{Hopkins:2012IMF}
{Hopkins} P.~F., 2012, arXiv:1204.2835

\bibitem[{{Hopkins} {et~al}\mbox{.}(2010{\natexlab{a}}){Hopkins}, Bundy,
  Croton, Hernquist, Kere{\v s}, Khochfar, Stewart, Wetzel, \&
  Younger}]{Hopkins:2010merger_rates}
{Hopkins} P.~F. {et~al.}, 2010{\natexlab{a}}, \apj, 715, 202

\bibitem[{{Hopkins} {et~al}\mbox{.}(2009{\natexlab{a}}){Hopkins}, {Cox},
  {Dutta}, {Hernquist}, {Kormendy}, \& {Lauer}}]{Hopkins:2009cusps}
{Hopkins} P.~F., {Cox} T.~J., {Dutta} S.~N., {Hernquist} L., {Kormendy} J.,
  {Lauer} T.~R., 2009{\natexlab{a}}, \apjs, 181, 135

\bibitem[{{Hopkins} {et~al}\mbox{.}(2012){Hopkins}, {Cox}, {Hernquist},
  {Narayanan}, {Hayward}, \& {Murray}}]{Hopkins:2012mergers}
{Hopkins} P.~F., {Cox} T.~J., {Hernquist} L., {Narayanan} D., {Hayward} C.~C.,
  {Murray} N., 2012, arXiv:1206.0011

\bibitem[{{Hopkins} {et~al}\mbox{.}(2008{\natexlab{a}}){Hopkins}, {Cox},
  {Kere{\v s}}, \& {Hernquist}}]{Hopkins:2008red_Es}
{Hopkins} P.~F., {Cox} T.~J., {Kere{\v s}} D., {Hernquist} L.,
  2008{\natexlab{a}}, \apjs, 175, 390

\bibitem[{{Hopkins} {et~al}\mbox{.}(2010{\natexlab{b}}){Hopkins}, {Croton},
  {Bundy}, {Khochfar}, {van den Bosch}, {Somerville}, {Wetzel}, {Kere{\v s}},
  {Hernquist}, {Stewart}, {Younger}, {Genel}, \&
  {Ma}}]{Hopkins:2010merger_rate_uncertainties}
{Hopkins} P.~F. {et~al.}, 2010{\natexlab{b}}, \apj, 724, 915

\bibitem[{{Hopkins} \& {Hernquist}(2010)}]{HH:2010}
{Hopkins} P.~F., {Hernquist} L., 2010, \mnras, 402, 985

\bibitem[{{Hopkins} {et~al}\mbox{.}(2008{\natexlab{b}}){Hopkins}, {Hernquist},
  {Cox}, {Dutta}, \& {Rothberg}}]{Hopkins:2008extra_light}
{Hopkins} P.~F., {Hernquist} L., {Cox} T.~J., {Dutta} S.~N., {Rothberg} B.,
  2008{\natexlab{b}}, \apj, 679, 156

\bibitem[{{Hopkins} {et~al}\mbox{.}(2008{\natexlab{c}}){Hopkins}, {Hernquist},
  {Cox}, \& {Kere{\v s}}}]{Hopkins:2008cosm_frame1}
{Hopkins} P.~F., {Hernquist} L., {Cox} T.~J., {Kere{\v s}} D.,
  2008{\natexlab{c}}, \apjs, 175, 356

\bibitem[{{Hopkins} {et~al}\mbox{.}(2009{\natexlab{b}}){Hopkins}, {Lauer},
  {Cox}, {Hernquist}, \& {Kormendy}}]{Hopkins:2009cores}
{Hopkins} P.~F., {Lauer} T.~R., {Cox} T.~J., {Hernquist} L., {Kormendy} J.,
  2009{\natexlab{b}}, \apjs, 181, 486

\bibitem[{{Hopkins} {et~al}\mbox{.}(2007){Hopkins}, Richards, \&
  Hernquist}]{Hopkins:2007}
{Hopkins} P.~F., Richards G.~T., Hernquist L., 2007, \apj, 654, 731

\bibitem[{{Hopkins} {et~al}\mbox{.}(2010{\natexlab{c}}){Hopkins}, Younger,
  Hayward, Narayanan, \& Hernquist}]{Hopkins:2010IR_LF}
{Hopkins} P.~F., Younger J.~D., Hayward C.~C., Narayanan D., Hernquist L.,
  2010{\natexlab{c}}, \mnras, 402, 1693

\bibitem[{Hoyle \& Lyttleton(1939)}]{Hoyle:1939}
Hoyle F., Lyttleton R.~A., 1939, Proc. Cam. Philos. Soc., 35, 405

\bibitem[{Hughes {et~al}\mbox{.}(1998)Hughes, Serjeant, Dunlop, Rowan-Robinson,
  Blain, Mann, Ivison, Peacock, Efstathiou, Gear, Oliver, Lawrence, Longair,
  Goldschmidt, \& Jenness}]{Hughes:1998}
Hughes D.~H. {et~al.}, 1998, \nat, 394, 241

\bibitem[{{Ilbert} {et~al}\mbox{.}(2010){Ilbert}, {Salvato}, {Le Floc'h},
  {Aussel}, {Capak}, {McCracken}, {Mobasher}, {Kartaltepe}, {Scoville},
  {Sanders}, {Arnouts}, {Bundy}, {Cassata}, {Kneib}, {Koekemoer}, {Le
  F{\`e}vre}, {Lilly}, {Surace}, {Taniguchi}, {Tasca}, {Thompson}, {Tresse},
  {Zamojski}, {Zamorani}, \& {Zucca}}]{Ilbert:2010}
{Ilbert} O. {et~al.}, 2010, \apj, 709, 644

\bibitem[{Iono {et~al}\mbox{.}(2009)Iono, Wilson, Yun, Baker, Petitpas, Peck,
  Krips, Cox, Matsushita, Mihos, \& Pihlstrom}]{Iono:2009}
Iono D. {et~al.}, 2009, \apj, 695, 1537

\bibitem[{{Ivison} {et~al}\mbox{.}(2007){Ivison}, Greve, Dunlop, Peacock,
  Egami, Smail, Ibar, van Kampen, Aretxaga, Babbedge, Biggs, Blain, Chapman,
  Clements, Coppin, Farrah, Halpern, Hughes, Jarvis, Jenness, Jones, Mortier,
  Oliver, Papovich, P{\'e}rez-Gonz{\'a}lez, Pope, Rawlings, Rieke,
  Rowan-Robinson, Savage, Scott, Seigar, Serjeant, Simpson, Stevens, Vaccari,
  Wagg, \& Willott}]{Ivison:2007}
{Ivison} R.~J. {et~al.}, 2007, \mnras, 380, 199

\bibitem[{{Ivison} {et~al}\mbox{.}(2002){Ivison}, {Greve}, {Smail}, {Dunlop},
  {Roche}, {Scott}, {Page}, {Stevens}, {Almaini}, {Blain}, {Willott}, {Fox},
  {Gilbank}, {Serjeant}, \& {Hughes}}]{Ivison:2002}
{Ivison} R.~J. {et~al.}, 2002, \mnras, 337, 1

\bibitem[{{Ivison} {et~al}\mbox{.}(2010){Ivison}, {Smail}, {Papadopoulos},
  {Wold}, {Richard}, {Swinbank}, {Kneib}, \& {Owen}}]{Ivison:2010}
{Ivison} R.~J., {Smail} I., {Papadopoulos} P.~P., {Wold} I., {Richard} J.,
  {Swinbank} A.~M., {Kneib} J., {Owen} F.~N., 2010, \mnras, 404, 198

\bibitem[{James {et~al}\mbox{.}(2002)James, Dunne, Eales, \&
  Edmunds}]{James:2002}
James A., Dunne L., Eales S., Edmunds M.~G., 2002, \mnras, 335, 753

\bibitem[{Jonsson {et~al}\mbox{.}(2006)Jonsson, Cox, Primack, \&
  Somerville}]{Jonsson:2006}
Jonsson P., Cox T.~J., Primack J.~R., Somerville R.~S., 2006, \apj, 637, 255

\bibitem[{Jonsson {et~al}\mbox{.}(2010)Jonsson, Groves, \&
  Cox}]{Jonsson:2010sunrise}
Jonsson P., Groves B.~A., Cox T.~J., 2010, \mnras, 403, 17

\bibitem[{Jonsson \& Primack(2010)}]{Jonsson:2010gpu}
Jonsson P., Primack J.~R., 2010, New Astron., 15, 509

\bibitem[{Katz {et~al}\mbox{.}(1996)Katz, Weinberg, \& Hernquist}]{Katz:1996}
Katz N., Weinberg D.~H., Hernquist L., 1996, \apjs, 105, 19

\bibitem[{{Kaviani} {et~al}\mbox{.}(2003){Kaviani}, {Haehnelt}, \&
  {Kauffmann}}]{Kaviani:2003}
{Kaviani} A., {Haehnelt} M.~G., {Kauffmann} G., 2003, \mnras, 340, 739

\bibitem[{Kennicutt(1998)}]{Kennicutt:1998}
Kennicutt R.~C., 1998, \apj, 498, 541

\bibitem[{{Kennicutt} {et~al}\mbox{.}(2003){Kennicutt}, {Armus}, {Bendo},
  {Calzetti}, {Dale}, {Draine}, {Engelbracht}, {Gordon}, {Grauer}, {Helou},
  {Hollenbach}, {Jarrett}, {Kewley}, {Leitherer}, {Li}, {Malhotra}, {Regan},
  {Rieke}, {Rieke}, {Roussel}, {Smith}, {Thornley}, \&
  {Walter}}]{Kennicutt:2003}
{Kennicutt} R.~C. {et~al.}, 2003, \pasp, 115, 928

\bibitem[{{Kennicutt}(1983)}]{Kennicutt:1983}
{Kennicutt}, Jr. R.~C., 1983, \apj, 272, 54

\bibitem[{Kere{\v s} {et~al}\mbox{.}(2005)Kere{\v s}, Katz, Weinberg, \&
  Dav{\'e}}]{Keres:2005}
Kere{\v s} D., Katz N., Weinberg D.~H., Dav{\'e} R., 2005, \mnras, 363, 2

\bibitem[{{Kere{\v s}} {et~al}\mbox{.}(2012){Kere{\v s}}, {Vogelsberger},
  {Sijacki}, {Springel}, \& {Hernquist}}]{Keres:2012}
{Kere{\v s}} D., {Vogelsberger} M., {Sijacki} D., {Springel} V., {Hernquist}
  L., 2012, \mnras, 425, 2027

\bibitem[{{Kewley} \& {Ellison}(2008)}]{Kewley:2008}
{Kewley} L.~J., {Ellison} S.~L., 2008, \apj, 681, 1183

\bibitem[{{Khochfar} \& {Burkert}(2006)}]{Khochfar:2006}
{Khochfar} S., {Burkert} A., 2006, \aap, 445, 403

\bibitem[{{Knudsen} {et~al}\mbox{.}(2008){Knudsen}, {van der Werf}, \&
  {Kneib}}]{Knudsen:2008}
{Knudsen} K.~K., {van der Werf} P.~P., {Kneib} J.-P., 2008, \mnras, 384, 1611

\bibitem[{{Kov{\'a}cs} {et~al}\mbox{.}(2006){Kov{\'a}cs}, Chapman, Dowell,
  Blain, Ivison, Smail, \& Phillips}]{Kovacs:2006}
{Kov{\'a}cs} A., Chapman S.~C., Dowell C.~D., Blain A.~W., Ivison R.~J., Smail
  I., Phillips T.~G., 2006, \apj, 650, 592

\bibitem[{{Kov{\'a}cs} {et~al}\mbox{.}(2010){Kov{\'a}cs}, {Omont}, {Beelen},
  {Lonsdale}, {Polletta}, {Fiolet}, {Greve}, {Borys}, {Cox}, {De Breuck},
  {Dole}, {Dowell}, {Farrah}, {Lagache}, {Menten}, {Bell}, \&
  {Owen}}]{Kovacs:2010}
{Kov{\'a}cs} A. {et~al.}, 2010, \apj, 717, 29

\bibitem[{Kroupa(2001)}]{Kroupa:2001}
Kroupa P., 2001, \mnras, 322, 231

\bibitem[{{Krumholz} \& {Thompson}(2007)}]{Krumholz:2007KS}
{Krumholz} M.~R., {Thompson} T.~A., 2007, \apj, 669, 289

\bibitem[{{Lacey} {et~al}\mbox{.}(2010){Lacey}, {Baugh}, {Frenk}, {Benson},
  {Orsi}, {Silva}, {Granato}, \& {Bressan}}]{Lacey:2010}
{Lacey} C.~G., {Baugh} C.~M., {Frenk} C.~S., {Benson} A.~J., {Orsi} A., {Silva}
  L., {Granato} G.~L., {Bressan} A., 2010, \mnras, 405, 2

\bibitem[{{Lacey} {et~al}\mbox{.}(2008){Lacey}, {Baugh}, {Frenk}, {Silva},
  {Granato}, \& {Bressan}}]{Lacey:2008}
{Lacey} C.~G., {Baugh} C.~M., {Frenk} C.~S., {Silva} L., {Granato} G.~L.,
  {Bressan} A., 2008, \mnras, 385, 1155

\bibitem[{{Lagache} {et~al}\mbox{.}(2003){Lagache}, {Dole}, \&
  {Puget}}]{Lagache:2003}
{Lagache} G., {Dole} H., {Puget} J.-L., 2003, \mnras, 338, 555

\bibitem[{{Larson}(1998)}]{Larson:1998}
{Larson} R.~B., 1998, \mnras, 301, 569

\bibitem[{{Larson}(2005)}]{Larson:2005}
{Larson} R.~B., 2005, \mnras, 359, 211

\bibitem[{Leitherer {et~al}\mbox{.}(1999)Leitherer, Schaerer, Goldader,
  Delgado, Robert, Kune, de~Mello, Devost, \& Heckman}]{Leitherer:1999}
Leitherer C. {et~al.}, 1999, \apjs, 123, 3

\bibitem[{{Lima} {et~al}\mbox{.}(2010){Lima}, {Jain}, {Devlin}, \&
  {Aguirre}}]{Lima:2010}
{Lima} M., {Jain} B., {Devlin} M., {Aguirre} J., 2010, \apjl, 717, L31

\bibitem[{{Lo Faro} {et~al}\mbox{.}(2009){Lo Faro}, {Monaco}, {Vanzella},
  {Fontanot}, {Silva}, \& {Cristiani}}]{LoFaro:2009}
{Lo Faro} B., {Monaco} P., {Vanzella} E., {Fontanot} F., {Silva} L.,
  {Cristiani} S., 2009, \mnras, 399, 827

\bibitem[{{Lonsdale} {et~al}\mbox{.}(2006){Lonsdale}, {Farrah}, \&
  {Smith}}]{Lonsdale:2006}
{Lonsdale} C.~J., {Farrah} D., {Smith} H.~E., 2006, {Ultraluminous Infrared
  Galaxies}, Springer Verlag, p. 285

\bibitem[{{Lu} {et~al}\mbox{.}(2012){Lu}, {Mo}, {Katz}, \&
  {Weinberg}}]{Lu:2012}
{Lu} Y., {Mo} H.~J., {Katz} N., {Weinberg} M.~D., 2012, \mnras, 2380

\bibitem[{{Lu} {et~al}\mbox{.}(2011){Lu}, {Mo}, {Weinberg}, \&
  {Katz}}]{Lu:2011a}
{Lu} Y., {Mo} H.~J., {Weinberg} M.~D., {Katz} N., 2011, \mnras, 416, 1949

\bibitem[{{Lucy}(1977)}]{Lucy:1977}
{Lucy} L.~B., 1977, \aj, 82, 1013

\bibitem[{{Magnelli} {et~al}\mbox{.}(2010){Magnelli}, {Lutz}, {Berta},
  {Altieri}, {Andreani}, {Aussel}, {Casta{\~n}eda}, {Cava}, {Cepa}, {Cimatti},
  {Daddi}, {Dannerbauer}, {Dominguez}, {Elbaz}, {F{\"o}rster Schreiber},
  {Genzel}, {Grazian}, {Gruppioni}, {Magdis}, {Maiolino}, {Nordon}, {P{\'e}rez
  Fournon}, {P{\'e}rez Garc{\'{\i}}a}, {Poglitsch}, {Popesso}, {Pozzi},
  {Riguccini}, {Rodighiero}, {Saintonge}, {Santini}, {Sanchez-Portal}, {Shao},
  {Sturm}, {Tacconi}, {Valtchanov}, {Wieprecht}, \&
  {Wiezorrek}}]{Magnelli:2010}
{Magnelli} B. {et~al.}, 2010, \aap, 518, L28

\bibitem[{{Magnelli} {et~al}\mbox{.}(2012){Magnelli}, {Lutz}, {Santini},
  {Saintonge}, {Berta}, {Albrecht}, {Altieri}, {Andreani}, {Aussel},
  {Bertoldi}, {B{\'e}thermin}, {Bongiovanni}, {Capak}, {Chapman}, {Cepa},
  {Cimatti}, {Cooray}, {Daddi}, {Danielson}, {Dannerbauer}, {Dunlop}, {Elbaz},
  {Farrah}, {F{\"o}rster Schreiber}, {Genzel}, {Hwang}, {Ibar}, {Ivison}, {Le
  Floc'h}, {Magdis}, {Maiolino}, {Nordon}, {Oliver}, {P{\'e}rez Garc{\'{\i}}a},
  {Poglitsch}, {Popesso}, {Pozzi}, {Riguccini}, {Rodighiero}, {Rosario},
  {Roseboom}, {Salvato}, {Sanchez-Portal}, {Scott}, {Smail}, {Sturm},
  {Swinbank}, {Tacconi}, {Valtchanov}, {Wang}, \& {Wuyts}}]{Magnelli:2012}
{Magnelli} B. {et~al.}, 2012, \aap, 539, A155

\bibitem[{{Maiolino} {et~al}\mbox{.}(2008){Maiolino}, {Nagao}, {Grazian},
  {Cocchia}, {Marconi}, {Mannucci}, {Cimatti}, {Pipino}, {Ballero}, {Calura},
  {Chiappini}, {Fontana}, {Granato}, {Matteucci}, {Pastorini}, {Pentericci},
  {Risaliti}, {Salvati}, \& {Silva}}]{Maiolino:2008}
{Maiolino} R. {et~al.}, 2008, \aap, 488, 463

\bibitem[{Marchesini {et~al}\mbox{.}(2009)Marchesini, van Dokkum, {F{\"o}rster
  Schreiber}, Franx, Labb{\'e}, \& Wuyts}]{Marchesini:2009}
Marchesini D., van Dokkum P.~G., {F{\"o}rster Schreiber} N.~M., Franx M.,
  Labb{\'e} I., Wuyts S., 2009, \apj, 701, 1765

\bibitem[{{Micha{\l}owski} {et~al}\mbox{.}(2012){Micha{\l}owski}, {Dunlop},
  {Cirasuolo}, {Hjorth}, {Hayward}, \& {Watson}}]{Michalowski:2012}
{Micha{\l}owski} M.~J., {Dunlop} J.~S., {Cirasuolo} M., {Hjorth} J., {Hayward}
  C.~C., {Watson} D., 2012, \aap, 541, A85

\bibitem[{{Micha{\l}owski} {et~al}\mbox{.}(2010){Micha{\l}owski}, {Hjorth}, \&
  {Watson}}]{Michalowski:2010masses}
{Micha{\l}owski} M.~J., {Hjorth} J., {Watson} D., 2010, \aap, 514, A67

\bibitem[{{Mihos} \& {Hernquist}(1996)}]{Mihos:1996}
{Mihos} J.~C., {Hernquist} L., 1996, \apj, 464, 641

\bibitem[{{Narayanan} {et~al}\mbox{.}(2012{\natexlab{a}}){Narayanan},
  {Bothwell}, \& {Dav{\'e}}}]{Narayanan:2012gas_frac}
{Narayanan} D., {Bothwell} M., {Dav{\'e}} R., 2012{\natexlab{a}},
  arXiv:1209.0771

\bibitem[{Narayanan {et~al}\mbox{.}(2011)Narayanan, {Cox}, {Hayward}, \&
  {Hernquist}}]{Narayanan:2011ks}
Narayanan D., {Cox} T.~J., {Hayward} C.~C., {Hernquist} L., 2011, \mnras, 412,
  287

\bibitem[{Narayanan {et~al}\mbox{.}(2009)Narayanan, Cox, Hayward, Younger, \&
  Hernquist}]{Narayanan:2009}
Narayanan D., Cox T.~J., Hayward C.~C., Younger J.~D., Hernquist L., 2009,
  \mnras, 400, 1919

\bibitem[{{Narayanan} {et~al}\mbox{.}(2008){Narayanan}, {Cox}, \&
  {Hernquist}}]{Narayanan:2008SFR_dense_gas}
{Narayanan} D., {Cox} T.~J., {Hernquist} L., 2008, \apjl, 681, L77

\bibitem[{Narayanan {et~al}\mbox{.}(2008)Narayanan, Cox, Shirley, Dav{\'e},
  Hernquist, \& Walker}]{Narayanan:2008CO_SFR}
Narayanan D., Cox T.~J., Shirley Y., Dav{\'e} R., Hernquist L., Walker C.~K.,
  2008, \apj, 684, 996

\bibitem[{{Narayanan} \& {Dav{\'e}}(2012)}]{Narayanan:2012IMF}
{Narayanan} D., {Dav{\'e}} R., 2012, \mnras, 3122

\bibitem[{Narayanan {et~al}\mbox{.}(2010{\natexlab{a}})Narayanan, {Dey},
  {Hayward}, {Cox}, {Bussmann}, {Brodwin}, {Jonsson}, {Hopkins}, {Groves},
  {Younger}, \& {Hernquist}}]{Narayanan:2010dog}
Narayanan D. {et~al.}, 2010{\natexlab{a}}, \mnras, 407, 1701

\bibitem[{Narayanan {et~al}\mbox{.}(2010{\natexlab{b}})Narayanan, Hayward, Cox,
  Hernquist, Jonsson, Younger, \& Groves}]{Narayanan:2010smg}
Narayanan D., Hayward C.~C., Cox T.~J., Hernquist L., Jonsson P., Younger
  J.~D., Groves B., 2010{\natexlab{b}}, \mnras, 401, 1613

\bibitem[{{Narayanan} {et~al}\mbox{.}(2012{\natexlab{b}}){Narayanan},
  {Krumholz}, {Ostriker}, \& {Hernquist}}]{Narayanan:2012X_CO_II}
{Narayanan} D., {Krumholz} M.~R., {Ostriker} E.~C., {Hernquist} L.,
  2012{\natexlab{b}}, \mnras, 421, 3127

\bibitem[{{Negrello} {et~al}\mbox{.}(2010){Negrello}, {Hopwood}, {De Zotti},
  {Cooray}, {Verma}, {Bock}, {Frayer}, {Gurwell}, {Omont}, {Neri},
  {Dannerbauer}, {Leeuw}, {Barton}, {Cooke}, {Kim}, {da Cunha}, {Rodighiero},
  {Cox}, {Bonfield}, {Jarvis}, {Serjeant}, {Ivison}, {Dye}, {Aretxaga},
  {Hughes}, {Ibar}, {Bertoldi}, {Valtchanov}, {Eales}, {Dunne}, {Driver},
  {Auld}, {Buttiglione}, {Cava}, {Grady}, {Clements}, {Dariush}, {Fritz},
  {Hill}, {Hornbeck}, {Kelvin}, {Lagache}, {Lopez-Caniego}, {Gonzalez-Nuevo},
  {Maddox}, {Pascale}, {Pohlen}, {Rigby}, {Robotham}, {Simpson}, {Smith},
  {Temi}, {Thompson}, {Woodgate}, {York}, {Aguirre}, {Beelen}, {Blain},
  {Baker}, {Birkinshaw}, {Blundell}, {Bradford}, {Burgarella}, {Danese},
  {Dunlop}, {Fleuren}, {Glenn}, {Harris}, {Kamenetzky}, {Lupu}, {Maddalena},
  {Madore}, {Maloney}, {Matsuhara}, {Michaowski}, {Murphy}, {Naylor}, {Nguyen},
  {Popescu}, {Rawlings}, {Rigopoulou}, {Scott}, {Scott}, {Seibert}, {Smail},
  {Tuffs}, {Vieira}, {van der Werf}, \& {Zmuidzinas}}]{Negrello:2010}
{Negrello} M. {et~al.}, 2010, Science, 330, 800

\bibitem[{{Negrello} {et~al}\mbox{.}(2007){Negrello}, {Perrotta},
  {Gonz{\'a}lez-Nuevo}, {Silva}, {de Zotti}, {Granato}, {Baccigalupi}, \&
  {Danese}}]{Negrello:2007}
{Negrello} M., {Perrotta} F., {Gonz{\'a}lez-Nuevo} J., {Silva} L., {de Zotti}
  G., {Granato} G.~L., {Baccigalupi} C., {Danese} L., 2007, \mnras, 377, 1557

\bibitem[{Neri {et~al}\mbox{.}(2003)Neri, Genzel, Ivison, Bertoldi, Blain,
  Chapman, Cox, Greve, Omont, \& Frayer}]{Neri:2003}
Neri R. {et~al.}, 2003, \apj, 597, L113

\bibitem[{{Paciga} {et~al}\mbox{.}(2009){Paciga}, {Scott}, \&
  {Chapin}}]{Paciga:2009}
{Paciga} G., {Scott} D., {Chapin} E.~L., 2009, \mnras, 395, 1153

\bibitem[{{Pearson} \& {Rowan-Robinson}(1996)}]{Pearson:1996}
{Pearson} C., {Rowan-Robinson} M., 1996, \mnras, 283, 174

\bibitem[{{Pope} {et~al}\mbox{.}(2006){Pope}, {Scott}, {Dickinson}, {Chary},
  {Morrison}, {Borys}, {Sajina}, {Alexander}, {Daddi}, {Frayer}, {MacDonald},
  \& {Stern}}]{Pope:2006}
{Pope} A. {et~al.}, 2006, \mnras, 370, 1185

\bibitem[{{Ricciardelli} {et~al}\mbox{.}(2010){Ricciardelli}, {Trujillo},
  {Buitrago}, \& {Conselice}}]{Ricciardelli:2010}
{Ricciardelli} E., {Trujillo} I., {Buitrago} F., {Conselice} C.~J., 2010,
  \mnras, 406, 230

\bibitem[{{Riechers} {et~al}\mbox{.}(2011{\natexlab{a}}){Riechers}, {Carilli},
  {Walter}, {Weiss}, {Wagg}, {Bertoldi}, {Downes}, {Henkel}, \&
  {Hodge}}]{Riechers:2011b}
{Riechers} D.~A. {et~al.}, 2011{\natexlab{a}}, \apjl, 733, L11

\bibitem[{{Riechers} {et~al}\mbox{.}(2011{\natexlab{b}}){Riechers}, {Hodge},
  {Walter}, {Carilli}, \& {Bertoldi}}]{Riechers:2011a}
{Riechers} D.~A., {Hodge} J., {Walter} F., {Carilli} C.~L., {Bertoldi} F.,
  2011{\natexlab{b}}, \apjl, 739, L31

\bibitem[{{Robertson} {et~al}\mbox{.}(2006){Robertson}, Hernquist, Cox, Matteo,
  Hopkins, Martini, \& Springel}]{Robertson:2006}
{Robertson} B., Hernquist L., Cox T.~J., Matteo T.~D., Hopkins P.~F., Martini
  P., Springel V., 2006, \apj, 641, 90

\bibitem[{{Roseboom} {et~al}\mbox{.}(2012){Roseboom}, {Ivison}, {Greve},
  {Amblard}, {Arumugam}, {Auld}, {Aussel}, {Bethermin}, {Blain}, {Bock},
  {Boselli}, {Brisbin}, {Buat}, {Burgarella}, {Castro-Rodr{\'{\i}}guez},
  {Cava}, {Chanial}, {Chapin}, {Chapman}, {Clements}, {Conley}, {Conversi},
  {Cooray}, {Dowell}, {Dunlop}, {Dwek}, {Eales}, {Elbaz}, {Farrah},
  {Franceschini}, {Glenn}, {Griffin}, {Halpern}, {Hatziminaoglou}, {Ibar},
  {Isaak}, {Lagache}, {Levenson}, {Lu}, {Madden}, {Maffei}, {Mainetti},
  {Marchetti}, {Marsden}, {Morrison}, {Mortier}, {Nguyen}, {O'Halloran},
  {Oliver}, {Omont}, {Page}, {Panuzzo}, {Papageorgiou}, {Pearson},
  {P{\'e}rez-Fournon}, {Pohlen}, {Rawlings}, {Raymond}, {Rigopoulou}, {Rizzo},
  {Rodighiero}, {Rowan-Robinson}, {Schulz}, {Scott}, {Seymour}, {Shupe},
  {Smith}, {Stevens}, {Symeonidis}, {Trichas}, {Tugwell}, {Vaccari},
  {Valtchanov}, {Vieira}, {Viero}, {Vigroux}, {Wardlow}, {Wang}, {Wright},
  {Xu}, \& {Zemcov}}]{Roseboom:2012}
{Roseboom} I.~G. {et~al.}, 2012, \mnras, 419, 2758

\bibitem[{{Savaglio} {et~al}\mbox{.}(2005){Savaglio}, {Glazebrook}, {Le
  Borgne}, {Juneau}, {Abraham}, {Chen}, {Crampton}, {McCarthy}, {Carlberg},
  {Marzke}, {Roth}, {J{\o}rgensen}, \& {Murowinski}}]{Savaglio:2005}
{Savaglio} S. {et~al.}, 2005, \apj, 635, 260

\bibitem[{Schmidt(1959)}]{Schmidt:1959}
Schmidt M., 1959, \apj, 129, 243

\bibitem[{{Scott} {et~al}\mbox{.}(2010){Scott}, {Yun}, {Wilson}, {Austermann},
  {Aguilar}, {Aretxaga}, {Ezawa}, {Ferrusca}, {Hatsukade}, {Hughes}, {Iono},
  {Giavalisco}, {Kawabe}, {Kohno}, {Mauskopf}, {Oshima}, {Perera}, {Rand},
  {Tamura}, {Tosaki}, {Velazquez}, {Williams}, \& {Zeballos}}]{Scott:2010}
{Scott} K.~S. {et~al.}, 2010, \mnras, 405, 2260

\bibitem[{{Scudder} {et~al}\mbox{.}(2012){Scudder}, {Ellison}, {Torrey},
  {Patton}, \& {Mendel}}]{Scudder:2012}
{Scudder} J.~M., {Ellison} S.~L., {Torrey} P., {Patton} D.~R., {Mendel} J.~T.,
  2012, arXiv:1207.4791

\bibitem[{{Shimizu} {et~al}\mbox{.}(2012){Shimizu}, {Yoshida}, \&
  {Okamoto}}]{Shimizu:2012}
{Shimizu} I., {Yoshida} N., {Okamoto} T., 2012, arXiv:1207.3856

\bibitem[{{Sijacki} {et~al}\mbox{.}(2012){Sijacki}, {Vogelsberger}, {Kere{\v
  s}}, {Springel}, \& {Hernquist}}]{Sijacki:2012}
{Sijacki} D., {Vogelsberger} M., {Kere{\v s}} D., {Springel} V., {Hernquist}
  L., 2012, \mnras, 424, 2999

\bibitem[{Silva {et~al}\mbox{.}(1998)Silva, Granato, Bressan, \&
  Danese}]{Silva:1998}
Silva L., Granato G.~L., Bressan A., Danese L., 1998, \apj, 509, 103

\bibitem[{Smail {et~al}\mbox{.}(2004)Smail, Chapman, Blain, \&
  Ivison}]{Smail:2004}
Smail I., Chapman S.~C., Blain A.~W., Ivison R.~J., 2004, \apj, 616, 71

\bibitem[{Smail {et~al}\mbox{.}(1997)Smail, Ivison, \& Blain}]{Smail:1997}
Smail I., Ivison R.~J., Blain A.~W., 1997, \apjl, 490, L5

\bibitem[{Smol\v{c}i\'{c} {et~al}\mbox{.}(2012)Smol\v{c}i\'{c}, {Aravena},
  {Navarrete}, {Schinnerer}, {Riechers}, {Bertoldi}, {Feruglio}, {Finoguenov},
  {Salvato}, {Sargent}, {McCracken}, {Albrecht}, {Karim}, {Capak}, {Carilli},
  {Cappelluti}, {Elvis}, {Ilbert}, {Kartaltepe}, {Lilly}, {Sanders}, {Sheth},
  {Scoville}, \& {Taniguchi}}]{Smolcic:2012}
Smol\v{c}i\'{c} V. {et~al.}, 2012, arXiv:1205.6470

\bibitem[{{Snyder} {et~al}\mbox{.}(2011){Snyder}, {Cox}, {Hayward},
  {Hernquist}, \& {Jonsson}}]{Snyder:2011}
{Snyder} G.~F., {Cox} T.~J., {Hayward} C.~C., {Hernquist} L., {Jonsson} P.,
  2011, \apj, 741, 77

\bibitem[{{Somerville} {et~al}\mbox{.}(2008){Somerville}, {Barden}, {Rix},
  {Bell}, {Beckwith}, {Borch}, {Caldwell}, {H{\"a}u{\ss}ler}, {Heymans},
  {Jahnke}, {Jogee}, {McIntosh}, {Meisenheimer}, {Peng}, {S{\'a}nchez},
  {Wisotzki}, \& {Wolf}}]{Somerville:2008disk_sizes}
{Somerville} R.~S. {et~al.}, 2008, \apj, 672, 776

\bibitem[{Springel(2005)}]{Springel:2005gadget}
Springel V., 2005, \mnras, 364, 1105

\bibitem[{{Springel}(2010)}]{Springel:2010arepo}
{Springel} V., 2010, \mnras, 401, 791

\bibitem[{Springel(2010)}]{Springel:2010}
Springel V., 2010, \araa, 48, 391

\bibitem[{Springel {et~al}\mbox{.}(2005)Springel, {Di Matteo}, \&
  Hernquist}]{Springel:2005feedback}
Springel V., {Di Matteo} T., Hernquist L., 2005, \mnras, 361, 776

\bibitem[{Springel \& Hernquist(2002)}]{Springel:2002}
Springel V., Hernquist L., 2002, \mnras, 333, 649

\bibitem[{Springel \& Hernquist(2003)}]{Springel:2003}
Springel V., Hernquist L., 2003, \mnras, 339, 289

\bibitem[{{Springel} {et~al}\mbox{.}(2001){Springel}, {Yoshida}, \&
  {White}}]{Springel:2001gadget}
{Springel} V., {Yoshida} N., {White} S.~D.~M., 2001, NewA, 6, 79

\bibitem[{Swinbank {et~al}\mbox{.}(2008)Swinbank, Lacey, Smail, Baugh, Frenk,
  Blain, Chapman, Coppin, Ivison, Gonzalez, \& Hainline}]{Swinbank:2008}
Swinbank A.~M. {et~al.}, 2008, \mnras, 391, 420

\bibitem[{Swinbank {et~al}\mbox{.}(2004)Swinbank, Smail, Chapman, Blain,
  Ivison, \& Keel}]{Swinbank:2004}
Swinbank A.~M., Smail I., Chapman S.~C., Blain A.~W., Ivison R.~J., Keel W.~C.,
  2004, \apj, 617, 64

\bibitem[{Tacconi {et~al}\mbox{.}(2008)Tacconi, Genzel, Smail, Neri, Chapman,
  Ivison, Blain, Cox, Omont, Bertoldi, Greve, Schreiber, Genel, Lutz, Swinbank,
  Shapley, Erb, Cimatti, Daddi, \& Baker}]{Tacconi:2008}
Tacconi L.~J. {et~al.}, 2008, \apj, 680, 246

\bibitem[{Tacconi {et~al}\mbox{.}(2006)Tacconi, Neri, Chapman, Genzel, Smail,
  Ivison, Bertoldi, Blain, Cox, Greve, \& Omont}]{Tacconi:2006}
Tacconi L.~J. {et~al.}, 2006, \apj, 640, 228

\bibitem[{{Targett} {et~al}\mbox{.}(2012){Targett}, {Dunlop}, {Cirasuolo},
  {McLure}, {Bruce}, {Fontana}, {Galametz}, {Paris}, {Dav{\'e}}, {Dekel},
  {Faber}, {Ferguson}, {Grogin}, {Kartaltepe}, {Kocevski}, {Koekemoer},
  {Kurczynski}, {Lai}, \& {Lotz}}]{Targett:2012}
{Targett} T.~A. {et~al.}, 2012, arXiv:1208.3464

\bibitem[{{Targett} {et~al}\mbox{.}(2011){Targett}, {Dunlop}, {McLure}, {Best},
  {Cirasuolo}, \& {Almaini}}]{Targett:2011}
{Targett} T.~A., {Dunlop} J.~S., {McLure} R.~J., {Best} P.~N., {Cirasuolo} M.,
  {Almaini} O., 2011, \mnras, 412, 295

\bibitem[{{Torrey} {et~al}\mbox{.}(2011){Torrey}, {Vogelsberger}, {Sijacki},
  {Springel}, \& {Hernquist}}]{Torrey:2011}
{Torrey} P., {Vogelsberger} M., {Sijacki} D., {Springel} V., {Hernquist} L.,
  2011, arXiv:1110.5635

\bibitem[{{Tremonti} {et~al}\mbox{.}(2004){Tremonti}, {Heckman}, {Kauffmann},
  {Brinchmann}, {Charlot}, {White}, {Seibert}, {Peng}, {Schlegel}, {Uomoto},
  {Fukugita}, \& {Brinkmann}}]{Tremonti:2004}
{Tremonti} C.~A. {et~al.}, 2004, \apj, 613, 898

\bibitem[{{van Dokkum}(2008)}]{vanDokkum:2008}
{van Dokkum} P.~G., 2008, \apj, 674, 29

\bibitem[{{van Dokkum} \& {Conroy}(2010)}]{vanDokkum:2010}
{van Dokkum} P.~G., {Conroy} C., 2010, \nat, 468, 940

\bibitem[{{van Dokkum} \& {Conroy}(2011)}]{vanDokkum:2011}
{van Dokkum} P.~G., {Conroy} C., 2011, \apjl, 735, L13

\bibitem[{{Vieira} {et~al}\mbox{.}(2010){Vieira}, {Crawford}, {Switzer}, {Ade},
  {Aird}, {Ashby}, {Benson}, {Bleem}, {Brodwin}, {Carlstrom}, {Chang}, {Cho},
  {Crites}, {de Haan}, {Dobbs}, {Everett}, {George}, {Gladders}, {Hall},
  {Halverson}, {High}, {Holder}, {Holzapfel}, {Hrubes}, {Joy}, {Keisler},
  {Knox}, {Lee}, {Leitch}, {Lueker}, {Marrone}, {McIntyre}, {McMahon}, {Mehl},
  {Meyer}, {Mohr}, {Montroy}, {Padin}, {Plagge}, {Pryke}, {Reichardt}, {Ruhl},
  {Schaffer}, {Shaw}, {Shirokoff}, {Spieler}, {Stalder}, {Staniszewski},
  {Stark}, {Vanderlinde}, {Walsh}, {Williamson}, {Yang}, {Zahn}, \&
  {Zenteno}}]{Vieira:2010}
{Vieira} J.~D. {et~al.}, 2010, \apj, 719, 763

\bibitem[{{Viola} {et~al}\mbox{.}(2008){Viola}, {Monaco}, {Borgani}, {Murante},
  \& {Tornatore}}]{Viola:2008}
{Viola} M., {Monaco} P., {Borgani} S., {Murante} G., {Tornatore} L., 2008,
  \mnras, 383, 777

\bibitem[{{Vogelsberger} {et~al}\mbox{.}(2011){Vogelsberger}, {Sijacki},
  {Kere{\v s}}, {Springel}, \& {Hernquist}}]{Vogelsberger:2011}
{Vogelsberger} M., {Sijacki} D., {Kere{\v s}} D., {Springel} V., {Hernquist}
  L., 2011, arXiv:1109.1281

\bibitem[{{Wang} {et~al}\mbox{.}(2011){Wang}, {Cowie}, {Barger}, \&
  {Williams}}]{Wang:2011}
{Wang} W.-H., {Cowie} L.~L., {Barger} A.~J., {Williams} J.~P., 2011, \apjl,
  726, L18

\bibitem[{{Wardlow} {et~al}\mbox{.}(2011){Wardlow}, {Smail}, {Coppin},
  {Alexander}, {Brandt}, {Danielson}, {Luo}, {Swinbank}, {Walter}, {Wei{\ss}},
  {Xue}, {Zibetti}, {Bertoldi}, {Biggs}, {Chapman}, {Dannerbauer}, {Dunlop},
  {Gawiser}, {Ivison}, {Knudsen}, {Kov{\'a}cs}, {Lacey}, {Menten}, {Padilla},
  {Rix}, \& {van der Werf}}]{Wardlow:2011}
{Wardlow} J.~L. {et~al.}, 2011, \mnras, 415, 1479

\bibitem[{Weingartner \& Draine(2001)}]{Weingartner:2001}
Weingartner J.~C., Draine B.~T., 2001, \apj, 548, 296

\bibitem[{{Wei{\ss}} {et~al}\mbox{.}(2009){Wei{\ss}}, {Kov{\'a}cs}, {Coppin},
  {Greve}, {Walter}, {Smail}, {Dunlop}, {Knudsen}, {Alexander}, {Bertoldi},
  {Brandt}, {Chapman}, {Cox}, {Dannerbauer}, {De Breuck}, {Gawiser}, {Ivison},
  {Lutz}, {Menten}, {Koekemoer}, {Kreysa}, {Kurczynski}, {Rix}, {Schinnerer},
  \& {van der Werf}}]{Weiss:2009}
{Wei{\ss}} A. {et~al.}, 2009, \apj, 707, 1201

\bibitem[{{Whitaker} {et~al}\mbox{.}(2012){Whitaker}, {van Dokkum}, {Brammer},
  \& {Franx}}]{Whitaker:2012}
{Whitaker} K.~E., {van Dokkum} P.~G., {Brammer} G., {Franx} M., 2012, \apjl,
  754, L29

\bibitem[{Wilson {et~al}\mbox{.}(2008)Wilson, Austermann, Perera, Scott, Ade,
  Bock, Glenn, Golwala, Kim, Kang, Lydon, Mauskopf, Predmore, Roberts, Souccar,
  \& Yun}]{Wilson:2008}
Wilson G.~W. {et~al.}, 2008, \mnras, 386, 807

\bibitem[{Wuyts {et~al}\mbox{.}(2010)Wuyts, Cox, Hayward, Franx, Hernquist,
  Hopkins, Jonsson, \& van Dokkum}]{Wuyts:2010}
Wuyts S., Cox T.~J., Hayward C.~C., Franx M., Hernquist L., Hopkins P.~F.,
  Jonsson P., van Dokkum P.~G., 2010, \apj, 722, 1666

\bibitem[{{Wuyts} {et~al}\mbox{.}(2011){Wuyts}, {F{\"o}rster Schreiber}, {van
  der Wel}, {Magnelli}, {Guo}, {Genzel}, {Lutz}, {Aussel}, {Barro}, {Berta},
  {Cava}, {Graci{\'a}-Carpio}, {Hathi}, {Huang}, {Kocevski}, {Koekemoer},
  {Lee}, {Le Floc'h}, {McGrath}, {Nordon}, {Popesso}, {Pozzi}, {Riguccini},
  {Rodighiero}, {Saintonge}, \& {Tacconi}}]{Wuyts:2011b}
{Wuyts} S. {et~al.}, 2011, \apj, 742, 96

\bibitem[{Wuyts {et~al}\mbox{.}(2009)Wuyts, Franx, Cox, {F{\"o}rster
  Schreiber}, Hayward, Hernquist, Hopkins, Labb{\'e}, Marchesini, Robertson,
  Toft, \& van Dokkum}]{Wuyts:2009b}
Wuyts S. {et~al.}, 2009, \apj, 700, 799

\bibitem[{{Younger} {et~al}\mbox{.}(2010){Younger}, {Fazio}, {Ashby}, {Civano},
  {Gurwell}, {Huang}, {Iono}, {Peck}, {Petitpas}, {Scott}, {Wilner}, {Wilson},
  \& {Yun}}]{Younger:2010}
{Younger} J.~D. {et~al.}, 2010, \mnras, 407, 1268

\bibitem[{{Younger} {et~al}\mbox{.}(2008){Younger}, {Fazio}, {Wilner}, {Ashby},
  {Blundell}, {Gurwell}, {Huang}, {Iono}, {Peck}, {Petitpas}, {Scott},
  {Wilson}, \& {Yun}}]{Younger:2008phys_scale}
{Younger} J.~D. {et~al.}, 2008, \apj, 688, 59

\bibitem[{{Younger} {et~al}\mbox{.}(2009){Younger}, Hayward, Narayanan, Cox,
  Hernquist, \& Jonsson}]{Younger:2009}
{Younger} J.~D., Hayward C.~C., Narayanan D., Cox T.~J., Hernquist L., Jonsson
  P., 2009, \mnras, 396, L66

\bibitem[{{Yun} {et~al}\mbox{.}(2012){Yun}, {Scott}, {Guo}, {Aretxaga},
  {Giavalisco}, {Austermann}, {Capak}, {Chen}, {Ezawa}, {Hatsukade}, {Hughes},
  {Iono}, {Johnson}, {Kawabe}, {Kohno}, {Lowenthal}, {Miller}, {Morrison},
  {Oshima}, {Perera}, {Salvato}, {Silverman}, {Tamura}, {Williams}, \&
  {Wilson}}]{Yun:2012}
{Yun} M.~S. {et~al.}, 2012, \mnras, 420, 957

\bibitem[{{Zemcov} {et~al}\mbox{.}(2010){Zemcov}, {Blain}, {Halpern}, \&
  {Levenson}}]{Zemcov:2010}
{Zemcov} M., {Blain} A., {Halpern} M., {Levenson} L., 2010, \apj, 721, 424

\end{thebibliography}

\label{lastpage}

\end{document}